\PassOptionsToPackage{dvipsnames}{xcolor}

\documentclass[authoryear,final,3p,times,twocolumn]{elsarticle}

\usepackage{xcolor}
\usepackage{colortbl}

\usepackage{amsmath,amssymb,amsfonts,mathtools}
\usepackage{amsthm}

\usepackage{graphicx}
\usepackage{caption}
\usepackage{subcaption}
\usepackage{adjustbox}
\usepackage{tikz}
\usetikzlibrary{arrows,arrows.meta,positioning,calc,shapes.geometric,fit,backgrounds}
\usepackage{booktabs}
\usepackage{multirow}
\usepackage{makecell}
\usepackage{array}
\usepackage{tabularx}
\usepackage{arydshln}
\usepackage{textcomp}

\usepackage{hhline}

\providecommand{\textproc}[1]{%
  \ifmmode
    \text{\textsc{\detokenize{#1}}}%
  \else
    \textsc{\detokenize{#1}}%
  \fi
}

\DeclareRobustCommand{\code}[1]{%
  \ifmmode
    \text{\normalfont\ttfamily\detokenize{#1}}%
  \else
    {\normalfont\ttfamily\detokenize{#1}}%
  \fi
}

\DeclareRobustCommand{\edge}[1]{\code{#1}}
\DeclareRobustCommand{\node}[1]{\code{#1}}

\usepackage{algorithm}
\usepackage[noend]{algpseudocode}

\floatname{algorithm}{Algorithm}

\algrenewcommand\algorithmicrequire{\textbf{Input:}}
\algrenewcommand\algorithmicensure{\textbf{Output:}}

\algrenewcommand\algorithmiccomment[1]{
  \hfill$\triangleright$\,\textit{#1}
}

\algrenewcommand\alglinenumber[1]{
  \scriptsize\color{gray!75!black}#1
}

\algnewcommand{\Stage}[1]{
  \Statex\vspace{1pt}
  \textbf{#1}
  \vspace{1pt}
}

\algnewcommand{\Continue}{
  \State \textbf{continue}
}

\AtBeginDocument{

}

\usepackage{float}
\usepackage{placeins}

\usepackage[most]{tcolorbox}
\usepackage[framemethod=TikZ]{mdframed}

\usepackage{listings}

\usepackage{xurl}

\usepackage[colorlinks=true,
    linkcolor=MidnightBlue,
    citecolor=ForestGreen,
    urlcolor=BrickRed
]{hyperref}

\usepackage{pifont}

\theoremstyle{plain}

\definecolor{titlegray}{HTML}{555555}
\definecolor{framegray}{gray}{0.6}
\definecolor{lightgray}{gray}{0.97}

\lstdefinelanguage{diff}{
  morecomment=[f][\color{gray}]{@@},
  morecomment=[f][\color{gray}]{---},
  morecomment=[f][\color{gray}]{+++},
  morecomment=[f][\color{red}]{-},
  morecomment=[f][\color{green!60!black}]{+},
}
\lstdefinestyle{diff}{
  language=diff,
  basicstyle=\ttfamily\footnotesize,
  columns=fullflexible,
  keepspaces=true,
  showstringspaces=false,
  breaklines=true
}
\definecolor{deepblue}{RGB}{0, 0, 139}

\definecolor{gainGreen}{RGB}{0, 120, 60}

\definecolor{cybg}{gray}{0.965}
\definecolor{cykw}{rgb}{0.13,0.13,0.62}
\definecolor{cycm}{rgb}{0.40,0.55,0.40}

\lstdefinelanguage{Cypher}{
  morekeywords={
    MATCH,OPTIONAL,MERGE,CREATE,DELETE,DETACH,SET,REMOVE,RETURN,WITH,
    WHERE,UNWIND,CALL,YIELD,IN,TRANSACTIONS,OF,ROWS,AS,ON,ORDER,BY,
    LIMIT,SKIP,COLLECT,DISTINCT,CONSTRAINT,INDEX,FOR,REQUIRE,IS,
    UNIQUE,AND,OR,NOT,CONTAINS,count,coalesce,labels,type,head,size,
    toLower
  },
  sensitive=false,
  morecomment=[l]{//},
  morestring=[b]"
}

\lstdefinestyle{cypher}{
  language=Cypher,
  backgroundcolor=\color{cybg},
  basicstyle=\ttfamily\scriptsize,
  keywordstyle=\color{cykw}\bfseries,
  commentstyle=\color{cycm}\itshape,
  breaklines=true,
  columns=fullflexible,
  keepspaces=true,
  showstringspaces=false,
  frame=single,
  framerule=0.3pt,
  xleftmargin=3pt,
  xrightmargin=2pt,
  aboveskip=4pt,
  belowskip=4pt
}

\newtcolorbox{boxnote}{
  colback=white,
  boxrule=0.4pt,
  borderline west={2pt}{0pt}{black!60},
}

\AtBeginDocument{
  
}

\tikzset{
  ast/.style={
    circle,
    draw=black!60,
    fill=white,
    line width=0.8pt,
    minimum size=25mm,
    inner sep=2pt,
    align=center,
    font=\small\sffamily
  }
}

\tikzset{
  ast root/.style={
    ast,
    minimum size=34mm,
    font=\normalsize\sffamily,
    fill=blue!2,
    draw=blue!25!black
  }
}

\tikzset{
  basic type ast/.style={
    ast,
    fill=black!1,
    draw=black!50
  }
}

\tikzset{
  identifier ast/.style={
    ast,
    fill=brown!2,
    draw=brown!25!black
  }
}

\tikzset{
  parameter ast/.style={
    ast,
    fill=blue!2,
    draw=blue!20!black
  }
}

\tikzset{
  return ast/.style={
    ast,
    fill=violet!2,
    draw=violet!20!black
  }
}

\tikzset{
  binaryop ast/.style={
    ast,
    fill=brown!2,
    draw=brown!20!black
  }
}

\tikzset{
  file/.style={
    rectangle,
    rounded corners=3mm,
    draw=green!35!black,
    fill=green!4,
    line width=0.8pt,
    minimum width=31mm,
    minimum height=16mm,
    align=center,
    font=\normalsize\sffamily
  }
}

\tikzset{
  relation/.style={
    -{Latex[length=2.8mm,width=2mm]},
    draw=black!70,
    line width=0.85pt,
    line cap=round
  }
}

\tikzset{
  edge label/.style={
    fill=white,
    inner xsep=3pt,
    inner ysep=1.5pt,
    align=center,
    font=\small\sffamily\bfseries,
    text depth=0pt
  }
}

\tikzset{
  commit node/.style={
    rectangle,
    rounded corners=3mm,
    draw=black,
    fill=white,
    line width=0.9pt,
    minimum width=35mm,
    minimum height=17mm,
    align=center,
    font=\sffamily\large
  }
}

\tikzset{
  file node/.style={
    rectangle,
    rounded corners=3mm,
    draw=black,
    fill=white,
    line width=0.9pt,
    minimum width=31mm,
    minimum height=15mm,
    align=center,
    font=\sffamily\large
  }
}

\tikzset{
  developer node/.style={
    circle,
    draw=black,
    fill=white,
    line width=0.9pt,
    minimum size=25mm,
    align=center,
    font=\sffamily\large
  }
}

\tikzset{
  issue node/.style={
    diamond,
    aspect=1.25,
    draw=black,
    fill=white,
    line width=0.9pt,
    minimum width=27mm,
    minimum height=19mm,
    align=center,
    inner sep=1pt,
    font=\sffamily\large
  }
}

\tikzset{
  relation/.style={
    -{Latex[length=3mm,width=2.2mm]},
    draw=black,
    line width=1pt,
    line cap=round
  }
}

\tikzset{
  relation label/.style={
    fill=white,
    inner xsep=3pt,
    inner ysep=1.5pt,
    align=center,
    font=\sffamily\bfseries\normalsize
  }
}
\usepackage{graphicx}
\usepackage{tikz}

\usetikzlibrary{
  arrows.meta,
  calc,
  fit,
  positioning,
  backgrounds
}

\definecolor{deltaAddC}{RGB}{0,140,0}
\definecolor{deltaRemC}{RGB}{190,0,0}
\definecolor{deltaUpdC}{RGB}{230,120,0}
\definecolor{deltaMovC}{RGB}{110,50,180}

\tikzset{
  deltaAst/.style={
    circle,
    draw=black,
    fill=white,
    line width=0.8pt,
    minimum size=8.8mm,
    inner sep=1pt,
    align=center,
    font=\scriptsize\sffamily
  }
}

\tikzset{
  deltaRoot/.style={
    deltaAst,
    minimum size=10mm
  }
}

\tikzset{
  deltaCommit/.style={
    rectangle,
    rounded corners=2mm,
    draw=black,
    fill=white,
    line width=0.8pt,
    minimum width=17mm,
    minimum height=9mm,
    align=center,
    font=\scriptsize\sffamily
  }
}

\tikzset{
  deltaEdge/.style={
    -{Latex[length=1.8mm,width=1.35mm]},
    draw=black,
    line width=0.75pt
  }
}

\tikzset{
  deltaUpdated/.style={
    deltaAst,
    draw=deltaUpdC,
    line width=1.1pt
  }
}

\tikzset{
  deltaRemoveBox/.style={
    draw=deltaRemC,
    dashed,
    rounded corners=1.5mm,
    line width=0.9pt,
    inner sep=2pt,
    fill=deltaRemC,
    fill opacity=0.12
  }
}

\tikzset{
  deltaAddBox/.style={
    draw=deltaAddC,
    dashed,
    rounded corners=1.5mm,
    line width=0.9pt,
    inner sep=2pt,
    fill=deltaAddC,
    fill opacity=0.12
  }
}

\tikzset{
  deltaMoveBox/.style={
    draw=deltaMovC,
    dashed,
    rounded corners=1.5mm,
    line width=0.9pt,
    inner sep=2pt,
    fill=deltaMovC,
    fill opacity=0.12
  }
}

\tikzset{
  deltaTopLabel/.style={
    fill=white,
    inner sep=0.7pt,
    font=\scriptsize\sffamily\bfseries
  }
}

\tikzset{
  deltaSideTitle/.style={
    font=\small\sffamily\bfseries
  }
}

\tikzset{
  deltaRelationLabel/.style={
    fill=white,
    inner sep=1pt,
    font=\scriptsize\sffamily\bfseries
  }
}

\tikzset{
  deltaLegend/.style={
    rectangle,
    rounded corners=1.3mm,
    draw=black!50,
    dashed,
    inner xsep=3pt,
    inner ysep=2pt,
    align=center,
    font=\scriptsize\sffamily
  }
}

\tikzset{
  deltaTreePanel/.style={
    rounded corners=2mm,
    draw=black!12,
    fill=black!3,
    fill opacity=0.5,
    line width=0.5pt
  }
}

\definecolor{HCFOneHue}{RGB}{0,114,178}
\definecolor{HCGHue}{RGB}{230,159,0}
\definecolor{HCAUCHue}{RGB}{0,158,115}

\journal{}

\begin{document}

\begin{frontmatter}



\title{KG-Commit: A Dynamic Knowledge Graph for Online Just-in-Time Software Defect Prediction}

\author[mce]{Mohsen Hesamolhokama\corref{cor1}}
\ead{hokama@ce.sharif.edu}

\author[mce]{Mohammad Sina Beyrami Aghbash\corref{cor1}}
\ead{mohammadsina.beyrami@sharif.edu}

\author[math]{Behnam Rohani}
\ead{behnam.rohani058@sharif.edu}

\author[mce]{Mohammadamin Fazli}
\ead{fazli@sharif.edu}

\author[mce]{Jafar Habibi}
\ead{jhabibi@sharif.edu}

\cortext[cor1]{These authors contributed equally to this work.}

\address[mce]{Department of Computer Engineering, Sharif University of Technology, Tehran, Iran}
\address[math]{Department of Mathematical Sciences, Sharif University of Technology, Tehran, Iran}

\begin{abstract}
Just-in-time software defect prediction (JIT-SDP) aims to identify risky commits as they arrive and provide developers with timely feedback. This need for low latency has led most approaches to rely on commit-level information and overlook the broader project context in which a change occurs. Incorporating this context is challenging because it requires both efficient retrieval for incoming commits and continual maintenance as the repository evolves. We introduce KG-Commit, a dynamic knowledge graph that incrementally maintains repository history, within-file code structure, and commit semantics as the project evolves. It also uses an AST-delta mechanism to track structural changes between file edits and relies on lightweight graph inference running entirely on CPU. Our evaluation on 11 Apache software projects against six baselines shows that KG-Commit achieves the highest aggregate Macro-F1 ($0.704$), G-Mean ($0.706$), and AUC ($0.809$) using our selected inference pipeline. Under a realistic online protocol, it outperforms LR, HGB, RF, and DeepJIT on all 11 projects, LApredict on 10, and JITLine on 9 projects in Macro-F1, with the aggregate paired difference significant in every case. KG-Commit processes each commit in approximately $1.33$~s, with a cost that remains stable as the graph grows and is compatible with commit rates observed in real-world projects. These findings show that rich project context can be efficiently maintained and exploited for online JIT-SDP.
\end{abstract}



\end{frontmatter}


\section{Introduction}
\label{sec:introduction}
Software underpins nearly every product and service that people rely on, and even a single defect can disrupt a critical system. Ensuring software quality is therefore a persistent concern for development teams, yet the time and resources available for testing and code review are limited. Software defect prediction (SDP) helps teams direct that effort to the right place by estimating which parts of a system are most likely to be faulty~\citep{menzies2007data,lessmann2008benchmarking,hall2012systematic}. Traditional SDP approaches operate at the level of files, modules, or releases~\citep{wang2016automatically,li2017software,zhou2022software}. At this granularity, predictions often arrive only after many changes have accumulated, making it difficult to identify the particular change that introduced a defect and to recover the context in which it was made. Just-in-time defect prediction (JIT-SDP) was introduced to close this gap~\citep{mockus2000predicting,kamei2013large}. It evaluates each commit as soon as it is submitted, so a risky change can be reviewed while its author still remembers the context. Because of this immediate feedback, JIT-SDP has become a natural fit for continuous integration and an active research topic~\citep{zhao2023systematic}.

For more than a decade, JIT-SDP has relied primarily on a compact set of hand-crafted metrics that summarize each commit. These metrics capture properties such as change size, code churn~\citep{nagappan2005use}, the extent to which a change spans files and subsystems, and the experience of its author~\citep{kamei2013large,mcintosh2018fix}. Standard classifiers then use these summaries to estimate defect risk. As deep learning became more prominent, research gradually shifted from hand-crafted commit features to deep representation learning.~\citet{yang2015deep} used a deep belief network to transform conventional metrics into richer features, while DeepJIT and CC2Vec learned representations from commit messages and code diffs~\citep{hoang2019deepjit,hoang2020cc2vec}. Later work extended this direction with line-level defect localization~\citep{pornprasit2021jitline,pornprasit2023deeplinedp}, pre-trained code encoders such as CodeBERT~\citep{ni2022best,jiang2025bicc}, and parameter-efficient adaptation of these models~\citep{abutalib2024parameter}.

What these models largely overlook is the broader project context in which the change occurs~\citep{bryan2023graph,zhao2023systematic}. Knowledge graphs are well suited to representing such structure because they organize information as entities connected by typed relations~\citep{hogan2021knowledge,ji2022survey} and support reasoning and learning over these relations~\citep{nickel2016review}. They have been applied successfully in areas such as recommendation, life sciences, and cybersecurity~\citep{wang2019kgat,maclean2021knowledge,zhao2024survey}. Their use in software engineering, however, remains relatively limited and has focused mainly on organizing software artifacts such as APIs and libraries~\citep{liu2023recommending}. Knowledge graphs remain even less explored in software defect prediction and, to the best of our knowledge, have not been applied to JIT-SDP. This gap is particularly notable because a commit is inherently relational. It modifies specific files and functions, interacts with other parts of the system, is associated with an author, and carries an intent expressed through its message. Dynamic knowledge graphs provide a natural way to capture these evolving relationships~\citep{zhang2024survey,cai2024survey}. By incorporating new events as they occur, they maintain an up-to-date representation of a changing domain~\citep{trivedi2017knowevolve,goel2020diachronic}. Since software evolves incrementally through a stream of commits, this representation is especially well aligned with the online nature of JIT-SDP.

In just-in-time prediction, however, accuracy is of little value if it arrives too late~\citep{kamei2013large}, so richer context must be incorporated at a computational cost that remains practical for an incoming stream of commits. This creates a tension between contextual richness and per-commit processing cost. Project-level context, such as how a changed function is called elsewhere or how a file depends on others, is known to improve defect prediction~\citep{zimmermann2009cross,nam2013transfer}, but reconstructing this information from scratch for every incoming commit can be expensive. To remain efficient, many approaches mainly rely on information contained within the commit itself, such as its message and code diff~\citep{hoang2019deepjit,hoang2020cc2vec,pornprasit2021jitline}. Others pursue efficiency more aggressively through a single inexpensive feature~\citep{zeng2021deep} or lightweight retrieval over historical commits~\citep{sahar2024irjit}. The open question is whether rich, project-specific context can be maintained at a cost compatible with the pace at which commits actually arrive.

Dynamic or real-time knowledge graphs offer a natural way to address this problem. Rather than reconstructing the project state from scratch, the graph is updated incrementally as each commit arrives, with each update affecting only the relevant portion of the graph~\citep{zhao2023incremental}. The accumulated project context can then be queried through a graph query language such as Cypher~\citep{francis2018cypher,angles2017foundations}. This replaces repeated reconstruction of project history with incremental per-commit work. Its practical suitability therefore depends on whether this processing cost remains stable as the project grows and is sufficiently small relative to the rate at which commits arrive.

Building on this idea, we introduce KG-Commit, a dynamic knowledge graph that evolves with the repository and provides project-specific context at inference time. KG-Commit is maintained incrementally in Neo4j and organizes information across three layers: (1) Core layer for repository entities and their relations, (2) AST layer for within-file program structure, and (3) Commit Semantic-Text Graph (CSTG) layer for the semantics of commit messages and code changes. We combine graph-based and feature-based inference over these representations to estimate defect risk and evaluate KG-Commit in terms of predictive performance, query and update cost, and suitability for online JIT-SDP. 

\subsection{Motivating Example}
\label{sec:motivating}
\begin{figure}[!htbp]
\centering
\includegraphics[width=0.90\columnwidth]{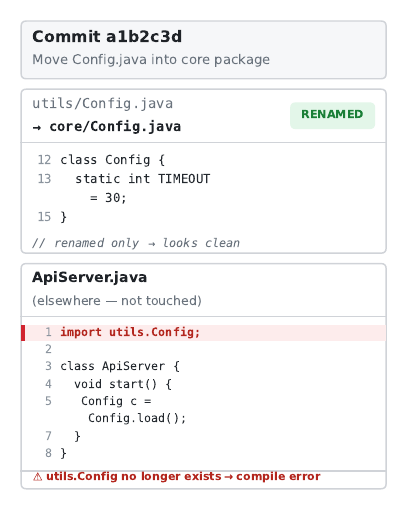}
\caption{A commit moves \texttt{Config.java} from package \texttt{com.app.utils} to \texttt{com.app.core}. The change has no effect inside the file, so it looks clean, but \texttt{ApiServer.java} elsewhere in the project still imports \texttt{com.app.utils.Config} and fails to compile. The defect is not visible from the commit alone.}
\label{fig:motivating}
\end{figure}

Consider a commit that only moves a file, for example relocating \texttt{Config.java} from the \texttt{com.app.utils} package to \texttt{com.app.core} (\autoref{fig:motivating}). Judged on its own, the commit looks harmless. It moves a file without changing its contents, so its change metrics are small and nothing inside the change points to a defect. A developer reviewing only the change, or a model that relies only on commit-level information, would therefore be likely to classify it as clean~\citep{hoang2019deepjit,zeng2021deep}. The commit is nonetheless bug-inducing, because another part of the project still refers to the file through its old package. The file \texttt{ApiServer.java}, which this commit does not touch, still contains \texttt{import com.app.utils.Config;}. After the move, this import refers to a class that no longer exists in that package, causing the build to fail. This defect cannot be detected from the commit in isolation. It becomes visible only when the rest of the project is taken into account. Reliably labeling such a commit therefore requires information about the project context that surrounds the change.

\subsection{Contributions}
The main contributions of this paper are as follows:
\begin{itemize}
\item \textbf{A Practical Solution to Rich Project Context in Online JIT-SDP.} KG-Commit addresses the difficulty of incorporating rich project context without sacrificing the speed required for JIT-SDP. To the best of our knowledge, it is the first knowledge graph designed for JIT-SDP. It represents repository history, code structure, and commit semantics in a Neo4j graph database.
        
\item \textbf{Incremental Maintenance of Within-File Structure through AST Deltas.} We introduce an AST-delta mechanism that tracks how a file's structure changes between edits by recording added, removed, updated, and moved AST nodes. This lets KG-Commit maintain fine-grained within-file structure incrementally as the project evolves.

\item \textbf{Strong Predictive Performance with Efficient and Scalable Deployment.} KG-Commit achieves the highest aggregate Macro-F1, G-Mean, and AUC against six baselines across 11 real-world projects under a fixed online protocol. It outperforms each baseline on at least 9 out of the 11 projects in Macro-F1, with statistically significant aggregate differences against all of them. KG-Commit processes each incoming commit in approximately $1.33$~s on CPU. Lifecycle analysis further shows that per-commit processing remains stable as the graph grows, supporting the practical use of KG-Commit throughout project evolution.
\end{itemize}

\subsection{Paper Outline}
The remainder of this paper is organized as follows.~\autoref{sec:related_work} reviews related work on JIT-SDP, code representation, and knowledge graphs.~\autoref{sec:methodology} presents KG-Commit, its layered representation, incremental maintenance, and inference methods.~\autoref{sec:setup} describes the experimental design and online evaluation protocol.~\autoref{sec:results} reports predictive performance, efficiency and scalability, layer contributions, and inference-channel analysis.~\autoref{sec:discussion} discusses the main findings, deployment considerations, sensitivity to operational constraints, limitations, and future work.~\autoref{sec:validity} addresses threats to validity, and~\autoref{sec:conclusion} concludes the paper.

\section{Related Work}
\label{sec:related_work}
Our work brings together three research lines: JIT-SDP, the representation of source code for defect prediction, and knowledge graphs. We review each line in turn and close by positioning our proposal, which represents the sequence of code changes as a dynamic knowledge graph and queries that graph to predict defects.

\subsection{Just-in-Time Software Defect Prediction}
\label{subsec:jitsdp}
Traditional defect prediction works at coarse levels such as the file, module, or the release~\citep{Menzies2010}. This approach delays feedback and forces developers to inspect large amounts of code. More recently,~\citet{hesamolhokama2025illusion} showed that file overlap and label persistence bias can inflate within-project SDP performance at file granularity. JIT-SDP instead targets the individual commits, so risky changes can be reviewed as soon as they are submitted. The idea goes back to the change-level risk models of~\citet{mockus2000predicting}, and its modern formulation was established by the large-scale empirical study of~\citet{kamei2013large}. That study introduced a set of change-level metrics that is now standard, covering the size, diffusion, history, and developer experience of a change, together with effort-aware evaluation that ranks changes by defect likelihood per unit of inspection effort. Training labels are commonly derived using the SZZ algorithm, which traces bug-fixing commits back to the changes that likely introduced the defect~\citep{sliwerski2005changes}. Early work formulated the problem as a binary classification task, predicting whether each change is clean or buggy~\citep{kim2008classifying}. Later studies examined how such models generalize across projects~\citep{fukushima2014empirical,kamei2016studying} and how they degrade over time, with~\citet{mcintosh2018fix} showing that fix-inducing changes are a moving target.

A first wave of learning-based approaches moved beyond hand-crafted metrics by learning richer representations from traditional change features.~\citet{yang2015deep}, for example, used a deep belief network to transform conventional metrics into higher-level features. Later work shifted toward end-to-end representation learning. DeepJIT~\citep{hoang2019deepjit} learns from commit messages and code changes, while CC2Vec~\citep{hoang2020cc2vec} learns representations of code changes using their associated log messages. Other studies have targeted finer-grained prediction, identifying defect-prone files or lines within a commit~\citep{pascarella2019fine,pornprasit2021jitline}, while complementary work has focused on explaining predictions to developers~\citep{pornprasit2021pyexplainer}. 

More recent approaches have incorporated pre-trained code models. JIT-Fine combines semantic representations from CodeBERT with expert-designed features~\citep{ni2022best}, bi-modal pre-training further improves the representation of code changes~\citep{jiang2025bicc}, and parameter-efficient tuning reduces the cost of adapting such encoders~\citep{abutalib2024parameter}. At the same time, several studies have questioned whether increasingly complex models are always necessary.~\citet{zeng2021deep} show that a simple model based on a single churn feature can match complex approaches, while other work finds benefits in combining expert knowledge with learned semantic representations~\citep{zhou2025bridging,chen2023boosting}. IRJIT similarly demonstrates that a lightweight information-retrieval approach can remain competitive while offering much lower prediction cost~\citep{sahar2024irjit}.

Two practical issues have become increasingly important in JIT-SDP. One is the online nature of the task. Commits arrive as a stream, labels become available only after a verification latency, and the class distribution drifts.~\citet{cabral2023towards} study reliable online JIT-SDP under these conditions, while~\citet{tabassum2023cross} extend the setting to cross-project online learning. The other issue is how a code change is represented. Flat metrics and token sequences largely ignore the relational structure surrounding a change.~\citet{bryan2023graph} address this limitation by constructing contribution graphs over developers and files. The survey by~\citet{zhao2023systematic} likewise identifies richer structural and semantic modeling as an important direction for future work, while methodological studies emphasize the need for rigorous model validation~\citep{tantithamthavorn2017empirical}. 

Our work builds on this perspective but shifts the focus from developer--file relationships to the changed code itself and its broader project context, represented as an evolving knowledge graph.

\subsection{Software Defect Prediction and Code Representation}
\label{subsec:coderep}
Defect prediction has long been studied at the file and module level. Early approaches relied on static code properties, such as size and complexity metrics~\citep{menzies2007data}, while a parallel line of work showed that change and process information, including code churn and properties of the change history, can be even more predictive~\citep{nagappan2005use,moser2008comparative,hassan2009predicting}. Subsequent empirical studies and benchmarks compared classifiers and datasets under common evaluation settings~\citep{lessmann2008benchmarking,hall2012systematic,dambros2012evaluating}. Another substantial body of work examined cross-project defect prediction, where models are transferred across projects when project-specific training data are limited~\citep{zimmermann2009cross,nam2013transfer}. Together, these studies established many of the features, datasets, and evaluation practices that later defect-prediction research builds upon.

A second line of work learns representations from code without relying on hand-crafted metrics. This direction is motivated by the observation that source code exhibits regular and predictable patterns similar to natural language~\citep{hindle2012naturalness}, making distributed representation learning a natural fit~\citep{mikolov2013distributed}.~\citet{wang2016automatically} learn semantic features from token sequences derived from abstract syntax trees (ASTs) using a deep belief network, while convolutional and tree-based models capture syntactic structure~\citep{mou2016convolutional,li2017software,tai2015improved}. 
Other approaches represent programs through AST fragments, including path-based models such as code2vec~\citep{alon2019code2vec} and code2seq~\citep{alon2019code2seq}, and statement-level encoders such as ASTNN~\citep{zhang2019astnn}. More recently, self-supervised pre-training has become a common foundation for code representation learning. Representative models include InferCode~\citep{bui2021infercode}, contextual code embeddings learned with large Transformers~\citep{kanade2020learning}, CodeBERT~\citep{feng2020codebert}, GraphCodeBERT~\citep{guo2021graphcodebert}, and CodeT5~\citep{wang2021codet5}. Related work also augments sequence models with global relational information to better capture dependencies that are difficult to express through token order alone~\citep{hellendoorn2020global}.

Since trees and token sequences capture only part of a program’s structure, a growing body of work represents code as graphs and learns over them with graph neural networks~\citep{scarselli2009graph,wu2021comprehensive}.~\citet{allamanis2018learning} augment ASTs with data-flow and apply gated graph neural networks~\citep{li2016gated}, while Devign~\citep{zhou2019devign} combines AST, control-flow, and data-flow information in a unified graph for vulnerability detection. Graph-based representations have also been adopted in defect prediction.~\citet{zhou2022software} combine ASTs with class-dependency networks,~\citet{liu2023semantic} enrich code representations with external knowledge, and~\citet{vsikic2022graph} apply graph neural networks over ASTs for module-level prediction. 

Related work further shows the value of augmenting syntax trees with flow information for code analysis tasks~\citep{wang2020detecting}. Recent surveys reflect this broader shift toward deep and structurally informed representations in defect prediction~\citep{zain2023application}. These approaches demonstrate the value of modeling program structure, but they largely remain confined to individual files or modules and do not simultaneously capture project-wide context and satisfy the online constraints of JIT-SDP.

\subsection{Knowledge Graphs and Their Applications}
\label{subsec:kg}
Knowledge graphs (KGs) represent information through entities and typed relations, providing a foundation for reasoning, retrieval, and learning~\citep{hogan2021knowledge,ji2022survey,nickel2016review}. Much of the literature focuses on learning continuous representations of these entities and relations so that symbolic knowledge can be incorporated into machine learning models~\citep{ji2022survey,wang2017knowledge}. Early approaches include tensor-factorization methods such as RESCAL~\citep{nickel2011three} and translational models such as TransE~\citep{bordes2013translating}, TransH~\citep{wang2014knowledge}, and TransR~\citep{lin2015learning}, which represent relations through transformations in the embedding space. Later models improve expressiveness through bilinear or complex-valued scoring functions, including DistMult~\citep{yang2015embedding}, ComplEx~\citep{trouillon2016complex}, and RotatE~\citep{sun2019rotate}. Neural approaches extend this line further, with models such as ConvE~\citep{dettmers2018convolutional}, relational graph convolutional networks~\citep{schlichtkrull2018modeling}, and composition-based message-passing methods~\citep{vashishth2020composition} learning over multi-relational graph structure.

Knowledge graphs have been applied across a wide range of domains. Large general-purpose resources such as YAGO, Freebase, and Wikidata demonstrated that relational knowledge can be organized and queried at web scale~\citep{suchanek2007yago,bollacker2008freebase,vrandecic2014wikidata}. In recommender systems, knowledge graphs help alleviate data sparsity and support more interpretable recommendations, as illustrated by KGAT~\citep{wang2019kgat}, RippleNet~\citep{wang2018ripplenet}, and the broader literature surveyed by~\citet{guo2022survey}. They have also been used to integrate biomedical evidence for drug discovery~\citep{maclean2021knowledge}, consolidate cyber-threat intelligence~\citep{zhao2024survey}, and incorporate structured knowledge into language models~\citep{zhang2019ernie}. 

Dynamic knowledge graphs further extend this framework by modeling how entities and relations evolve over time~\citep{zhang2024survey,cai2024survey,trivedi2017knowevolve,goel2020diachronic,jin2020renet,lacroix2020tensor}. Within software engineering, knowledge graphs have mainly been used to
organize and connect software artifacts such as APIs and libraries
\citep{wang2023application,liu2023recommending}. For example,~\citet{liu2023recommending} embed an API knowledge graph to recommend analogous APIs. In contrast, knowledge graphs remain largely unexplored across software defect prediction. To the best of our knowledge, no prior work has used knowledge graphs or comparable relational representations to incorporate project-wide context into JIT-SDP.

\section{Methodology}
\label{sec:methodology}

\subsection{Motivation}
\label{sec:motivation}
JIT-SDP must quickly provide feedback as a change is submitted~\citep{kamei2013large}. Existing models therefore rely largely on features extracted from the current commit, such as code diffs and churn metrics~\citep{kamei2013large,hoang2019deepjit,nagappan2005use}, but these features omit broader repository context, including information around changed lines, developer experience, and developer--file relationships~\citep{kondo2020context,cho2022developer,bryan2023graph}. Maintaining and efficiently integrating such context is challenging in a just-in-time setting~\citep{kondo2020context,bryan2023graph}. KG-Commit instead maintains it incrementally in Neo4j. Each commit updates only the affected graph neighborhood, so the accumulated project context is available at prediction time through bounded Cypher queries.

\subsection{Problem Formulation}
\label{sec:problem_formulation}
Let a software project $P$ be represented as a chronologically ordered sequence of commits $C = \{c^{(1)}, c^{(2)}, \dots, c^{(n)}\}$. Each commit $c^{(i)} \in C$ represents an evolutionary step that alters the state of the project from $s^{(i-1)}$ to $s^{(i)}$. Each commit $c^{(i)}$ targets a set of pre-existing files $F^{(i)}_{\text{prev}} = \{f_1, f_2, \dots, f_m\}$ and produces a set of modified or newly created files $F^{(i)}_{\text{new}} = \{f'_1, f'_2, \dots, f'_p\}$. Therefore, a commit may add, remove, modify, or rename/move one or more files in the project.

\subsubsection{Defect Modeling}
\label{sec:defect-modeling}
A commit $c^{(i)}$ is defined as bug-inducing if and only if it introduces a defect into the codebase that requires a subsequent fix. Formally, we define the ground-truth binary label $y^{(i)} \in \{0, 1\}$ for each commit with index $i \ge 1$ as
\begin{align}
y^{(i)} = \begin{cases} 1 & \text{if } \exists f \in F^{(i)}_{\text{new}} \text{ such that } f \text{ is defective} \\ 0 & \text{otherwise} \end{cases}
\end{align}
In our experiments, we use the labels provided by the ApacheJIT dataset~\citep{keshavarz2022apachejit}, where bug-inducing commits are identified using an SZZ-style algorithm.
\subsubsection{Operational Constraints}
\label{sec:operational-constraints}
To simulate real-world deployment settings, the online formulation introduces two constraints: a warm-up period $t$ and a gap $g$~\citep{lee2024neurojit,song2023validity}. The warm-up period requires $t$ commits to be processed before evaluation begins, restricting the target commit $c^{(j)}$ to indices where $j > t$. The gap $g$ accounts for the operational delay in defect identification (e.g., via the SZZ algorithm), meaning that when commit $c^{(j)}$ arrives, historical ground-truth labels are only resolved and available up to commit $c^{(j-g-1)}$.

\subsubsection{Online Predictive Objective}
\label{sec:online-objective}
Given a target commit $c^{(j)}$, where $j > t$, the objective is to estimate the conditional probability $\hat{y}^{(j)}$ that the commit is bug-inducing based on the current project state $s^{(j-1)}$:

\begin{align}
\hat{y}^{(j)} = \hat{\mathbb{P}}\left( y^{(j)} = 1 \;\middle|\; c^{(j)}, s^{(j-1)} \right)
\end{align}

\subsubsection{Knowledge Graph State Representation and Evolution}
\label{sec:kg_evolution}
To avoid recomputing deep repository histories on demand, the abstract project state and its evolutionary history are captured by an incrementally updated knowledge graph $\mathcal{K}^{(j)}$. Upon the arrival of a target commit $c^{(j)}$, its corresponding entities and relations are immediately updated and inserted into the graph before classification occurs. This state evolution follows a recurrence relation defined by a state transition function $\Phi$:
\begin{align}
\mathcal{K}^{(j)} = \Phi\left(\mathcal{K}^{(j-1)}, c^{(j)}, y^{(j-g-1)}\right)
\end{align}
where $c^{(j)}$ provides the immediate modifications for the incoming commit, and $y^{(j-g-1)}$ injects the historical ground-truth label newly exposed by the expiration of the gap window. By unrolling this recurrence, $\mathcal{K}^{(j)}$ preserves the complete history of previous project states and commits:
\begin{align}
\mathcal{K}^{(j)} = \text{Encode}\left( \{s^{(i)}\}_{i=1}^{j}, \; \{c^{(i)}\}_{i=1}^{j}, \; \{y^{(i)}\}_{i=1}^{j-g-1} \right),
\end{align}
in the sense that all this information can later be accessed or reconstructed. In our case, this is made possible by using appropriate Cypher queries against the graph database.

Because $\mathcal{K}^{(j)}$ embeds this entire history alongside the unlabeled properties of the recent commits, the predictive objective is resolved by conditioning on the new graph state:

\begin{align}
\hat{y}^{(j)} = \hat{\mathbb{P}}\left( y^{(j)} = 1 \;\middle|\; \mathcal{K}^{(j)} \right)
\end{align}

\subsection{KG-Commit}
\label{sec:kg-commit}
\begin{figure*}[!htbp]
    \centering
    \includegraphics[width=0.8\textwidth]
        {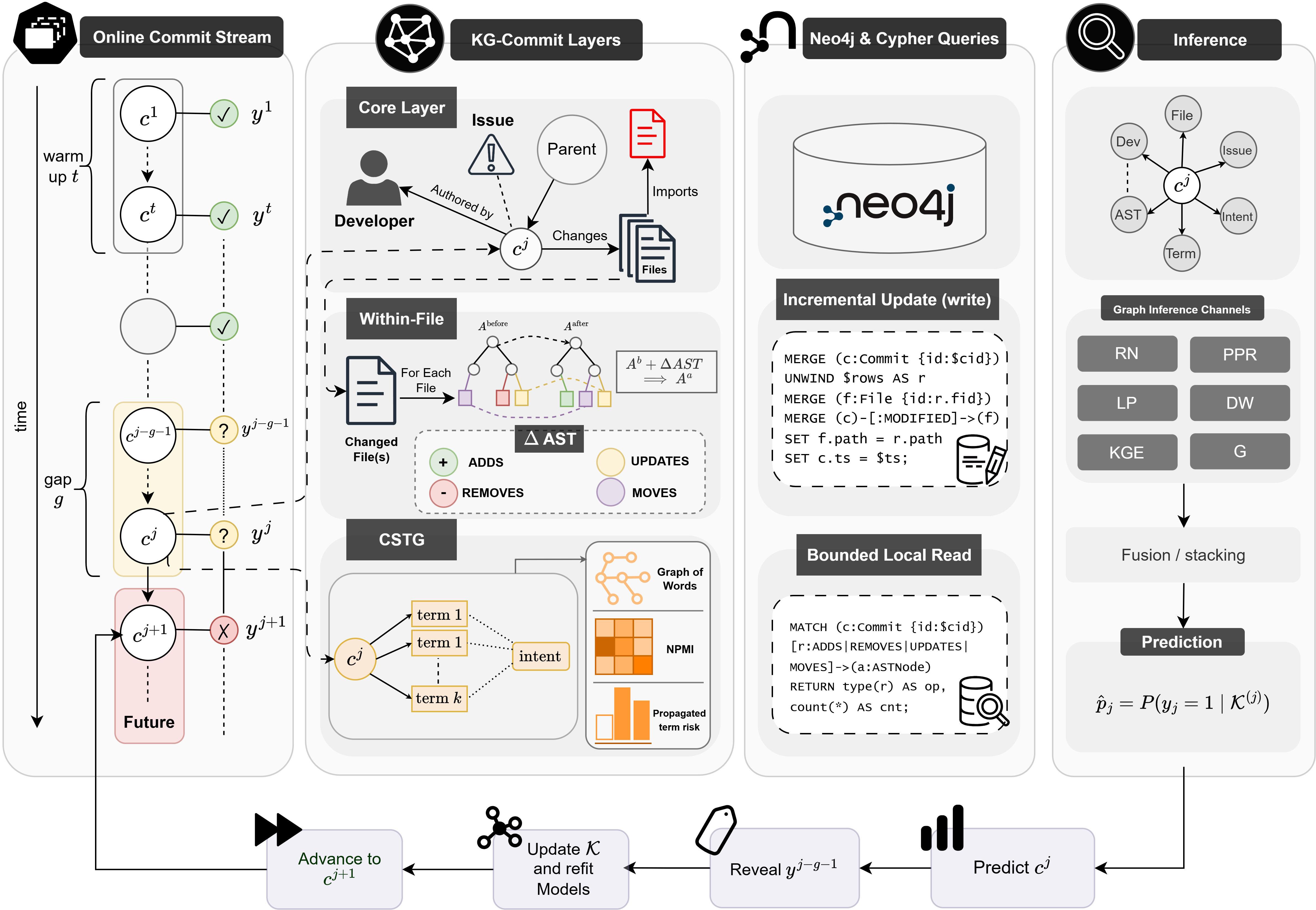}
    \caption{Overview of KG-Commit's online workflow.}
    \label{fig:kgcommit-overview}
\end{figure*}

KG-Commit models the evolving repository as a three-layer knowledge graph whose entity and relation schema is summarized in~\autoref{tab:schema}, while the overall online workflow is illustrated in~\autoref{fig:kgcommit-overview}. The Core layer captures repository-level development context, the AST layer represents within-file program structure through an AST and updates it incrementally using delta extraction, and the Commit Semantic-Text Graph (CSTG) adds information from commit messages and diff text through graph-of-words statistics, term associations, and propagated defect risks. Commits are processed chronologically and inserted into Neo4j. The graph is updated through Cypher operations, and only information available at that point in time is queried for prediction. The resulting graph is then used by the inference channels through a compact projection, with selected context taken from the full graph, and their outputs are combined to estimate the defect risk of the current commit.

\begin{table}[!htbp]
\centering
\caption{Typed schema of KG-Commit.}
\label{tab:schema}

\scriptsize
\setlength{\tabcolsep}{3pt}
\renewcommand{\arraystretch}{1.0}

\begin{tabularx}{\columnwidth}{
    >{\raggedright\arraybackslash}p{0.29\columnwidth}
    >{\raggedright\arraybackslash}p{0.30\columnwidth}
    >{\raggedright\arraybackslash}X
}
\toprule
\textbf{Relation} &
\textbf{Source $\rightarrow$ Target} &
\textbf{Description}
\\
\midrule

\multicolumn{3}{l}{\textbf{Core layer}}\\
\multicolumn{3}{l}{
\textit{Entities:}
\node{Commit}, \node{Developer}, \node{File}, \node{Issue}
}\\
\midrule

\edge{AUTHORED_BY} &
\node{Commit} $\rightarrow$ \node{Developer} &
Records commit authorship.
\\

\edge{PARENT_OF} &
\node{Commit} $\rightarrow$ \node{Commit} &
Represents repository history.
\\

\edge{FIXES_ISSUE} &
\node{Commit} $\rightarrow$ \node{Issue} &
Links a commit to an issue.
\\[5pt]

\makecell[l]{\edge{ADDED}/\edge{MODIFIED}/\\
\edge{DELETED}
} &
\node{Commit} $\rightarrow$ \node{File} &
Records file-level changes.
\\[5pt]

\makecell[l]{\edge{RENAMED_FROM}/\\
\edge{RENAMED_TO}
} &
\node{Commit} $\rightarrow$ \node{File} &
Records old/new paths of a rename.
\\

\edge{IMPORTS} &
\node{File} $\rightarrow$ \node{File} &
Plain import.
\\
\midrule

\multicolumn{3}{l}{\textbf{AST layer}}\\
\multicolumn{3}{l}{
\textit{Entities:}
\node{ASTNode}, \node{Package}
}\\
\midrule
\edge{HAS_AST} &
\node{File} $\rightarrow$ \node{ASTNode} &
Attaches a file to its AST root.
\\

\edge{AST_CHILD} &
\node{ASTNode} $\rightarrow$ \node{ASTNode} &
Ordered AST parent--child relation.
\\

\makecell[l]{\edge{ADDS}/\edge{REMOVES}/\\
\edge{UPDATES}/\edge{MOVES}
} &
\node{Commit} $\rightarrow$ \node{ASTNode} &
Records the structural delta.
\\[5pt]

\edge{RESOLVES_TO} &
\node{ASTNode} $\rightarrow$ \node{File} &
Resolves a static-member import to its owning type file.
\\

\makecell[l]{\edge{RESOLVES_TO_}\\
\edge{PACKAGE}} &
\node{ASTNode} $\rightarrow$ \node{Package} &
Resolves a wildcard import to its package.
\\

\edge{CONTAINS} &
\node{Package} $\rightarrow$ \node{File} &
Records files belonging to a package.
\\
\midrule

\multicolumn{3}{l}{\textbf{CSTG layer}}\\
\multicolumn{3}{l}{
\textit{Entities:}
\node{Term}, \node{Intent}
}\\
\midrule

\edge{MENTIONS} &
\node{Commit} $\rightarrow$ \node{Term} &
Records terms occurring in a change.
\\

\edge{COOCCURS} &
\node{Term} $\rightarrow$ \node{Term} &
NPMI-based term association.
\\

\edge{HAS_INTENT} &
\node{Commit} $\rightarrow$ \node{Intent} &
Assigns the commit's intent.
\\

\edge{GROUNDS_IN} &
\node{Term} $\rightarrow$ \node{ASTNode} &
Grounds code terms in matching AST leaves.
\\

\bottomrule
\end{tabularx}
\end{table}

\subsubsection{Cypher Queries}
KG-Commit is stored in Neo4j and accessed through Cypher, its declarative graph query language~\citep{francis2018cypher}. Cypher specifies graph patterns through labeled nodes and typed relationships, such as \texttt{(c:Commit)} and \texttt{-[:MODIFIED]->}. During construction, \texttt{MERGE} avoids duplicate nodes and relationships by creating them only when they are not already present, while \texttt{UNWIND} enables multiple relationships to be processed in a single database request.~\autoref{fig:core-query} shows a Core-layer query that upserts a \texttt{Commit} node and batch-creates its \texttt{MODIFIED} relationships to \texttt{File} nodes. At prediction time, \texttt{MATCH} retrieves the local neighborhood required by the inference methods.

\begin{figure}[!htbp]
\centering
\begin{minipage}{0.95\linewidth}
\begin{lstlisting}[
style=cypher
]
// (1) upsert an entity node by primary key
MERGE (e:Commit {id: $pk_value})
  ON CREATE SET e += $props
  ON MATCH  SET e += $props

// (2) batch-create the process edges from one source node
MATCH (src:Commit {id: $src_id})
UNWIND $targets AS t
  MERGE (tgt:File {id: t.target_id})
  MERGE (src)-[r:MODIFIED]->(tgt)
    ON CREATE SET r += t.props
\end{lstlisting}
\end{minipage}
\caption{Core-layer Cypher query: (1) upserts a \texttt{Commit} node by its identifier, and (2) batch-creates \texttt{MODIFIED} relationships to the corresponding \texttt{File} nodes.}
\label{fig:core-query}
\end{figure}

\subsubsection{Layer 1: Core Components}
\label{sec:kg-core}
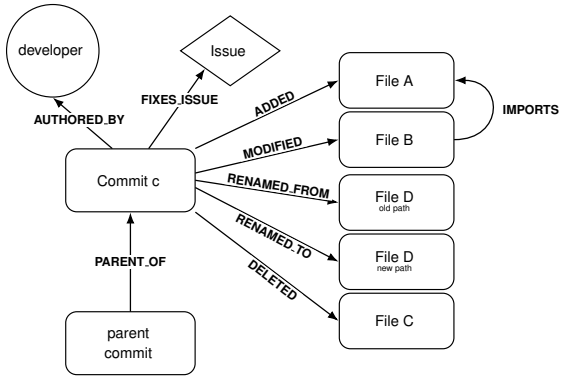
\begin{figure}[!htbp]
\centering

\resizebox{0.95\columnwidth}{!}{
\begin{tikzpicture}

\node[commit node] (commit) at (0,0)
{
  Commit c
};

\node[commit node] (parent) at (0,-4.4)
{
  \shortstack{
    parent\\[2pt]
    commit
  }
};

\node[developer node] (developer) at (-2.1,3.5)
{
  developer
};

\node[issue node] (issue) at (2.7,3.5)
{
  Issue
};

\node[file node] (fileA) at (7.2,2.7)
{
  File A
};

\node[file node] (fileB) at (7.2,1.1)
{
  File B
};

\node[file node] (fileOld) at (7.2,-0.6)
{
  \shortstack{
    File D\\[-1pt]
    \scriptsize old path
  }
};

\node[file node] (fileNew) at (7.2,-2.2)
{
  \shortstack{
    File D\\[-1pt]
    \scriptsize new path
  }
};

\node[file node] (fileC) at (7.2,-3.8)
{
  File C
};

\draw[relation]
  (parent.north)
  --
  node[
    relation label,
    pos=0.50
  ] {PARENT\_OF}
  (commit.south);

\draw[relation]
  ([xshift=-5mm]commit.north)
  --
  node[
    relation label,
    above=-5pt,
    pos=0.55
  ] {AUTHORED\_BY}
  (developer.south);

\draw[relation]
  ([xshift=5mm]commit.north)
  --
  node[
    relation label,
    above=-2pt,
    pos=0.55
  ] {FIXES\_ISSUE}
  (issue.south west);

\draw[relation]
  (commit.north east)
  --
  node[
    relation label,
    sloped,
    above,
    pos=0.57
  ] {ADDED}
  (fileA.west);

\draw[relation]
  ([yshift=2mm]commit.east)
  --
  node[
    relation label,
    sloped,
    above,
    pos=0.55
  ] {MODIFIED}
  (fileB.west);

\draw[relation]
  (commit.east)
  --
  node[
    relation label,
    sloped,
    above,
    pos=0.57
  ] {RENAMED\_FROM}
  (fileOld.west);

\draw[relation]
  ([yshift=-2mm]commit.east)
  --
  node[
    relation label,
    sloped,
    below,
    pos=0.57
  ] {RENAMED\_TO}
  (fileNew.west);

\draw[relation]
  (commit.south east)
  --
  node[
    relation label,
    sloped,
    below,
    pos=0.57
  ] {DELETED}
  (fileC.west);

\draw[relation]
  (fileB.east)
  .. controls +(1.35,0) and +(1.35,0) ..
  node[
    relation label,
    right=5pt,
    pos=0.50
  ] {IMPORTS}
  (fileA.east);

\end{tikzpicture}%
}

\caption{Illustration of the Core layer.}
\label{fig:core-layer}

\end{figure}

The Core layer captures the repository’s high-level entities and their basic relations (see~\autoref{fig:core-layer}). Each commit becomes a \node{Commit} node, and its bug-inducing label is attached when it becomes available; it is linked to its author (\edge{AUTHORED_BY}), its children (\edge{PARENT_OF}), the issues referenced in its message (\edge{FIXES_ISSUE}), and the files it changes, with the edge type recording the kind of change (\edge{ADDED}, \edge{MODIFIED}, \edge{DELETED}, \edge{RENAMED_FROM}, or \edge{RENAMED_TO}). The layer also captures cross-file dependencies through \edge{IMPORTS}. These edges are maintained over time so that both current and historical dependencies remain represented.

\subsubsection{Layer 2: Within-File AST Subgraph}
\label{sec:kg-ast}
The AST layer captures the syntactic structure of Java files using a deterministic AST representation. Each file is parsed with \texttt{javalang}, and its parse tree is traversed in pre-order to construct the corresponding subgraph. For a file with repository-relative path $rel$, the $i$-th visited node is assigned the identifier ${rel}\code{::A}{[i]}$. Each \node{ASTNode} records its \texttt{javalang} type, syntactic category, and a compact value when applicable, such as an identifier, literal, or operator; non-blank string children are represented as \code{Identifier} leaves. To ensure that the same source produces the same graph across runs, ordered child containers keep their original order, whereas unordered containers are canonically sorted. The resulting AST is attached to its \node{File} through \edge{HAS_AST}, while \edge{AST_CHILD} relations encode the parent--child relations and child positions.

\paragraph{AST matching and delta extraction}
For a modified file, let $A^b$ and $A^a$ be the stored and newly parsed ASTs. The matcher constructs a partial injective correspondence $\mu:V(A^b)\rightharpoonup V(A^a)$ in three stages. First, each node is assigned a bottom-up fingerprint: Let
$u_1,\ldots,u_k$ be the ordered children of $v$, and define $h(v)= H\!\left(\ell(v)\Vert h(u_1)\Vert\cdots\Vert h(u_k)\right)$ where $\ell(v)=\code{ast_type}(v)\Vert\code{value}(v)$. Here, $\Vert$ denotes an unambiguous concatenation and $H$ is the MD5
message-digest algorithm~\citep{rivest1992md5}. This bottom-up construction is commonly known as subtree hashing or syntax-tree fingerprinting~\citep{chilowicz2009syntax}. Equal fingerprints identify candidate identical subtrees. Larger candidate subtrees are processed first, with source position used to break ties, following the general strategy of prioritizing large isomorphic subtrees in AST differencing~\citep{falleri2014fine}. Second, matches are propagated to unmatched parents of the same AST type until a fixpoint is reached. Third, remaining nodes are matched by $(\code{ast_type},\code{line},\code{column})$ whenever this tuple is unique. The unmatched and changed nodes define the delta:
\begin{align}
  R &= V(A^b)\setminus\operatorname{dom}(\mu),\\
  A &= V(A^a)\setminus\operatorname{ran}(\mu),\\
  U &=
  \left\{
    (b,a)\in\mu \;\middle|\;
    \begin{aligned}
      &\code{ast_type}(b)\neq\code{ast_type}(a)\\
      &{}\lor\ \code{value}(b)\neq\code{value}(a)
    \end{aligned}
  \right\},\\
  M &=
  \left\{
    (b,a)\in\mu \;\middle|\;
    \begin{aligned}
      &\operatorname{par}(b)\in\operatorname{dom}(\mu)\\
      &{}\land\
      \mu(\operatorname{par}(b))
      \neq\operatorname{par}(a)
    \end{aligned}
  \right\}.
\end{align}

\begin{figure}[!htbp]
    \centering
    \resizebox{0.9\columnwidth}{!}{
    \begin{tikzpicture}[x=1cm,y=1cm]

    \node[deltaSideTitle] at (-2.95,7.65)
      {Before $A^b$};

    \node[deltaSideTitle] at (2.95,7.65)
      {After $A^a$};

    \node[deltaCommit] (commit) at (0,6.95)
    {
      \shortstack{
        \textbf{Commit}\\
        $c^{(j)}$
      }
    };

    \node[
      deltaTopLabel,
      text=deltaRemC
    ] (remLbl) at (-2.15,5.90)
      {REMOVES};

    \node[
      deltaTopLabel,
      text=deltaUpdC
    ] (updLbl) at (-0.72,5.90)
      {UPDATES};

    \node[
      deltaTopLabel,
      text=deltaAddC
    ] (addLbl) at (0.72,5.90)
      {ADDS};

    \node[
      deltaTopLabel,
      text=deltaMovC
    ] (movLbl) at (2.15,5.90)
      {MOVES};

    \draw[
      -{Latex[length=1.8mm,width=1.35mm]},
      draw=deltaRemC,
      line width=0.8pt
    ]
      ([xshift=-6mm]commit.south)
      --
      (remLbl.north);

    \draw[
      -{Latex[length=1.8mm,width=1.35mm]},
      draw=deltaUpdC,
      line width=0.8pt
    ]
      ([xshift=-2mm]commit.south)
      --
      (updLbl.north);

    \draw[
      -{Latex[length=1.8mm,width=1.35mm]},
      draw=deltaAddC,
      line width=0.8pt
    ]
      ([xshift=2mm]commit.south)
      --
      (addLbl.north);

    \draw[
      -{Latex[length=1.8mm,width=1.35mm]},
      draw=deltaMovC,
      line width=0.8pt
    ]
      ([xshift=6mm]commit.south)
      --
      (movLbl.north);

    \node[deltaRoot] (Bb0) at (-2.95,4.65)
      {\shortstack{Block\\B0}};

    \node[deltaAst] (BA1) at (-4.90,3.05)
      {\shortstack{Assign\\A1}};

    \node[deltaAst] (BA2) at (-2.80,3.05)
      {\shortstack{Assign\\A2}};

    \node[deltaAst] (BR1) at (-1.00,3.05)
      {\shortstack{Return\\R1}};

    \draw[deltaEdge] (Bb0) -- (BA1);
    \draw[deltaEdge] (Bb0) -- (BA2);
    \draw[deltaEdge] (Bb0) -- (BR1);

    \node[deltaAst] (BIdx) at (-5.45,1.35)
      {\shortstack{Id\\x}};

    \node[deltaUpdated] (BLit1) at (-4.35,1.35)
      {\shortstack{Lit\\1}};

    \draw[deltaEdge] (BA1) -- (BIdx);
    \draw[deltaEdge] (BA1) -- (BLit1);

    \node[deltaAst] (BIdy) at (-3.35,1.35)
      {\shortstack{Id\\y}};

    \node[deltaAst] (BLit2) at (-2.25,1.35)
      {\shortstack{Lit\\2}};

    \draw[deltaEdge] (BA2) -- (BIdy);
    \draw[deltaEdge] (BA2) -- (BLit2);

    \node[deltaAst] (BP1) at (-1.00,1.35)
      {\shortstack{Print\\P1}};

    \node[deltaAst] (BRemId) at (-1.00,-0.25)
      {\shortstack{Id\\y}};

    \draw[deltaEdge] (BR1) -- (BP1);
    \draw[deltaEdge] (BP1) -- (BRemId);

    \node[deltaAst] (BOp) at (-2.80,-1.95)
      {\shortstack{BinOp\\Op$_b$}};

    \node[deltaAst] (BOpA) at (-3.35,-3.55)
      {\shortstack{Id\\a}};

    \node[deltaAst] (BOpB) at (-2.25,-3.55)
      {\shortstack{Id\\b}};

    \draw[deltaEdge]
      (BA2.south)
      --
      (BOp.north);

    \draw[deltaEdge] (BOp) -- (BOpA);
    \draw[deltaEdge] (BOp) -- (BOpB);

    \node[deltaRoot] (Ab0) at (2.95,4.65)
      {\shortstack{Block\\B0}};

    \node[deltaAst] (AA1) at (0.90,3.05)
      {\shortstack{Assign\\A1}};

    \node[deltaAst] (AA2) at (2.90,3.05)
      {\shortstack{Assign\\A2}};

    \node[deltaAst] (AR1) at (4.90,3.05)
      {\shortstack{Return\\R1}};

    \draw[deltaEdge] (Ab0) -- (AA1);
    \draw[deltaEdge] (Ab0) -- (AA2);
    \draw[deltaEdge] (Ab0) -- (AR1);

    \node[deltaAst] (AIdx) at (0.35,1.35)
      {\shortstack{Id\\x}};

    \node[deltaUpdated] (ALit42) at (1.45,1.35)
      {\shortstack{Lit\\42}};

    \draw[deltaEdge] (AA1) -- (AIdx);
    \draw[deltaEdge] (AA1) -- (ALit42);

    \node[deltaAst] (AIdy) at (2.35,1.35)
      {\shortstack{Id\\y}};

    \node[deltaAst] (ALit2) at (3.45,1.35)
      {\shortstack{Lit\\2}};

    \draw[deltaEdge] (AA2) -- (AIdy);
    \draw[deltaEdge] (AA2) -- (ALit2);

    \node[deltaAst] (AOp) at (4.15,-1.00)
      {\shortstack{BinOp\\Op$_b$}};

    \node[deltaAst] (AOpA) at (3.60,-2.60)
      {\shortstack{Id\\a}};

    \node[deltaAst] (AOpB) at (4.70,-2.60)
      {\shortstack{Id\\b}};

    \draw[deltaEdge] (AR1) -- (AOp);
    \draw[deltaEdge] (AOp) -- (AOpA);
    \draw[deltaEdge] (AOp) -- (AOpB);

    \node[deltaAst] (AP2) at (6.00,-1.00)
      {\shortstack{Print\\P2}};

    \node[deltaAst] (AAddId) at (6.00,-2.60)
      {\shortstack{Id\\x}};

    \draw[deltaEdge] (AR1) -- (AP2);
    \draw[deltaEdge] (AP2) -- (AAddId);

    \begin{pgfonlayer}{background}

    \coordinate (BeforePanelNW) at (-5.95,5.15);
    \coordinate (BeforePanelSE) at (-0.40,-4.05);
    
    \coordinate (AfterPanelNW)  at ( -0.05,5.15);
    \coordinate (AfterPanelSE)  at ( 6.45,-4.05);
    
    \node[
      deltaTreePanel,
      fit=(BeforePanelNW)(BeforePanelSE)
    ] (BeforePanel) {};
    
    \node[
      deltaTreePanel,
      fit=(AfterPanelNW)(AfterPanelSE)
    ] (AfterPanel) {};
    
      \node[
        deltaRemoveBox,
        fit=(BP1)(BRemId)
      ] (Rbox) {};

      \node[
        deltaMoveBox,
        fit=(BOp)(BOpA)(BOpB)
      ] (BMoveBox) {};

      \node[
        deltaMoveBox,
        fit=(AOp)(AOpA)(AOpB)
      ] (AMoveBox) {};

      \node[
        deltaAddBox,
        fit=(AP2)(AAddId)
      ] (Ibox) {};

      \draw[
        -{Latex[length=1.9mm,width=1.4mm]},
        draw=deltaUpdC,
        line width=1pt,
        densely dashed
      ]
        (BLit1.north east)
        ..
        controls (-3.20,2.35) and (0.80,2.35)
        ..
        (ALit42.north west);

    \end{pgfonlayer}

    \node[
      deltaRelationLabel,
      text=deltaUpdC
    ] at (-0.75,2.33)
      {$U$};

    \draw[
      -{Latex[length=1.9mm,width=1.4mm]},
      draw=deltaMovC,
      line width=1pt,
      densely dashed
    ]
      (BMoveBox.south east)
      ..
      controls (-1.20,-4.55) and (2.20,-4.55)
      ..
      (AMoveBox.south west);

    \node[
      deltaRelationLabel,
      text=deltaMovC
    ] at (0.55,-4.53)
      {$M$};

    \node[deltaLegend] at (0,-5.35)
    {
      \textcolor{deltaAddC}{\textbf{A}} = adds
      \qquad
      \textcolor{deltaRemC}{\textbf{R}} = removes
      \qquad
      \textcolor{deltaUpdC}{\textbf{U}} = updates
      \qquad
      \textcolor{deltaMovC}{\textbf{M}} = moves
    };

    \end{tikzpicture}%
    }
    \caption{Simplified illustration of AST deltas between the before tree $A^b$ and the after tree $A^a$.}
    \label{fig:delta-ast}
\end{figure}
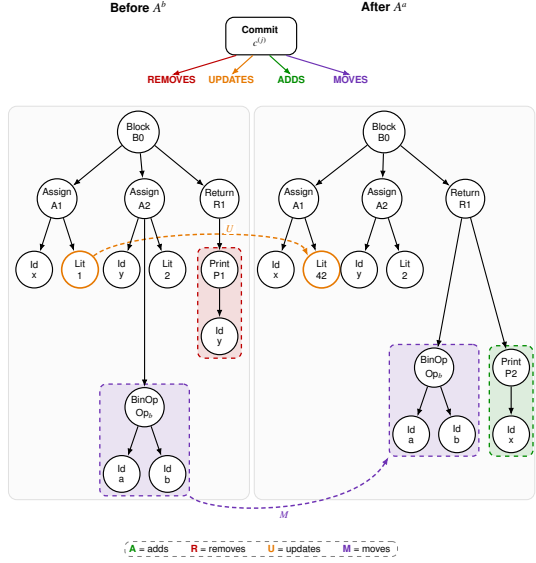

Here, $R$, $A$, $U$, and $M$ denote \edge{REMOVES}, \edge{ADDS}, \edge{UPDATES}, and \edge{MOVES} respectively.~\autoref{fig:delta-ast} illustrates the resulting correspondence and edit classes.

\paragraph{Online delta maintenance}
Commits are processed in chronological order while the system preserves, for each tracked file, its current AST and the mapping from parse identifiers to persistent graph identifiers. Added files are materialized once, files first encountered as modifications are initialized from their parent revision, and deleted files have their live nodes marked inactive. During a modification, matched nodes keep their identifiers, inserted nodes receive new identifiers, and removed nodes remain queryable through \code{alive}$=\textsc{false}$. The live \edge{AST_CHILD} relations are updated in place, while the commit stores only the resulting \edge{ADDS}, \edge{REMOVES}, \edge{UPDATES}, and \edge{MOVES} relations. Thus, storage and update costs follow the amount of syntactic change.

\paragraph{Alternative within-file representations}
Besides AST, we evaluate control-flow graph (CFG), data-flow graph (DFG), and program dependence graph (PDG) as alternative within-file representations. Each representation is mapped to the same \edge{ADDS}, \edge{REMOVES}, \edge{UPDATES}, and \edge{MOVES} delta vocabulary used for AST, where CFG, DFG, and PDG use a graph-specific ordinal matcher rather than the tree-specific AST matcher (see~\autoref{app:alternative-subgraphs} for details).

\subsubsection{Layer 3: Commit Semantic-Text Graph (CSTG)}
\label{sec:kg-cstg}
\begin{figure}[!htbp]
    \centering
    \includegraphics[width=0.9\columnwidth]
        {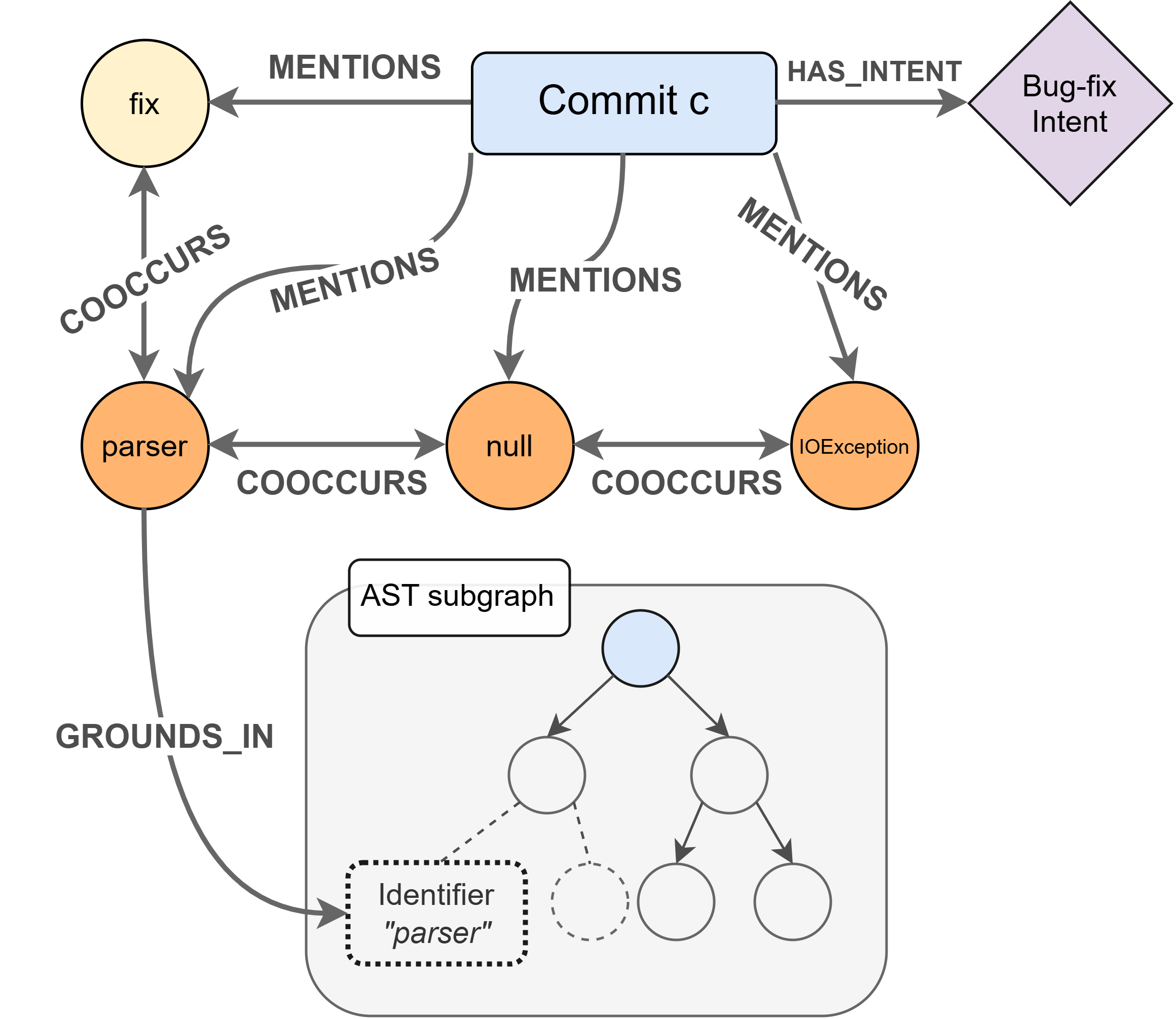}
    \caption{CSTG schema for a commit. Darker term nodes indicate higher propagated defect risk.}
    \label{fig:cstg-schema}
\end{figure}

The third layer represents the semantics of a change using its commit message and diff text. For each commit $c$, we concatenate the message with lexical tokens from changed lines to form $x_c$. The text is tokenized and lower-cased, camel-case identifiers are split, and defect-relevant identifiers such as \texttt{*Exception} and \texttt{*Error} are preserved. Terms are assigned types in $\{\text{code},\text{bug},\text{action},\text{error},\text{natural-language}\}$. The extraction rules for these types are summarized in~\autoref{tab:cstg-term-types}. Each commit yields a graph-of-words whose terms are linked to the commit through weighted \edge{MENTIONS} relations. Across previously observed commits, associated terms are connected through \edge{COOCCURS}, and each term receives a defect-risk estimate obtained from past observations and propagated through this term-association graph. The resulting CSTG schema is illustrated in~\autoref{fig:cstg-schema}.

\begin{table}[!htbp]
\centering
\caption{Term types used by CSTG and their extraction rules.}
\label{tab:cstg-term-types}
\small
\setlength{\tabcolsep}{4pt}
\renewcommand{\arraystretch}{0.8}
\begin{tabularx}{0.9\columnwidth}{lX}
\toprule
\textbf{Type} & \textbf{Extraction rule} \\
\midrule

\textit{error}
&
Identifiers denoting exception or error types, including identifiers ending in \texttt{Exception} or \texttt{Error}.
\\
\midrule
\textit{code}
&
Program identifiers recognized from camel-case, snake-case, or dotted forms.
\\
\midrule
\textit{bug}
&
Terms belonging to a fixed defect vocabulary, including concepts such as
\texttt{npe}, \texttt{leak}, \texttt{race}, \texttt{deadlock}, and
\texttt{regression}.
\\
\midrule
\textit{action}
&
Terms belonging to a fixed change-action vocabulary, such as
\texttt{add}, \texttt{fix}, \texttt{refactor}, and \texttt{revert}.
\\
\midrule
\textit{natural-language}
&
Remaining lexical terms, including subwords obtained by splitting compound program identifiers.
\\

\bottomrule
\end{tabularx}
\end{table}

Commits are also linked to \node{Intent} nodes through \edge{HAS_INTENT}. Each commit receives exactly one intent from $\{\textsc{fix},\textsc{feat},\textsc{refactor},\textsc{test},\textsc{docs},\textsc{perf},\textsc{revert},\textsc{other}\}$. Intent is assigned by a deterministic keyword rule over the lower-cased commit message; occurrences of the keywords associated with each class are counted, the highest-scoring class is selected, and \textsc{other} is returned when no keyword matches. Repeated occurrences contribute repeatedly to the class score. The complete taxonomy is given in~\autoref{tab:cstg-intent-taxonomy}. The \edge{GROUNDS_IN} relation provides the bridge from CSTG to the AST layer (\autoref{sec:kg-ast}). It is created only for \textit{code} \node{Term} nodes and connects them to live AST leaves whose stored value exactly equals the term text. Deleted AST nodes are excluded, and the number of matching leaves is capped for highly common identifiers.

\begin{table}[!htbp]
\centering
\caption{Intent taxonomy and keyword rules used by CSTG. Truncated stems such as \texttt{optimi} and \texttt{deprecat} intentionally match multiple inflected forms.}
\label{tab:cstg-intent-taxonomy}
\small
\setlength{\tabcolsep}{5pt}
\renewcommand{\arraystretch}{0.8}
\begin{tabularx}{0.9\columnwidth}{lX}
\toprule
\textbf{Intent} &
\textbf{Keywords} \\
\midrule

\textsc{fix}
&

\texttt{fix}, \texttt{bug}, \texttt{issue}, \texttt{error},
\texttt{fault}, \texttt{defect}, \texttt{npe}, \texttt{crash},
\texttt{fail}, \texttt{correct}, \texttt{resolve}, \texttt{patch},
\texttt{wrong}, \texttt{broken}
\\

\midrule

\textsc{feat}
&

\texttt{add}, \texttt{feature}, \texttt{implement}, \texttt{introduce},
\texttt{support}, \texttt{new}, \texttt{allow}, \texttt{enable},
\texttt{provide}
\\

\midrule

\textsc{refactor}
&

\texttt{refactor}, \texttt{cleanup}, \texttt{clean up},
\texttt{simplify}, \texttt{rename}, \texttt{reorganize},
\texttt{restructure}, \texttt{tidy}, \texttt{inline}, \texttt{extract},
\texttt{deprecat}
\\

\midrule

\textsc{test}
&

\texttt{test}, \texttt{junit}, \texttt{assert}, \texttt{testcase},
\texttt{coverage}, \texttt{spec}
\\

\midrule

\textsc{docs}
&

\texttt{doc}, \texttt{documentation}, \texttt{javadoc},
\texttt{readme}, \texttt{comment}, \texttt{license}
\\

\midrule

\textsc{perf}
&

\texttt{perf}, \texttt{performance}, \texttt{optimi},
\texttt{speed}, \texttt{faster}, \texttt{cache}, \texttt{latency}
\\

\midrule

\textsc{revert}
&

\texttt{revert}, \texttt{rollback}, \texttt{roll back},
\texttt{undo}, \texttt{back out}
\\

\midrule

\textsc{other}
&
No keyword from the preceding classes occurs.
\\

\bottomrule
\end{tabularx}
\end{table}

For each commit $c$, the distinct normalized terms $V_c$ form a weighted graph-of-words in which terms co-occurring within a window of $w$ tokens are connected~\citep{rousseau2013graph}, preserving local dependencies that a bag-of-words model discards~\citep{rousseau2015main}. Each term receives a weighted TextRank score $\mathrm{TR}_c(t)$ with damping factor $d$~\citep{mihalcea2004textrank}, and its \edge{MENTIONS} weight is $m_c(t)=\mathrm{TR}_c(t)\log((|C_{\mathrm{tr}}|+1)/(\mathrm{df}(t)+1))$, combining within-commit centrality with corpus-level rarity as a TextRank-based variant of graph term-weight IDF~\citep{rousseau2013graph}. 

\begin{figure}[!htbp]
    \centering
    \includegraphics[width=0.98\columnwidth]
        {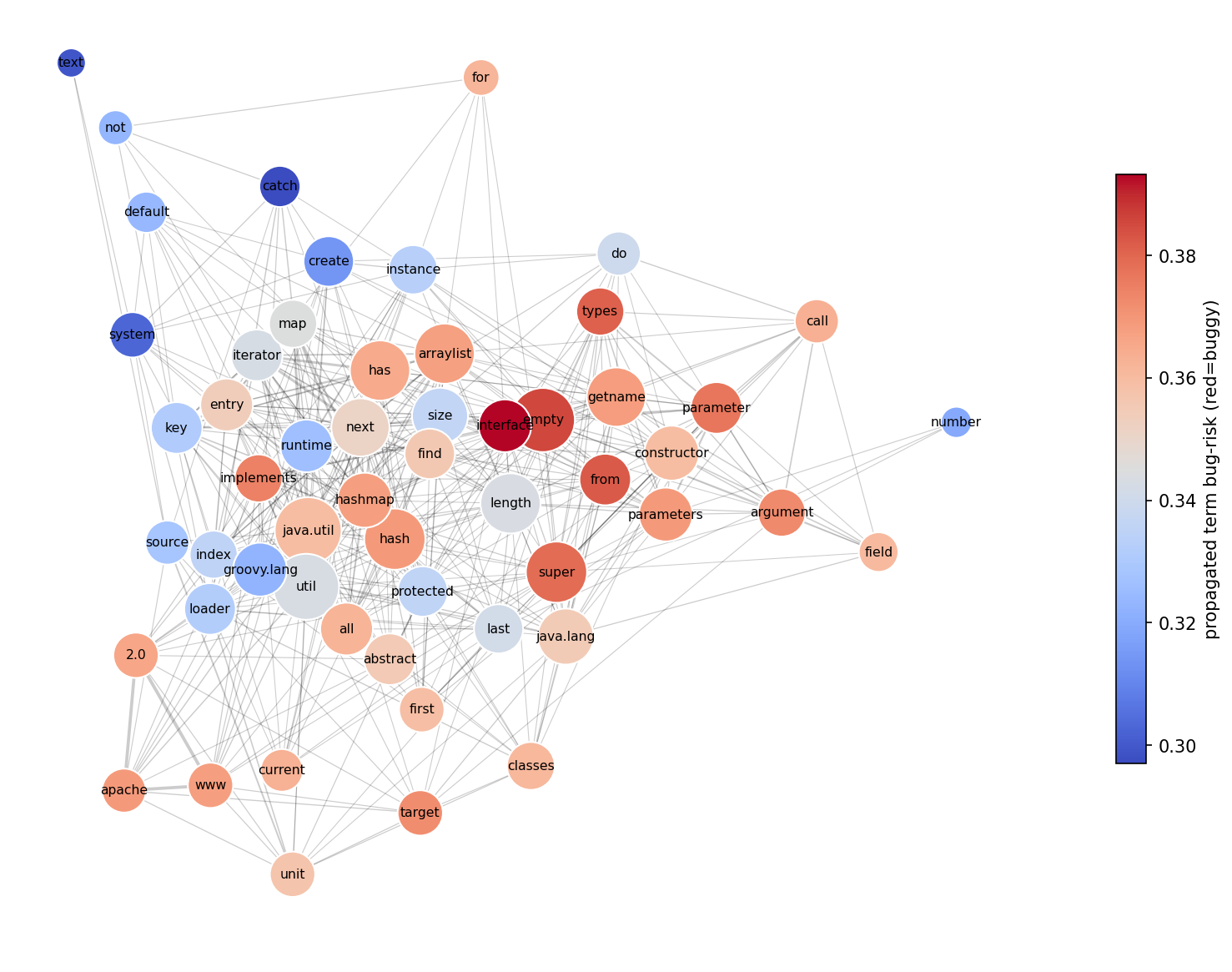}
    \caption{Example of global CSTG term graph. Edges denote retained positive NPMI associations, and node color indicates propagated term risk $r^{\star}(t)$.}
    \label{fig:cstg-term-graph}
\end{figure}

Across training commits, term associations are measured by NPMI~\citep{church1990word,bouma2009normalized}; a \edge{COOCCURS} edge is kept only when its support is at least $n_{\min}$ and its NPMI is at least $\tau_{\mathrm T}$. Each term is initialized with smoothed defect risk $r^{(0)}(t)=(\mathrm{bugdf}(t)+\lambda\bar y)/(\mathrm{df}(t)+\lambda)$ and propagated over the row-normalized term graph as $r^\star=(1-\alpha_{\mathrm T})r^{(0)}+\alpha_{\mathrm T}P_{\mathrm T}r^\star$, imposing smoothness over associated terms~\citep{belkin2006manifold}. CSTG finally summarizes this information through the weighted textual-risk prior $\rho(c)=\sum_t m_c(t)r^\star(t)/\sum_t m_c(t)$ and type-specific semantic masses $M_\kappa(c)=\sum_{\operatorname{type}(t)=\kappa}m_c(t)$ for the five term types defined above. An example of the resulting global term graph is shown in~\autoref{fig:cstg-term-graph}.

\subsubsection{Inference over KG-Commit}
\label{sec:kg-infer}
We use five graph-based inference channels: Relational Neighbor (RN)~\citep{macskassy2003probabilistic}, Personalized PageRank (PPR)~\citep{haveliwala2002topic}, clamped Label Propagation (LP)~\citep{zhu2003semi}, DeepWalk matrix factorization (DW)~\citep{perozzi2014deepwalk}, and DistMult knowledge-graph embedding (KGE)~\citep{yang2015embedding}. A sixth, feature-based $G$ channel uses the semantic statistics derived from components in~\autoref{sec:kg-cstg} (see~\autoref{app:g-channel-features} for full construction details).

\subsubsection{Projection for Lightweight Graph Inference}
\label{sec:kg-projection}
The five graph-based inference methods operate on a compact projection of KG-Commit. At commit $j$, this projection summarizes the neighborhood of each commit in the full graph as a weighted bipartite graph $\mathcal{P}^{(j)}$. RN, PPR, LP, DW, and KGE operate on this smaller graph, keeping graph inference practical on CPU. The construction of the projection and the corresponding inference procedures are described in~\autoref{app:commit-hub-projection}.

The projection does not include every relation in the full graph. For example, cross-file \edge{IMPORTS} relations remain in $\mathcal{K}^{(j)}$ and are not part of $\mathcal{P}^{(j)}$. We therefore also derive a small traversal/global-context (TGC) vector from the full graph. For a commit $c$, we denote this vector by
\[
    \mathbf{T}(c)=
    \left[
        \mathbf{T}_{0}(c),
        \mathbf{T}_{1,\mathrm{dep}}(c),
        \mathbf{T}_{1,\mathrm{dry}}(c),
        \mathbf{T}_{P}(c),
        \mathbf{M}_{1}(c)
    \right].
\]
The grouped vector covers the files changed by the commit ($\mathbf{T}_{0}(c)$), their untouched importers and importees, split by direction ($\mathbf{T}_{1,\mathrm{dep}}(c)$ and $\mathbf{T}_{1,\mathrm{dry}}(c)$), files in the same package ($\mathbf{T}_{P}(c)$), and their architectural position in the dependency graph ($\mathbf{M}_{1}(c)$), summarized using PageRank and $k$-core~\citep{brin1998anatomy,batagelj2003core}. The extracted values are cached and used together with the projection-based inference scores. The individual TGC features and their computation are described in~\autoref{app:tgc-features}.

\paragraph{Score fusion} We use logistic-regression stacking~\citep{wolpert1992stacked} to combine the different inference outputs. Let $\mathrm{F}\subseteq\{\mathrm{RN},\mathrm{PPR},\mathrm{LP},\mathrm{DW},\mathrm{KGE}\}$ denote a selected set of graph-based channels, with $s_m(c)$ denoting the score of channel $m$. We use $\mathrm{F}$ for their stacked combination, $\mathrm{G}$ for the CSTG channel, $\mathrm{F}+\mathrm{G}$ for their joint model, and $\mathrm{F}+\mathrm{G}+\mathrm{T}$ when the TGC features $\mathbf{T}(c)$ are also included. The full model is
\[
\widehat{y}(c)=
\sigma\left(
\theta_0
+\sum_{m\in F}\theta_m s_m(c)
+\theta_G G(c)
+\boldsymbol{\theta}_{T}^{\top}\mathbf{T}(c)
\right),
\]
where $\theta_0$ is the intercept, $\theta_m$ and $\theta_G$ are the coefficients of the graph and CSTG scores, and $\boldsymbol{\theta}_{T}$ contains the coefficients of the TGC features. Removing the corresponding terms gives the $F$ and $F+G$ variants. For deployment, a fixed overall subset $F_{\mathrm{ov}}\subseteq F$ may be used across projects (see~\autoref{sec:setup-protocol-section}).

\section{Experimental Design}
\label{sec:setup}
The objective of this study is to determine whether KG-Commit can make rich, project-wide context practically usable for online JIT-SDP. To achieve this objective, we formulate four main research questions as follows:

\subsection{Research Questions}

\begin{itemize}
    \item \textbf{RQ1.} \emph{What is the predictive performance of KG-Commit in Just-in-Time Software Defect Prediction?}
    \vspace{1em}
    \item \textbf{RQ2.} \emph{How well does KG-Commit maintain computational efficiency and scalability across the real-time evolution of different software projects?}
    \vspace{1em}
    \item \textbf{RQ3.} \emph{How do architectural choices regarding within-file subgraph representations and distinct KG-Commit layers affect the predictive performance?}
    \vspace{1em}
    \item \textbf{RQ4.} \emph{How does KG-Commit's predictive performance change across different inference channels?}
\end{itemize}

\subsection{Dataset}
\label{sec:dataset}
We perform the evaluation on 11 software projects from ApacheJIT~\citep{keshavarz2022apachejit}. The selected projects, shown in~\autoref{tab:projects}, contain 78,206 evaluated commits, of which 22,740 are bug-inducing and 55,466 are clean, corresponding to an overall bug-inducing rate of approximately 29.1\%. Labels are produced using the SZZ algorithm~\citep{sliwerski2005changes} and refined by linking fixes to issue reports and filtering trivial edits with GumTree over abstract syntax trees, following established filtering procedures~\citep{mcintosh2018fix}. Each commit is described by fourteen fields: a commit identifier, the binary bug-inducing label, and twelve change metrics that follow the definitions of~\citet{kamei2013large}. Because these metrics are highly right-skewed, we apply a logarithmic transform. 

\begin{table}[!htbp]
\centering
\caption{ApacheJIT projects used in the evaluation~\citep{keshavarz2022apachejit}. Percentages give the ratio of bug-inducing commits to total evaluated commits.}
\label{tab:projects}
\small
\setlength{\tabcolsep}{4pt}
\begin{tabular}{lrrr}
\toprule
Project & Bug-inducing & Clean & Total \\
\midrule
ActiveMQ  & 1{,}404 (23\%) & 4{,}722  & 6{,}126 \\
Camel     & 3{,}078 (14\%) & 19{,}617 & 22{,}695 \\
Cassandra & 3{,}117 (38\%) & 5{,}042  & 8{,}159 \\
Flink     & 2{,}811 (25\%) & 8{,}648  & 11{,}459 \\
Groovy    & 1{,}614 (20\%) & 6{,}445  & 8{,}059 \\
HBase     & 3{,}782 (43\%) & 4{,}945  & 8{,}727 \\
Hive      & 4{,}223 (62\%) & 2{,}619  & 6{,}842 \\
Kafka     & 1{,}115 (47\%) & 1{,}269  & 2{,}384 \\
Spark     & 632 (43\%)     & 833      & 1{,}465 \\
Zeppelin  & 622 (43\%)     & 829      & 1{,}451 \\
Zookeeper & 342 (41\%)     & 497      & 839 \\
\midrule
\textbf{Total}
& \textbf{22{,}740 (29.1\%)}
& \textbf{55{,}466}
& \textbf{78{,}206} \\
\bottomrule
\end{tabular}
\end{table}

\subsection{Baselines}
\label{sec:setup-baselines-section}

\paragraph{Logistic Regression (LR)} It is the classical change-metric baseline used in JIT-SDP~\citep{kamei2013large}. It operates on the twelve ApacheJIT change metrics after the transformations described in~\autoref{sec:dataset} and uses $\ell_2$-regularized logistic regression with balanced class weights.

\paragraph{Histogram Gradient Boosting (HGB)} It uses the same twelve change metrics but replaces the linear decision function with a histogram-based gradient-boosting classifier. It provides a non-linear change-metric baseline while keeping the input information identical to LR. Class weighting is used to account for label imbalance.

\paragraph{Random Forest (RF)} RF uses the same twelve ApacheJIT change metrics and the same feature-extraction pipeline as LR and HGB, differing only in the classifier.

\paragraph{LApredict~\citep{zeng2021deep}} This baseline deliberately uses only the number of added lines. We implement it using the same class-weighted logistic-regression family as LR.

\paragraph{DeepJIT~\citep{hoang2019deepjit}} Our implementation follows the official DeepJIT architecture, with separate convolutional encoders for the commit message and changed code, a hierarchical convolution over changed code lines, concatenation of the message and code representations, a 512-unit fully connected layer, and a sigmoid output. The implementation is adapted to our online evaluation setting.

\paragraph{JITLine~\citep{pornprasit2021jitline}} This method combines the twelve expert change metrics with token information extracted from the code diff and trains a random-forest classifier on the resulting representation. The implementation is adapted to our online evaluation setting.

\subsection{Experimental Protocol}
\label{sec:setup-protocol-section} 
Evaluation follows an online protocol. For each incoming commit, a prediction is produced using only the information at that point in the stream. The labels enter the historical state only after they become available under the gap parameter $g$ introduced in~\autoref{sec:operational-constraints}. Let $K$ denote the warm-up fraction and $t=\lfloor KN\rfloor$ for a project with $N$ commits. Commits before $t$ initialize the graph. The refit window $M$ determines how often model components are refitted, while commits are still processed one by one. We use a shared setting of $K=0.05$ and $g=50$ for KG-Commit and baselines, and $M=200$ for all learnable models that need a refit. Threshold-dependent metrics use an online operating point. The threshold is initialized at $0.5$ and, after the first $300$ predictions, is re-estimated every $150$ commits to maximize Macro-F1 on past commits.

Throughout~\autoref{sec:results}, $F_{\mathrm{ov}}$ denotes the fixed graph fusion ${\mathrm{RN},\mathrm{PPR}}\subseteq F$ (\autoref{sec:kg-infer}); the rationale for this choice is examined in~\autoref{sec:rq4}. The primary KG-Commit configuration augments $F_{\mathrm{ov}}$ with the semantic CSTG channel $G$ and the traversal/global-context feature block $T$, forming $F_{\mathrm{ov}}{+}G{+}T$ (\autoref{sec:kg-projection}). Unlike the graph channels, $G$ requires sufficient history to construct a vocabulary and is therefore unreliable during cold start. The S@200 policy addresses this problem by using $F_{\mathrm{ov}}$ alone for the first 200 evaluated commits after warm-up and then switching once to $F_{\mathrm{ov}}{+}G$ for the remainder of the stream, as discussed further in~\autoref{sec:switch-s200}. The addition of $T$ does not change the definition of this switch. S@200 continues to denote the one-time introduction of the history-dependent configuration and is distinct from the periodic refit interval $M=200$. Unless an experiment explicitly performs a controlled ablation or sensitivity analysis, the reported KG-Commit result corresponds to $F_{\mathrm{ov}}{+}G{+}T$. For compactness, tables and figures may denote this configuration as $F{+}G{+}T$, with $F=F_{\mathrm{ov}}$.

\paragraph{Reproducibility of Results Across Different Runs} We assessed seed sensitivity across five runs. The $F_{\mathrm{ov}}{+}G$ backbone reproduces to numerical precision, and the added TGC extraction and \texttt{liblinear} stacking head are deterministic for a fixed commit stream, so $T$ does not introduce an additional source of random variation. Graph-only fusions containing learned channels show greater run-to-run variation, as examined in~\autoref{sec:rq4}. LR and LApredict are deterministic, HGB shows negligible variation ($3.9\times10^{-4}$ in Macro-F1), and JITLine-online is evaluated with a fixed random state. Only RF and DeepJIT exhibit noticeable seed sensitivity, with mean Macro-F1 standard deviations of $0.0097$ and $0.0102$, respectively. Given this limited variation,~\autoref{sec:results} reports aggregate performance only.

\subsection{Experimental Settings}
\label{sec:setup-settings-section}

\autoref{tab:hyperparams} summarizes the model configurations and hyperparameters used throughout the experiments, while~\autoref{tab:environments} reports the corresponding execution environments and software stacks. The reported settings are fixed across projects, and parameters not listed in the tables keep their implementation defaults. KG-Commit and all non-neural baselines are executed in the same CPU environment. DeepJIT~\citep{hoang2019deepjit}, which requires repeated training of convolutional neural components, is implemented and executed separately in a Kaggle notebook using a single NVIDIA Tesla T4 GPU. Because DeepJIT is executed on different hardware, its runtime is reported for completeness only and should be excluded from hardware-matched runtime comparisons.

\begin{table}[!htbp]
\centering
\caption{Model hyperparameters and configurations used for KG-Commit and the evaluated baselines.}
\label{tab:hyperparams}
\footnotesize
\setlength{\tabcolsep}{3pt}
\renewcommand{\arraystretch}{0.8}
\adjustbox{max width=0.9\columnwidth}{
\begin{tabularx}{\columnwidth}{lX}
\toprule
\textbf{Component / Parameter} & \textbf{Value} \\
\midrule

\textit{KG-Commit graph inference} & \\

\quad \textit{Relational Neighbor (RN)} & \\
\qquad Neighborhood depth & $1$ hop \\
\qquad Learned embedding & None \\

\quad \textit{Personalized PageRank (PPR)} & \\
\qquad Restart probability $\alpha$ & $0.15$ \\
\qquad Power iterations & $40$ \\

\quad \textit{Label Propagation (LP)} & \\
\qquad Propagation iterations & $3$ \\

\quad \textit{DeepWalk (DW)} & \\
\qquad Matrix representation & PPMI \\
\qquad Factorization & Truncated SVD \\
\qquad Embedding dimension & $64$ \\

\quad \textit{DistMult (KGE)} & \\
\qquad Embedding dimension & $32$ \\
\qquad Optimizer & SGD \\
\qquad Learning rate $\eta$ & $0.05$ \\
\qquad Epochs per refit & $3$ \\
\qquad Negative samples per positive & $3$ \\
\qquad Triple-update chunk size & $2{,}048$ \\

\quad \textit{Classifier heads} & \\
\qquad Classifier & Logistic Regression \\
\qquad Solver & \texttt{lbfgs} \\
\qquad Class weighting & \texttt{balanced} \\
\qquad Maximum iterations & $1000$\\

\midrule

\textit{CSTG semantic channel} & \\
\quad TextRank damping factor & $0.85$ \\
\quad TextRank iterations & $30$ \\
\quad Minimum term co-occurrence & $2$ \\
\quad Risk-prior smoothing constant & $5$ \\
\quad Hashed-text dimension & $2^{18}$ \\
\quad Typed semantic-mass dimension & $5$ \\

\midrule

\textit{Traversal / global-context channel (TGC)} & \\
\quad Feature groups & $\mathbf{T}_{0}$, $\mathbf{T}_{1,\mathrm{dep}}$, $\mathbf{T}_{1,\mathrm{dry}}$, $\mathbf{T}_{P}$, $\mathbf{M}_{1}$ \\
\quad Dependency traversal depth & $1$ hop, direction-separated \\
\quad Architectural position & PageRank, $k$-core \\
\quad Feature scaling & raw and expanding-window $z$-score \\
\quad $z$-score warm-up & $30$ commits \\
\quad Tier extraction & once per project, cached \\
\quad Stacking head & Logistic Regression \\
\quad Penalty & $\ell_1$, $C = 0.2$ \\
\quad Solver & \texttt{liblinear} \\
\quad Class weighting & \texttt{balanced} \\
\quad Feature standardization & \texttt{StandardScaler} \\

\midrule

\textit{Baseline models} & \\

\quad \textit{Logistic Regression (LR)} & \\
\qquad Penalty & $\ell_2$ \\
\qquad Class weighting & \texttt{balanced} \\

\quad \textit{Histogram Gradient Boosting (HGB)} & \\
\qquad Classifier & \texttt{HistGradientBoosting} \\
\qquad Class weighting & \texttt{balanced} \\

\quad \textit{LApredict} & \\
\qquad Classifier & Logistic Regression \\
\qquad Class weighting & \texttt{balanced} \\

\quad \textit{JITLine} & \\
\qquad Classifier & Random Forest \\

\quad{DeepJIT} & \\
\qquad Optimizer & Adam \\
\qquad \texttt{embedding\_dim} & $64$ \\
\qquad \texttt{filter\_sizes} & $(1,2,3)$ \\
\qquad \texttt{num\_filters} & $64$ \\
\qquad \texttt{hidden\_units} & $512$ \\
\qquad \texttt{dropout\_keep\_prob} & $0.5$ \\
\qquad \texttt{learning\_rate} & $10^{-4}$ \\
\qquad \texttt{l2\_reg\_lambda} / weight decay & $10^{-5}$ \\
\qquad \texttt{epochs} & $10$ \\

\bottomrule
\end{tabularx}
}
\end{table}

\begin{table}[!htbp]
\centering
\caption{Execution environments, hardware resources, and software versions used in the experiments.}
\label{tab:environments}
\footnotesize
\setlength{\tabcolsep}{3pt}
\renewcommand{\arraystretch}{0.8}
\adjustbox{max width=0.9\columnwidth}{
\begin{tabularx}{\columnwidth}{lX}
\toprule
\textbf{Component / Resource} & \textbf{Value} \\
\midrule

\textit{KG-Commit execution environment} & \\
\quad Processor & Intel Core i7-13620H \\
\quad CPU cores / threads & $10$ / $16$ \\
\quad CPU base frequency & $2.4$\,GHz \\
\quad Memory & $16$\,GB RAM \\
\quad GPU usage & None (CPU-only) \\

\midrule

\textit{KG-Commit software stack} & \\
\quad Python & 3.13.5 \\
\quad Neo4j Community & 2026.05 \\
\quad \texttt{neo4j} Python driver & 6.2.0 \\
\quad Git & 2.45.1 \\
\quad javalang & 0.13.0 \\
\quad NumPy & 2.2.6 \\
\quad SciPy & 1.16.2 \\
\quad scikit-learn & 1.7.2 \\
\quad pandas & 2.3.2 \\
\quad networkx & 3.5 \\

\midrule

\textit{DeepJIT execution environment} & \\
\quad Platform & Kaggle Notebook \\
\quad Accelerator & NVIDIA Tesla T4 \\
\quad GPU compute capability & $7.5$ (\texttt{sm\_75}) \\

\midrule

\textit{DeepJIT software stack} & \\
\quad Python & 3.12.13 \\
\quad PyTorch & 2.10.0+\texttt{cu128} \\
\quad CUDA runtime & 12.8 \\
\bottomrule
\end{tabularx}
}
\end{table}

\subsection{Evaluation Metrics}
\label{sec:setup-metrics-section}
Given the class imbalance of bug-inducing commits, we primarily report Macro-F1, AUC, and G-Mean $=\sqrt{\mathrm{TPR}\cdot\mathrm{TNR}}$ as class-balanced measures. We also use two effort-aware metrics with inspection effort defined as $\mathtt{la}+\mathtt{ld}$, including the normalized $P_{\mathrm{opt}}$, which measures how closely the effort-ranked inspection order approaches the optimal buggy-first, least-effort-first ranking, and $\mathrm{ACC@20\%LOC}$~\citep{kamei2013large}, which measures the proportion of bug-inducing commits found within the top 20\% of cumulative changed lines.

Average rows in~\autoref{sec:results} report an unweighted macro-average over projects and a micro-average over their pooled predictions. For runtime and lifecycle measurements in \autoref{sec:rq2}, we report the mean, standard deviation (SD), median, and selected percentiles where appropriate. P05, P50, and P95 denote the 5th, 50th, and 95th percentiles, respectively, representing the lower tail, median, and upper tail of the observed distribution.

\section{Experimental Results}
\label{sec:results}

\subsection{RQ1: Predictive Performance against Baselines}
\label{sec:rq1}
\begin{figure}[!htbp]
\centering
\begin{subfigure}[t]{0.9\columnwidth}
    \centering
    \includegraphics[width=\linewidth]{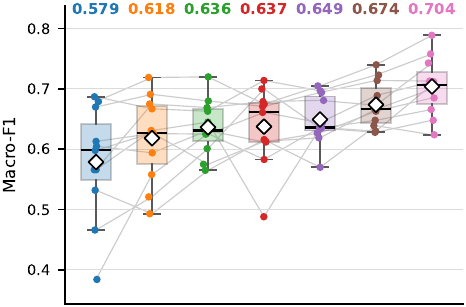}
    \caption{Macro-F1.}
    \label{fig:rq1_box_macro_f1}
\end{subfigure}

\vspace{0.4ex}

\begin{subfigure}[t]{0.9\columnwidth}
    \centering
    \includegraphics[width=\linewidth]{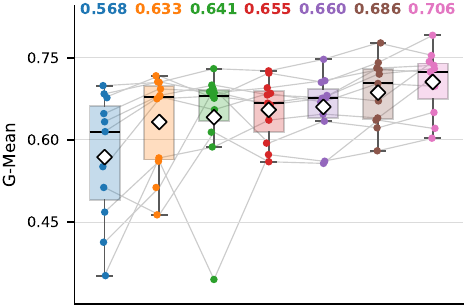}
    \caption{G-Mean.}
    \label{fig:rq1_box_gmean}
\end{subfigure}

\vspace{0.4ex}

\begin{subfigure}[t]{0.9\columnwidth}
    \centering
    \includegraphics[width=\linewidth]{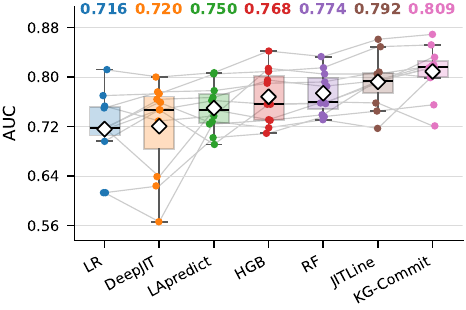}
    \caption{AUC.}
    \label{fig:rq1_box_auc}
\end{subfigure}

\vspace{0.4ex}

\caption{Project-level distributions of the performance metrics under the online protocol. Each boxplot summarizes results over the 11 ApacheJIT projects (\autoref{sec:dataset}). Aggregate means are annotated above the boxplots, and individual project values are overlaid to show cross-project variation.}
\label{fig:rq1_boxplots}
\end{figure}

\begin{figure*}[!htbp]
    \centering
    \includegraphics[width=0.9\textwidth]{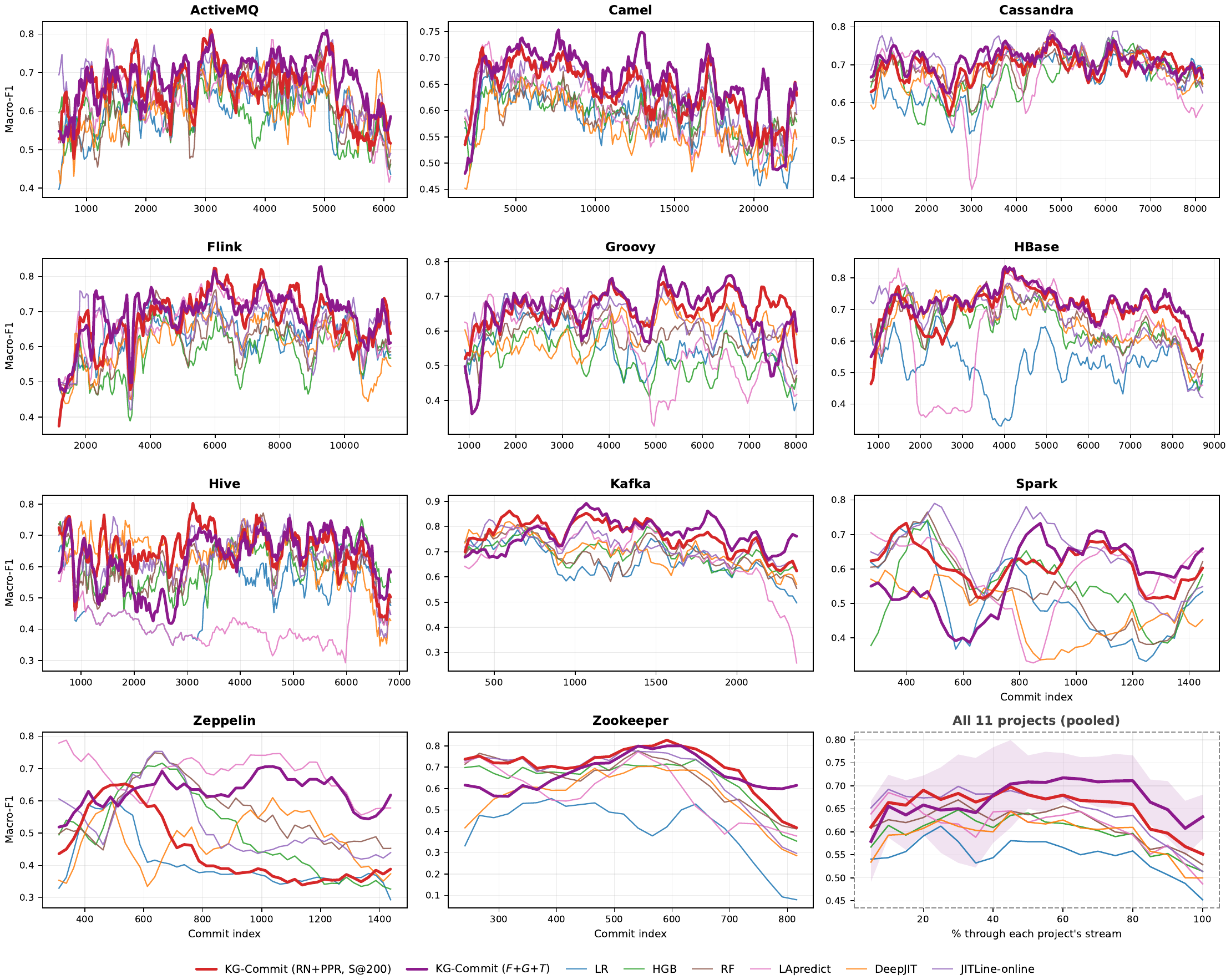}
    \caption{Running online Macro-F1 of KG-Commit compared with the online implementation of baselines. The rolling window is $w{=}150$ at stride $25$.}
    \label{fig:rq1_streams}
\end{figure*}

KG-Commit achieves the highest aggregate Macro-F1, G-Mean, and AUC among the evaluated methods (\autoref{tab:rq1_perf_g50}), with their project-level distributions shown in~\autoref{fig:rq1_boxplots}. It improves Macro-F1 over RF ($0.649\rightarrow0.704$), LApredict ($0.637\rightarrow0.704$), HGB ($0.636\rightarrow0.704$), DeepJIT ($0.618\rightarrow0.704$), LR ($0.579\rightarrow0.704$), and JITLine-online ($0.674\rightarrow0.704$). The pairwise Macro-F1 summary in~\autoref{tab:rq1_perf_g50} shows W/T/L records of 11/0/0 against RF, HGB, DeepJIT, and LR, 10/0/1 against LApredict, and 9/0/2 against JITLine-online. These differences are statistically significant across all six baselines. For G-Mean and AUC, the paired differences are also significant against the other five baselines, but not against JITLine-online ($p=.116$ and $p=.067$, respectively). The online evolution of Macro-F1 across the project streams is shown in~\autoref{fig:rq1_streams}.

The effort-aware comparison with JITLine-online also favors KG-Commit (\autoref{tab:rq1_effort_g50}). KG-Commit achieves a higher Macro-Avg $P_{\mathrm{opt}}$ ($0.889$ versus $0.878$) and ACC@20\%LOC ($0.653$ versus $0.633$), with W/T/L records of 9/1/1 and 10/0/1 respectively. The paired differences are statistically significant for both $P_{\mathrm{opt}}$ ($p=.004$) and ACC@20\%LOC ($p=.002$). At the project level, the relative performance varies across the two measures, but the aggregate results indicate that KG-Commit's predictive gains remain evident when inspection effort is taken into account.

\begin{table*}[!htbp]
\centering
\tiny
\setlength{\tabcolsep}{2pt}
\caption{Performance of KG-Commit and baselines under the online protocol (\autoref{sec:setup-protocol-section}) over ApacheJIT projects (\autoref{sec:dataset}). The best model for each metric is shown in \textbf{bold} within every project and aggregate row. The final rows summarize pairwise differences between KG-Commit ($F{+}G{+}T$) and each baseline; W/T/L denotes project-level wins/ties/losses, and Wilcoxon $p$ values are from two-sided paired tests over the 11 project scores.}
\label{tab:rq1_perf_g50}
\resizebox{\textwidth}{!}{
\begin{tabular}{l*{21}{c}}
\toprule
\multirow{2}{*}{Project}
& \multicolumn{3}{c}{KG-Commit ($F{+}G{+}T$, S@200)}
& \multicolumn{3}{c}{JITLine-online}
& \multicolumn{3}{c}{RF}
& \multicolumn{3}{c}{LApredict}
& \multicolumn{3}{c}{HGB}
& \multicolumn{3}{c}{DeepJIT}
& \multicolumn{3}{c}{LR} \\
\cmidrule(lr){2-4}
\cmidrule(lr){5-7}
\cmidrule(lr){8-10}
\cmidrule(lr){11-13}
\cmidrule(lr){14-16}
\cmidrule(lr){17-19}
\cmidrule(lr){20-22}
& Macro-F1 & G-Mean & AUC
& Macro-F1 & G-Mean & AUC
& Macro-F1 & G-Mean & AUC
& Macro-F1 & G-Mean & AUC
& Macro-F1 & G-Mean & AUC
& Macro-F1 & G-Mean & AUC
& Macro-F1 & G-Mean & AUC \\
\midrule

ActiveMQ
& \textbf{0.712} & \textbf{0.730} & \textbf{0.816}
& 0.689 & 0.720 & 0.793
& 0.644 & 0.677 & 0.757
& 0.661 & 0.700 & 0.762
& 0.625 & 0.682 & 0.756
& 0.626 & 0.681 & 0.747
& 0.613 & 0.633 & 0.696 \\

Camel
& \textbf{0.685} & 0.698 & \textbf{0.832}
& 0.663 & \textbf{0.703} & 0.790
& 0.627 & 0.669 & 0.758
& 0.612 & 0.655 & 0.691
& 0.631 & 0.685 & 0.756
& 0.594 & 0.679 & 0.747
& 0.605 & 0.615 & 0.715 \\

Cassandra
& 0.713 & 0.724 & 0.799
& \textbf{0.714} & \textbf{0.728} & \textbf{0.808}
& 0.694 & 0.708 & 0.792
& 0.671 & 0.686 & 0.778
& 0.680 & 0.695 & 0.790
& 0.691 & 0.706 & 0.776
& 0.670 & 0.684 & 0.750 \\

Flink
& \textbf{0.758} & 0.754 & \textbf{0.852}
& 0.723 & \textbf{0.777} & 0.849
& 0.697 & 0.747 & 0.815
& 0.714 & 0.730 & 0.805
& 0.663 & 0.726 & 0.809
& 0.667 & 0.705 & 0.773
& 0.687 & 0.699 & 0.770 \\

Groovy
& \textbf{0.700} & \textbf{0.736} & \textbf{0.820}
& 0.666 & 0.731 & 0.795
& 0.619 & 0.676 & 0.739
& 0.613 & 0.688 & 0.724
& 0.575 & 0.664 & 0.730
& 0.627 & 0.693 & 0.759
& 0.598 & 0.648 & 0.718 \\

HBase
& \textbf{0.743} & \textbf{0.743} & \textbf{0.815}
& 0.673 & 0.671 & 0.794
& 0.681 & 0.681 & 0.785
& 0.679 & 0.680 & 0.767
& 0.669 & 0.667 & 0.795
& 0.675 & 0.675 & 0.764
& 0.566 & 0.551 & 0.716 \\

Hive
& \textbf{0.666} & \textbf{0.603} & \textbf{0.818}
& 0.648 & 0.580 & 0.805
& 0.635 & 0.561 & 0.805
& 0.488 & 0.345 & 0.737
& 0.645 & 0.573 & 0.814
& 0.631 & 0.561 & 0.729
& 0.566 & 0.468 & 0.753 \\

Kafka
& \textbf{0.789} & \textbf{0.791} & \textbf{0.869}
& 0.740 & 0.741 & 0.861
& 0.705 & 0.705 & 0.833
& 0.675 & 0.676 & 0.807
& 0.720 & 0.722 & 0.842
& 0.719 & 0.717 & 0.800
& 0.679 & 0.677 & 0.812 \\

Spark
& 0.624 & 0.621 & 0.721
& \textbf{0.639} & \textbf{0.636} & 0.758
& 0.570 & 0.557 & \textbf{0.760}
& 0.616 & 0.614 & 0.747
& 0.565 & 0.560 & 0.731
& 0.493 & 0.463 & 0.639
& 0.532 & 0.513 & 0.718 \\

Zeppelin
& 0.648 & 0.650 & \textbf{0.755}
& 0.628 & 0.622 & 0.745
& 0.635 & 0.634 & 0.736
& \textbf{0.700} & \textbf{0.693} & 0.702
& 0.601 & 0.594 & 0.709
& 0.521 & 0.513 & 0.624
& 0.466 & 0.413 & 0.613 \\

Zookeeper
& \textbf{0.707} & \textbf{0.714} & \textbf{0.802}
& 0.630 & 0.640 & 0.717
& 0.636 & 0.647 & 0.731
& 0.583 & 0.587 & 0.728
& 0.626 & 0.636 & 0.718
& 0.558 & 0.567 & 0.566
& 0.384 & 0.352 & 0.613 \\

\midrule

\textbf{Macro-Avg}
& \textbf{0.704} & \textbf{0.706} & \textbf{0.809}
& 0.674 & 0.687 & 0.792
& 0.649 & 0.660 & 0.774
& 0.637 & 0.641 & 0.750
& 0.636 & 0.655 & 0.768
& 0.618 & 0.633 & 0.720
& 0.579 & 0.568 & 0.716 \\

\textbf{Micro-Avg}
& \textbf{0.709} & \textbf{0.713} & \textbf{0.823}
& 0.680 & 0.704 & 0.803
& 0.653 & 0.676 & 0.777
& 0.637 & 0.652 & 0.743
& 0.641 & 0.674 & 0.775
& 0.633 & 0.670 & 0.751
& 0.612 & 0.613 & 0.729 \\

\midrule

\textbf{Mean $\Delta$}
& \multicolumn{3}{c}{--}
& +.030 & +.020 & +.017
& +.055 & +.046 & +.035
& +.067 & +.065 & +.059
& +.068 & +.051 & +.041
& +.086 & +.073 & +.089
& +.125 & +.137 & +.093 \\

\textbf{W/T/L}
& \multicolumn{3}{c}{--}
& \textcolor{ForestGreen}{9}/0/\textcolor{BrickRed}{2}
& \textcolor{ForestGreen}{7}/0/\textcolor{BrickRed}{4}
& \textcolor{ForestGreen}{9}/0/\textcolor{BrickRed}{2}
& \textcolor{ForestGreen}{11}/0/0
& \textcolor{ForestGreen}{11}/0/0
& \textcolor{ForestGreen}{10}/0/\textcolor{BrickRed}{1}
& \textcolor{ForestGreen}{10}/0/\textcolor{BrickRed}{1}
& \textcolor{ForestGreen}{10}/0/\textcolor{BrickRed}{1}
& \textcolor{ForestGreen}{10}/0/\textcolor{BrickRed}{1}
& \textcolor{ForestGreen}{11}/0/0
& \textcolor{ForestGreen}{11}/0/0
& \textcolor{ForestGreen}{10}/0/\textcolor{BrickRed}{1}
& \textcolor{ForestGreen}{11}/0/0
& \textcolor{ForestGreen}{11}/0/0
& \textcolor{ForestGreen}{11}/0/0
& \textcolor{ForestGreen}{11}/0/0
& \textcolor{ForestGreen}{11}/0/0
& \textcolor{ForestGreen}{11}/0/0 \\

\textbf{Wilcoxon $p$}
& \multicolumn{3}{c}{--}
& $.005$ & $.116$ & $.067$
& $<.001$ & $<.001$ & $.019$
& $.010$ & $.011$ & $.003$
& $<.001$ & $<.001$ & $.005$
& $<.001$ & $<.001$ & $<.001$
& $<.001$ & $<.001$ & $<.001$ \\

\bottomrule
\end{tabular}}
\end{table*}

\begin{table}[!htbp]
\centering
\scriptsize
\setlength{\tabcolsep}{2pt}
\renewcommand{\arraystretch}{1.0}
\caption{Effort-aware comparison of KG-Commit and JITLine-online. For each KG-Commit result, parentheses report the absolute difference from JITLine-online. The best value for each metric is shown in \textbf{bold}. The final rows summarize project-level differences, wins/ties/losses, and two-sided paired Wilcoxon tests over the 11 projects.}
\label{tab:rq1_effort_g50}
\begin{tabular}{@{}lcccc@{}}
\toprule
\multirow{2}{*}{Project}
& \multicolumn{2}{c}{\shortstack{KG-Commit($F{+}G{+}T$)}}
& \multicolumn{2}{c}{JITLine-online} \\
\cmidrule(lr){2-3}\cmidrule(lr){4-5}
& $P_{\mathrm{opt}}$ & ACC@20
& $P_{\mathrm{opt}}$ & ACC@20 \\
\midrule

ActiveMQ
& \textbf{0.857}\,\textcolor{ForestGreen}{(.010$\uparrow$)}
& \textbf{0.624}\,\textcolor{ForestGreen}{(.023$\uparrow$)}
& 0.847 & 0.601 \\

Camel
& \textbf{0.837}\,\textcolor{ForestGreen}{(.024$\uparrow$)}
& \textbf{0.625}\,\textcolor{ForestGreen}{(.034$\uparrow$)}
& 0.813 & 0.591 \\

Cassandra
& \textbf{0.914}\,\textcolor{gray}{(.000$=$)}
& \textbf{0.696}\,\textcolor{ForestGreen}{(.003$\uparrow$)}
& \textbf{0.914} & 0.693 \\

Flink
& \textbf{0.860}\,\textcolor{ForestGreen}{(.007$\uparrow$)}
& \textbf{0.591}\,\textcolor{ForestGreen}{(.005$\uparrow$)}
& 0.853 & 0.586 \\

Groovy
& \textbf{0.863}\,\textcolor{ForestGreen}{(.013$\uparrow$)}
& \textbf{0.678}\,\textcolor{ForestGreen}{(.035$\uparrow$)}
& 0.850 & 0.643 \\

HBase
& \textbf{0.927}\,\textcolor{ForestGreen}{(.007$\uparrow$)}
& \textbf{0.716}\,\textcolor{ForestGreen}{(.010$\uparrow$)}
& 0.920 & 0.706 \\

Hive
& 0.952\,\textcolor{BrickRed}{(.002$\downarrow$)}
& 0.737\,\textcolor{BrickRed}{(.001$\downarrow$)}
& \textbf{0.954} & \textbf{0.738} \\

Kafka
& \textbf{0.907}\,\textcolor{ForestGreen}{(.007$\uparrow$)}
& \textbf{0.632}\,\textcolor{ForestGreen}{(.008$\uparrow$)}
& 0.900 & 0.624 \\

Spark
& \textbf{0.841}\,\textcolor{ForestGreen}{(.004$\uparrow$)}
& \textbf{0.531}\,\textcolor{ForestGreen}{(.014$\uparrow$)}
& 0.837 & 0.517 \\

Zeppelin
& \textbf{0.915}\,\textcolor{ForestGreen}{(.012$\uparrow$)}
& \textbf{0.682}\,\textcolor{ForestGreen}{(.028$\uparrow$)}
& 0.903 & 0.654 \\

Zookeeper
& \textbf{0.901}\,\textcolor{ForestGreen}{(.030$\uparrow$)}
& \textbf{0.676}\,\textcolor{ForestGreen}{(.065$\uparrow$)}
& 0.871 & 0.611 \\

\midrule
\textbf{Macro-Avg}
& \textbf{0.889}\,\textcolor{ForestGreen}{(.011$\uparrow$)}
& \textbf{0.653}\,\textcolor{ForestGreen}{(.020$\uparrow$)}
& 0.878 & 0.633 \\

\textbf{Micro-Avg}
& \textbf{0.877}\,\textcolor{ForestGreen}{(.012$\uparrow$)}
& \textbf{0.653}\,\textcolor{ForestGreen}{(.019$\uparrow$)}
& 0.865 & 0.634 \\

\midrule

\textbf{Mean $\Delta$}
& -- & --
& +.010 & +.020 \\

\textbf{W/T/L}
& -- & --
& \textcolor{ForestGreen}{9}/1/\textcolor{BrickRed}{1}
& \textcolor{ForestGreen}{10}/0/\textcolor{BrickRed}{1} \\

\textbf{Wilcoxon $p$}
& -- & --
& $.004$ & $.002$ \\

\bottomrule
\end{tabular}
\end{table}

\subsection{RQ2: Efficiency and Scalability}
\label{sec:rq2}
We compare the computational cost of processing each incoming commit in \autoref{tab:rq2_deployment}. For KG-Commit, routine per-commit processing includes graph ingestion, context retrieval, and prediction. For the baselines, it includes the corresponding featurization and prediction steps. Periodic model refitting is reported separately because it is triggered only every $M=200$ commits.

KG-Commit's routine cost is dominated by graph ingestion, with a median of approximately $1.33$~s per commit. The isolated graph-retrieval and $F_{\mathrm{ov}}=\mathrm{RN}{+}\mathrm{PPR}$ inference components require only $0.116$~ms and $0.325$~ms respectively. The added TGC values in \autoref{sec:kg-projection} are cached and read as a compact feature vector (\autoref{app:tgc-features}); they do not introduce an additional full-graph Neo4j traversal. The evaluated baselines require approximately $7.9$--$8.5$~ms per commit before periodic refitting. KG-Commit therefore incurs higher per-commit computation in exchange for maintaining rich project-wide context. Whether this additional cost is operationally important depends on the rate at which commits arrive. As discussed in \autoref{sec:commit-spacing}, the observed commit arrival time is much larger than KG-Commit's processing time across all 11 projects.

\begin{table}[!htbp]
\centering
\scriptsize
\setlength{\tabcolsep}{1.2pt}
\renewcommand{\arraystretch}{0.8}

\caption{Median runtime costs across all projects. Routine processing includes the work performed for every incoming commit. Periodic refit is reported separately and amortized over $M=200$ commits. \emph{Feat.}, \emph{retr.}, and \emph{pred.} denote featurization, retrieval, and prediction. $^{*}$Includes computation of the 12 change metrics.}
\label{tab:rq2_deployment}

\begin{tabular*}{0.9\columnwidth}{
    @{}l @{\extracolsep{\fill}}
    r r r r r r r@{}
}
\toprule
& \multicolumn{1}{c}{\makecell{\textbf{KG-}\\\textbf{Commit}}}
& \multicolumn{1}{c}{\makecell{\textbf{LR}}}
& \multicolumn{1}{c}{\makecell{\textbf{HGB}}}
& \multicolumn{1}{c}{\makecell{\textbf{RF}}}
& \multicolumn{1}{c}{\makecell{\textbf{LA}\\\textbf{predict}}}
& \multicolumn{1}{c}{\makecell{\textbf{DeepJIT}}}
& \multicolumn{1}{c}{\makecell{\textbf{JITLine}\\\textbf{online}}} \\
\midrule

\multicolumn{8}{@{}l@{}}{\emph{Routine per-commit processing (ms/commit)}} \\
\midrule

Ingest
& 1,330
& -- & -- & -- & -- & -- & -- \\

Feat.
& --
& 7.912$^{*}$
& 7.912$^{*}$
& 7.912$^{*}$
& 7.885
& 7.912$^{*}$
& 8.213 \\

Retr.
& 0.116
& -- & -- & -- & -- & -- & -- \\

Pred.
& 0.325
& 0.001
& 0.009
& 0.245
& 0.001
& 0.363
& 0.298 \\

\textbf{Total}
& \textbf{$\approx$1,330.44}
& 7.913
& 7.921
& 8.157
& 7.886
& 8.275
& 8.511 \\

\midrule
\multicolumn{8}{@{}l@{}}{\emph{Periodic refit (amortized over $M=200$ commits)}} \\
\midrule

Refit
& 1.05
& 0.04
& 2.83
& 2.08
& 0.02
& 53.04
& 9.85 \\

\bottomrule
\end{tabular*}
\end{table}

\autoref{tab:rq2_kgonly} reports the graph scale, inference latency, and ingest cost of KG-Commit for each project. Graph-inference component $F_{\mathrm{ov}}{=}\mathrm{RN}{+}\mathrm{PPR}$, has a median inference time of $0.325$~ms and a median $p_{95}$ of $0.399$~ms. The overall runtime of KG-Commit is instead dominated by graph ingestion,
whose median cost is approximately $1.33$~s per commit. DeepWalk and KGE are faster to apply, at approximately $0.001$~ms, but require periodic refitting, whereas RN and PPR operate directly on the graph. The AST delta is highly skewed: the median commit changes $282$ edges, compared with $5{,}107$ at the $p_{95}$, with a Gini coefficient of $0.78$. Ingest cost is therefore driven mainly by a small number of large commits rather than the accumulated project history. 

\begin{table*}[!htbp]
\centering\scriptsize\setlength{\tabcolsep}{3.2pt}
\caption{Internal cost profile of KG-Commit by project. $N_c$ is the number of commit nodes, and \emph{c/s} is the processing throughput in commits per second. The inference columns isolate the five projection-based graph channels and $F_{\mathrm{ov}}=\mathrm{RN}{+}\mathrm{PPR}$ (\autoref{sec:kg-projection}); $F$-$p_{95}$ is the 95th-percentile latency of this graph component. The TGC block used by the final $F_{\mathrm{ov}}{+}G{+}T$ configuration is cached separately (\autoref{app:tgc-features}). Under \textsc{Ingest}, \emph{ms} is graph-update time per commit, \emph{edges} is the total AST delta (\autoref{sec:kg-ast}), and $p_{50}$, $p_{95}$, and Gini describe the AST-delta distribution.}

\label{tab:rq2_kgonly}
\resizebox{0.95\textwidth}{!}{
\begin{tabular}{l rr | rrrrrrr | rr rrr}
\toprule
& \multicolumn{2}{c|}{\textsc{Graph}} & \multicolumn{7}{c|}{\textsc{Inference} time (ms/commit)} & \multicolumn{5}{c}{\textsc{Ingest} \& per-commit AST delta} \\
\cmidrule(lr){2-3}\cmidrule(lr){4-10}\cmidrule(l){11-15}
& $N_c$ & c/s & RN & PPR & LP & DW & KGE & $\mathrm{F}_{\mathrm{ov}}$ & Fp95 & ms & edges & p50 & p95 & Gini \\
\midrule
ActiveMQ & 6,126 & 3,080 & 0.078 & 0.247 & 0.027 & 0.001 & 0.001 & \textbf{0.325} & 0.399 & 1,277 & 3.27M & 97 & 1,848 & 0.80 \\
Camel & 22,695 & 740 & 0.362 & 0.989 & 0.100 & 0.002 & 0.001 & \textbf{1.352} & 1.509 & 1,163 & 12.38M & 133 & 2,044 & 0.78 \\
Cassandra & 8,159 & 1,691 & 0.152 & 0.440 & 0.046 & 0.001 & 0.001 & \textbf{0.591} & 0.775 & 1,519 & 11.84M & 283 & 6,438 & 0.78 \\
Flink & 11,459 & 1,066 & 0.242 & 0.697 & 0.069 & 0.001 & 0.001 & \textbf{0.938} & 1.078 & 1,330 & 16.67M & 371 & 6,515 & 0.75 \\
Groovy & 8,059 & 3,219 & 0.103 & 0.208 & 0.038 & 0.001 & 0.000 & \textbf{0.311} & 0.367 & 823 & 3.61M & 84 & 1,464 & 0.83 \\
HBase & 8,727 & 1,547 & 0.161 & 0.485 & 0.049 & 0.001 & 0.001 & \textbf{0.646} & 0.796 & 1,648 & 14.01M & 293 & 6,578 & 0.81 \\
Hive & 6,842 & 1,846 & 0.133 & 0.409 & 0.041 & 0.001 & 0.001 & \textbf{0.542} & 0.677 & 2,177 & 12.20M & 383 & 7,621 & 0.79 \\
Kafka & 2,384 & 5,855 & 0.046 & 0.124 & 0.016 & 0.001 & 0.001 & \textbf{0.171} & 0.336 & 2,324 & 3.17M & 375 & 6,083 & 0.74 \\
Spark & 1,465 & 14,022 & 0.017 & 0.054 & 0.010 & 0.001 & 0.001 & \textbf{0.071} & 0.088 & 1,238 & 0.82M & 153 & 2,310 & 0.76 \\
Zeppelin & 1,451 & 11,652 & 0.023 & 0.063 & 0.011 & 0.001 & 0.001 & \textbf{0.086} & 0.116 & 1,243 & 1.63M & 213 & 5,107 & 0.79 \\
Zookeeper & 839 & 18,494 & 0.015 & 0.039 & 0.008 & 0.001 & 0.001 & \textbf{0.054} & 0.062 & 1,497 & 0.78M & 282 & 3,917 & 0.72 \\
\midrule
\textbf{Median} & \multicolumn{1}{c}{6,842} & 3,080 & & & & & & \textbf{0.325} & & 1,330 & & & & 0.78 \\
\bottomrule
\end{tabular}
}
\end{table*}

\begin{figure}[!htbp]
    \centering
    \captionsetup{font=footnotesize}
    \captionsetup[subfigure]{font=scriptsize}

    \begin{subfigure}[t]{0.45\columnwidth}
        \centering
        \includegraphics[width=\linewidth]
            {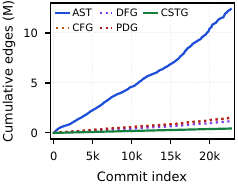}
        \caption{Cumulative edges by representation.}
        \label{fig:rq2_lifecycle_growth}
    \end{subfigure}
    \hfill
    \begin{subfigure}[t]{0.45\columnwidth}
        \centering
        \includegraphics[width=\linewidth]
            {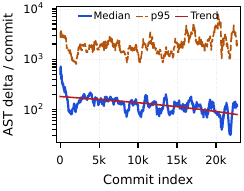}
        \caption{Rolling median, p95, and trend of the per-commit AST delta.}
        \label{fig:rq2_lifecycle_ast_delta}
    \end{subfigure}

    \vspace{2pt}

    \begin{subfigure}[t]{0.45\columnwidth}
        \centering
        \includegraphics[width=\linewidth]
            {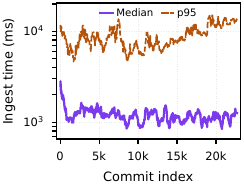}
        \caption{Rolling median and p95 per-commit ingest time.}
        \label{fig:rq2_lifecycle_ingest}
    \end{subfigure}
    \hfill
    \begin{subfigure}[t]{0.45\columnwidth}
        \centering
        \includegraphics[width=\linewidth]
            {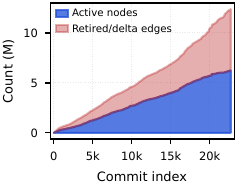}
        \caption{Active and retired graph structure.}
        \label{fig:rq2_lifecycle_churn}
    \end{subfigure}

    \vspace{2pt}

    \begin{subfigure}[t]{0.45\columnwidth}
        \centering
        \includegraphics[width=\linewidth]
            {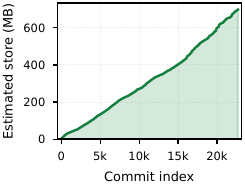}
        \caption{Estimated graph-store footprint.}
        \label{fig:rq2_lifecycle_storage}
    \end{subfigure}

    \vspace{2pt}

    \begin{subfigure}[t]{0.95\columnwidth}
        \centering
        \includegraphics[width=\linewidth]
            {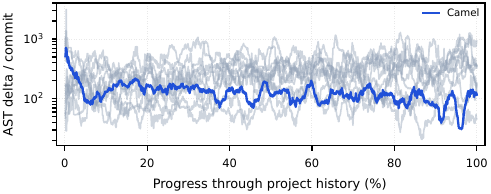}
        \caption{Rolling-median AST delta across all 11 projects. Camel is highlighted.}
        \label{fig:rq2_lifecycle_projects}
    \end{subfigure}

    \caption{Lifecycle and scaling of KG-Commit.~\autoref{fig:rq2_lifecycle_growth}--\ref{fig:rq2_lifecycle_storage} report Camel, the largest project history.~\autoref{fig:rq2_lifecycle_projects} compares normalized per-commit AST deltas across all 11 projects.}
    \label{fig:rq2_lifecycle}
\end{figure}

\paragraph{Lifecycle and Scaling}
\autoref{fig:rq2_lifecycle} shows how the graph and its deployment costs grow with project history.~\autoref{fig:rq2_lifecycle_growth} shows the accumulated edges, while~\autoref{fig:rq2_lifecycle_ast_delta} and~\autoref{fig:rq2_lifecycle_ingest} show that per-commit work does not increase with it. In the final fifth of Camel's history, the median AST delta and ingest time are $0.55\times$ and $0.86\times$ their values in the first fifth, and median ingest time remains approximately $1.16$\,s. Ingest cost therefore depends mainly on the incoming change instead of the existing graph size.~\autoref{fig:rq2_lifecycle_churn} and~\autoref{fig:rq2_lifecycle_storage} show that soft deletion preserves historical structure while storage grows linearly.~\autoref{fig:rq2_lifecycle_projects} and the corresponding analyses of the other projects show the same trends. We present Camel because it is the most demanding case, with $22{,}695$ commits, $12.4$ million cumulative AST edges, and an estimated footprint of $699$\,MB.

\subsection{RQ3: Contribution of KG-Commit Layers to  Predictive Performance}
\label{sec:rq3}
\begin{figure}[!htbp]
    \centering
    \includegraphics[width=0.9\columnwidth]
    {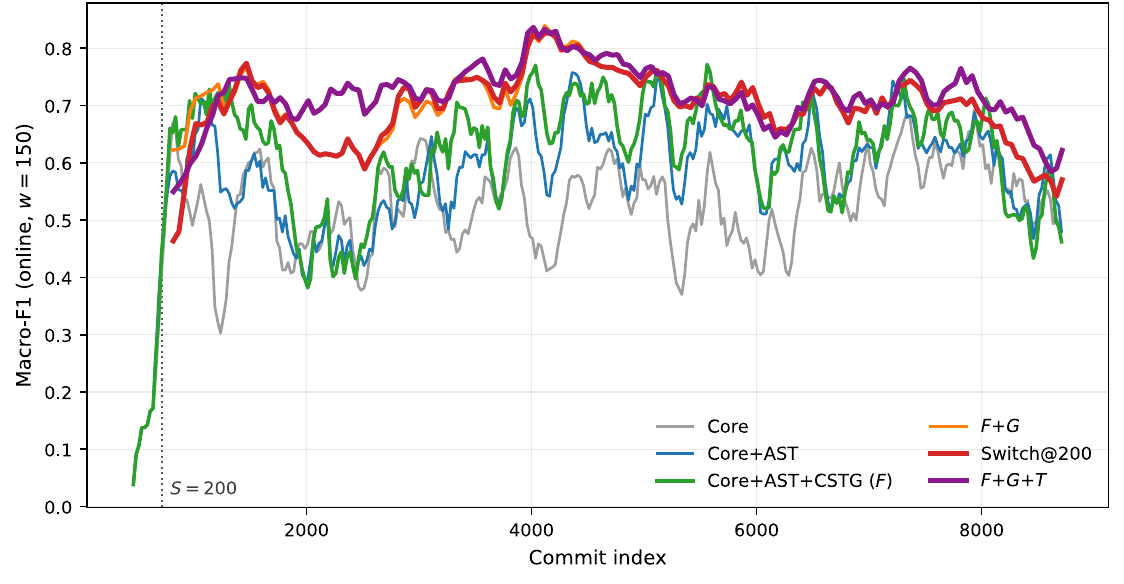}
    \caption{Online Macro-F1 by representation layer on HBase. The curves compare Core, Core$+$AST, Core$+$AST$+$CSTG, along with the fusions $F_{\mathrm{ov}}$, $F_{\mathrm{ov}}{+}G$, S@200 policy, and $F_{\mathrm{ov}}{+}G+T$. The dotted line marks the switch point $S{=}200$.}
    \label{fig:rq3_layerstream}
\end{figure}

\begin{figure}[!htbp]
    \centering
    \includegraphics[width=0.9\columnwidth]
    {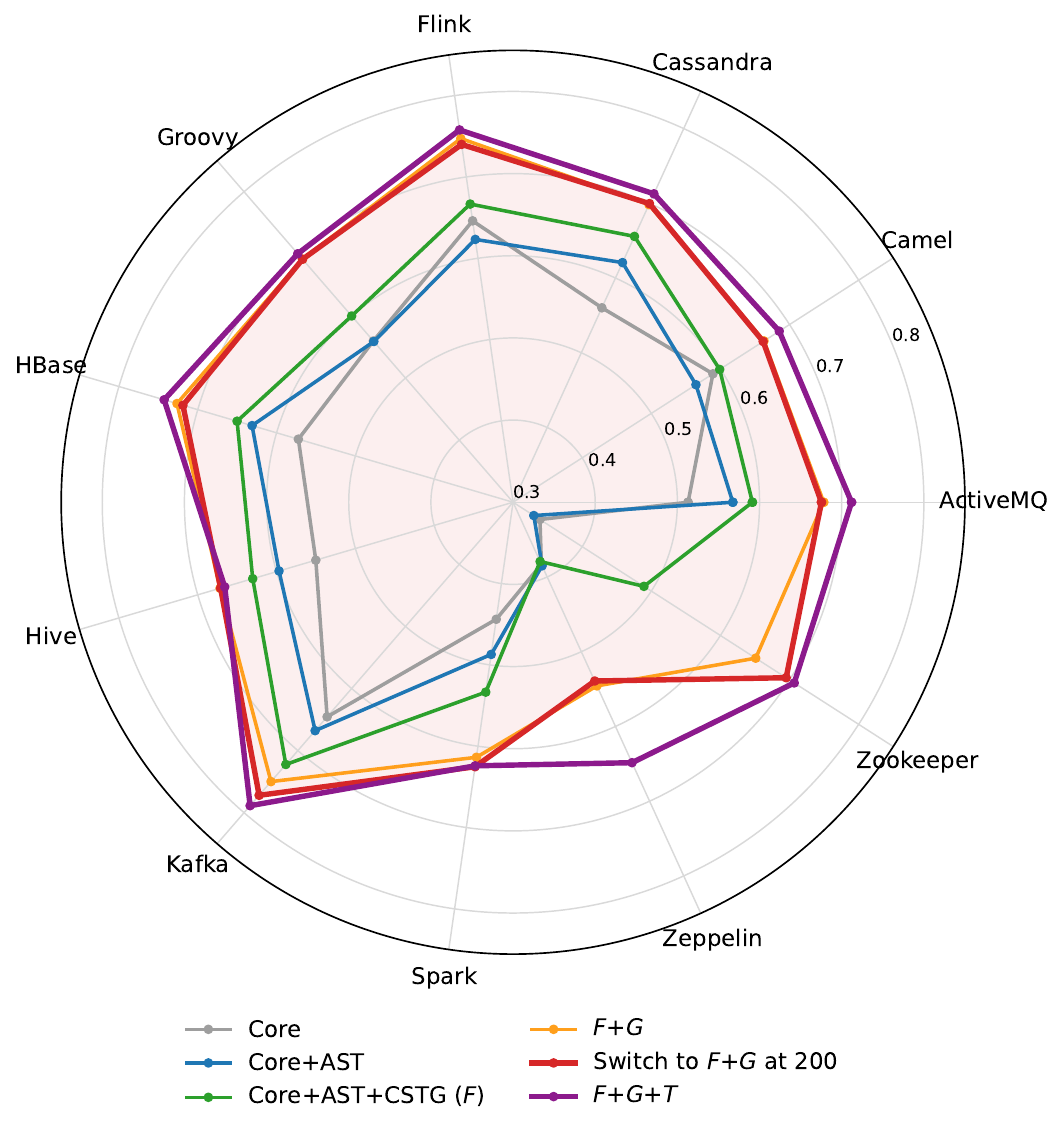}
    \caption{Macro-F1 score by KG-Commit representation layer across all 11 projects, together with the corresponding additional configurations in \autoref{sec:kg-projection}.}
    \label{fig:rq3_radar}
\end{figure}

\autoref{tab:rq3_layers_fov} examines the contribution of KG-Commit's representation layers while holding inference fixed at $F_{\mathrm{ov}}$. The graph is progressively enriched from Core to AST and then CSTG, so differences in performance reflect changes in the representation itself. Layer~2 provides little benefit under $F_{\mathrm{ov}}$, whereas adding CSTG consistently improves $F_{\mathrm{ov}}$ results across all projects, increasing the median Macro-F1 from $0.552$ for Core to $0.644$ for CSTG. The same pattern can be seen throughout the commit stream in~\autoref{fig:rq3_layerstream} and across projects in~\autoref{fig:rq3_radar}.

The limited gain from AST suggests that the generic inference methods in \autoref{sec:kg-infer} do not fully exploit its fine-grained structure. CSTG, in contrast, introduces semantic entities and relations that are more accessible to RN and PPR. This result is separate from the CSTG feature channel $G$ and the traversal/global-context feature block $T$, which augment the graph-based inference through $F{+}G$ and $F{+}G{+}T$, shown in~\autoref{fig:rq3_layerstream} and~\autoref{fig:rq3_radar}.

\begin{table*}[!htbp]
\centering
\caption{Online performance across all 11 projects under fusion $F_{\mathrm{ov}}$. Columns follow the KG-Commit hierarchy: Core (Layer~1,~\autoref{sec:kg-core}), four within-file graph representations (Layer~2,~\autoref{sec:kg-ast}), and CSTG (Layer~3,~\autoref{sec:kg-cstg}). \textbf{Bold} values mark the best representation for each project and metric.}

\label{tab:rq3_layers_fov}
\scriptsize
\setlength{\tabcolsep}{2.5pt}
\renewcommand{\arraystretch}{0.95}
\begin{tabular*}{0.9\textwidth}{
@{}l @{\extracolsep{\fill}}
rrr rrr rrr rrr rrr rrr@{}
}
\toprule
\textbf{Project} &
\multicolumn{3}{c}{\textsc{Layer 1}} &
\multicolumn{12}{c}{\textsc{Layer 2: within-file subgraph}} &
\multicolumn{3}{c}{\textsc{Layer 3}} \\
\cmidrule(lr){2-4}\cmidrule(lr){5-16}\cmidrule(lr){17-19}
&
\multicolumn{3}{c}{Core} &
\multicolumn{3}{c}{+AST} &
\multicolumn{3}{c}{+CFG} &
\multicolumn{3}{c}{+DFG} &
\multicolumn{3}{c}{+PDG} &
\multicolumn{3}{c}{+CSTG} \\
\cmidrule(lr){2-4}\cmidrule(lr){5-7}\cmidrule(lr){8-10}\cmidrule(lr){11-13}\cmidrule(lr){14-16}\cmidrule(lr){17-19}
&
$F_1$ & $G$ & AUC &
$F_1$ & $G$ & AUC &
$F_1$ & $G$ & AUC &
$F_1$ & $G$ & AUC &
$F_1$ & $G$ & AUC &
$F_1$ & $G$ & AUC \\
\midrule
\texttt{ActiveMQ} & .572 & .636 & .697 & .565 & .649 & .721 & .558 & .643 & .717 & .551 & .637 & .718 & .551 & .637 & .718 & \textbf{.642} & \textbf{.706} & \textbf{.778} \\
\texttt{Camel} & .579 & .668 & .725 & .595 & .684 & .743 & .603 & .689 & .754 & .610 & .687 & .764 & .599 & .692 & .754 & \textbf{.637} & \textbf{.730} & \textbf{.792} \\
\texttt{Cassandra} & .552 & .559 & .660 & .583 & .591 & .685 & .581 & .590 & .681 & .584 & .592 & .691 & .580 & .588 & .680 & \textbf{.644} & \textbf{.658} & \textbf{.749} \\
\texttt{Flink} & .647 & .637 & .742 & .607 & .654 & .753 & .614 & .651 & .741 & .606 & .646 & .733 & .616 & .653 & .741 & \textbf{.685} & \textbf{.731} & \textbf{.818} \\
\texttt{Groovy} & .596 & .649 & .721 & .566 & .661 & .748 & .593 & .679 & .747 & .598 & .678 & .738 & .594 & .681 & .747 & \textbf{.656} & \textbf{.729} & \textbf{.801} \\
\texttt{HBase} & .503 & .465 & .678 & .593 & .579 & .732 & .640 & .638 & .733 & .627 & .622 & .735 & .626 & .621 & .732 & \textbf{.679} & \textbf{.680} & \textbf{.793} \\
\texttt{Hive} & .520 & .412 & .640 & .554 & .456 & .691 & .565 & .474 & .689 & .558 & .461 & .679 & .581 & .498 & .689 & \textbf{.638} & \textbf{.573} & \textbf{.769} \\
\texttt{Kafka} & .619 & .618 & .686 & .602 & .585 & .712 & .688 & .686 & .746 & .661 & .655 & .734 & .685 & .684 & .745 & \textbf{.723} & \textbf{.724} & \textbf{.809} \\
\texttt{Spark} & .439 & .392 & .630 & .380 & .309 & .517 & .490 & .454 & .614 & .420 & .357 & .583 & .465 & .420 & .602 & \textbf{.574} & \textbf{.566} & \textbf{.666} \\
\texttt{Zeppelin} & .478 & .422 & .590 & .450 & .382 & .634 & .461 & .412 & .615 & .410 & .344 & .608 & .443 & .387 & .611 & \textbf{.525} & \textbf{.491} & \textbf{.654} \\
\texttt{Zookeeper} & .543 & .551 & .637 & .511 & .499 & .521 & .511 & .492 & .584 & .506 & .486 & .626 & .503 & .481 & .583 & \textbf{.651} & \textbf{.661} & \textbf{.683} \\
\bottomrule
\end{tabular*}
\end{table*}

The four Layer-2 alternatives perform similarly, and their ordering varies across projects (\autoref{tab:rq3_layers_fov}). Among these Layer-2 candidates, we select AST because its incremental delta can be computed using canonical sub-tree hashing (\autoref{sec:kg-ast}), whereas alternative choices require an ordinal fallback matcher (\autoref{app:alternative-subgraphs}). AST also contains more nodes and edges (\autoref{sec:rq2}, \autoref{fig:rq2_lifecycle_growth}), potentially capturing richer structures that $F_{\mathrm{ov}}$ may not fully exploit, a limitation we revisit in~\autoref{sec:limits}.

Performance also varies across projects and inference methods. For example, Macro-F1 for $F_{\mathrm{ov}}$ with CSTG ranges from $0.525$ on Zeppelin to $0.723$ on Kafka (\autoref{tab:rq3_layers_fov}). The results can vary across the individual inference methods, with Cassandra ranging from $0.327$ for LP to $0.643$ for KGE (see~\autoref{app:rq3_single_inference}). We study the differences among inference channels in~\autoref{sec:rq4}.

\subsection{RQ4: Inference Channels and Fusion}
\label{sec:rq4}
PPR provides the strongest single-channel Macro-F1 of $.643$, as shown in~\autoref{tab:combos_all_projects}. Combining additional channels brings modest gains, with the best fusion reaching $0.653$. We fixed $F_{\mathrm{ov}}=\mathrm{RN{+}PPR}$ before computing these aggregate results with the rationale of combining RN's local neighborhood evidence with PPR's global propagation (\autoref{sec:kg-infer}), hence avoiding any advantage from project-specific fusion tuning. The performance cost is small, as $F_{\mathrm{ov}}$ achieves $0.641$ Macro-F1, only $0.012$ below the best, while remaining competitive in G-Mean and AUC. Additionally, RN and PPR are computed directly from the graph topology and require no fitting, periodic refitting, random initialization, or seed control, unlike fusions that include learned embedding channels. Consequently, $F_{\mathrm{ov}}$ is reproducible and cheap to serve with a median inference cost of $0.325$~ms per commit.

\begin{figure}[!htbp]
    \centering
    \includegraphics[width=0.9\columnwidth]{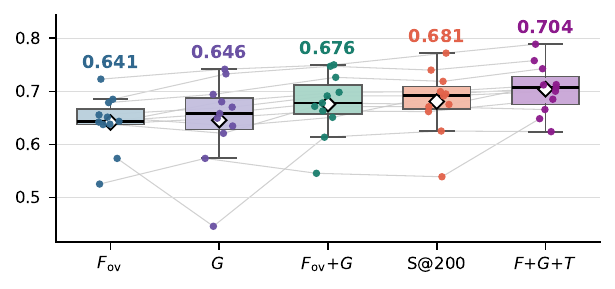}
    \caption{Project-level Macro-F1 for $F_{\mathrm{ov}}$, $G$, $F_{\mathrm{ov}}{+}G$, S@200 policy, and $F_{\mathrm{ov}}{+}G+T$. Boxes show the interquartile range, horizontal lines the medians, diamonds the means, and points the projects; faint lines connect results from the same project.}
    \label{fig:rq4_f1}
\end{figure}

\autoref{fig:rq4_f1} illustrates how $F_{\mathrm{ov}}$ and $G$ complement each other. Their fusion raises mean Macro-F1 to $0.676$, compared with $0.641$ for $F_{\mathrm{ov}}$ and $0.646$ for $G$ alone. S@200 slightly improves the mean further to $0.681$ and narrows the central spread, although its main benefit is during cold start, when it delays the CSTG channel until the project has accumulated 200 commits (see~\autoref{sec:switch-s200}). Adding $T$ further increases the mean Macro-F1 to $0.704$, giving the highest performance among the evaluated configurations.

\begin{table}[!htbp]
\centering\scriptsize\setlength{\tabcolsep}{3pt}
\renewcommand{\arraystretch}{0.8}
\caption{All 31 fixed fusion combinations across the 11 projects. Each cell reports the project-level mean $\pm$ the half-width of its Student-$t$ 95\% confidence interval. The fixed overall fusion
$F_{\mathrm{ov}}=\mathrm{RN{+}PPR}$ is highlighted in gray.}
\label{tab:combos_all_projects}
\resizebox{0.9\columnwidth}{!}{
\begin{tabular}{lccc}
\toprule
Combination & Macro-F1 & G-Mean & AUC \\ 
\midrule
LP & $.446 \pm .064$ & $.433 \pm .103$ & $.632 \pm .047$ \\
RN & $.410 \pm .040$ & $.383 \pm .071$ & $.593 \pm .039$ \\
DW & $.600 \pm .039$ & $.619 \pm .049$ & $.691 \pm .031$ \\
KGE & $.609 \pm .031$ & $.623 \pm .050$ & $.688 \pm .038$ \\
PPR & $.643 \pm .033$ & $.652 \pm .050$ & $.758 \pm .025$ \\
RN+LP & $.553 \pm .067$ & $.562 \pm .088$ & $.695 \pm .063$ \\
LP+KGE & $.578 \pm .059$ & $.587 \pm .085$ & $.698 \pm .068$ \\
RN+KGE & $.572 \pm .063$ & $.579 \pm .090$ & $.685 \pm .072$ \\
LP+DW & $.605 \pm .045$ & $.627 \pm .061$ & $.712 \pm .051$ \\
RN+DW & $.599 \pm .044$ & $.621 \pm .058$ & $.706 \pm .042$ \\
PPR+LP & $.632 \pm .034$ & $.651 \pm .052$ & $.744 \pm .046$ \\
DW+KGE & $.623 \pm .043$ & $.635 \pm .057$ & $.707 \pm .044$ \\
PPR+KGE& $.648 \pm .025$ & $.661 \pm .042$ & $.744 \pm .029$ \\
\rowcolor{gray!15}
RN+PPR & $.641 \pm .036$ & $.659 \pm .055$ & $.756 \pm .040$ \\
PPR+DW & $.649 \pm .038$ & $.663 \pm .050$ & $.739 \pm .037$ \\
RN+LP+KGE & $.575 \pm .063$ & $.582 \pm .089$ & $.708 \pm .064$ \\
RN+LP+DW & $.609 \pm .048$ & $.629 \pm .064$ & $.718 \pm .050$ \\
LP+DW+KGE & $.619 \pm .046$ & $.634 \pm .063$ & $.722 \pm .055$ \\
RN+DW+KGE & $.615 \pm .045$ & $.631 \pm .061$ & $.723 \pm .048$ \\
PPR+LP+KGE & $.635 \pm .035$ & $.651 \pm .054$ & $.745 \pm .045$ \\
RN+PPR+LP & $.640 \pm .041$ & $.656 \pm .057$ & $.754 \pm .044$ \\
RN+PPR+KGE & $.640 \pm .035$ & $.656 \pm .054$ & $.754 \pm .039$ \\
PPR+LP+DW & $.644 \pm .039$ & $.664 \pm .054$ & $.748 \pm .044$ \\
PPR+DW+KGE & $.651 \pm .037$ & $.663 \pm .051$ & $.744 \pm .039$ \\
RN+PPR+DW & $.652 \pm .039$ & $.670 \pm .052$ & $.754 \pm .041$ \\
RN+LP+DW+KGE & $.617 \pm .048$ & $.631 \pm .066$ & $.729 \pm .053$ \\
RN+PPR+LP+KGE & $.640 \pm .041$ & $.652 \pm .058$ & $.754 \pm .044$ \\
PPR+LP+DW+KGE & $.646 \pm .039$ & $.662 \pm .056$ & $.750 \pm .045$ \\
RN+PPR+LP+DW & $.650 \pm .042$ & $.667 \pm .056$ & $.756 \pm .044$ \\
RN+PPR+DW+KGE & $.653 \pm .038$ & $.670 \pm .052$ & $.756 \pm .042$ \\
RN+PPR+LP+DW+KGE & $.652 \pm .042$ & $.667 \pm .056$ & $.758 \pm .044$ \\
\bottomrule
\end{tabular}
}
\end{table}

\section{Discussion}
\label{sec:discussion}

\subsection{On the Necessity of Realistic Online Evaluation}
\label{sec:discuss_online}
Some JIT-SDP studies use conventional cross-validation~\citep{qiao2019effort}, which does not preserve the chronological order of commits. Accordingly, such time-agnostic splitting can allow later commits to appear in the training data when earlier commits are used for testing, potentially leading to optimistic performance estimates, as shown in~\autoref{tab:protocol_comparison}. Other studies adopt time-aware evaluation~\citep{mcintosh2018fix,pornprasit2021jitline,pornprasit2021pyexplainer}, where earlier commits are used for training and later commits for testing. Although this preserves temporal order, the model is typically fixed throughout an entire train/test block. Such an evaluation requires a sufficiently large historical training block, making it unsuitable for projects in the early-to-mid development cycle. It also does not reproduce the per-commit processing and maintenance cycle of a deployed system or account for label-verification latency through the gap parameter~\citep{tan2015online,cabral2019class,song2023procedure, cabral2023towards}. Consequently, challenges such as delayed feedback and concept drift remain largely unexamined~\citep{mcintosh2018fix, cabral2023towards}. Block-based evaluation also cannot reveal potential catastrophic forgetting that may arise under sequential updating ~\citep{cabral2023towards,kirkpatrick2017overcoming}.

In our comparison, block-based time-aware evaluation underestimates performance (\autoref{tab:protocol_comparison}) since holding the model fixed over many commits prevents continual refitting and adaptation as the project evolves. To capture this adaptation while avoiding temporal leakage, we use the online protocol described in~\autoref{sec:setup-protocol-section}, which reflects the deployment constraints in~\autoref{sec:operational-constraints}. Its small warm-up ratio of $.05$ also enables evaluation from the early stages of a project.

\begin{table}[!htbp]
\centering
\caption{Mean baseline performance across 11 projects under online, block-based time-aware, and random evaluation.}
\label{tab:protocol_comparison}

\scriptsize
\setlength{\tabcolsep}{2.6pt}
\renewcommand{\arraystretch}{0.95}

\begin{tabular*}{0.98\columnwidth}{
    @{}l @{\extracolsep{\fill}}
    r r r
    r r r
    r r r@{}
}
\toprule
& \multicolumn{3}{c}{Macro-F1}
& \multicolumn{3}{c}{G-Mean}
& \multicolumn{3}{c}{AUC} \\
\cmidrule(lr){2-4}
\cmidrule(lr){5-7}
\cmidrule(l){8-10}

\textbf{Model}
& \textbf{On.} & \textbf{Off.} & \textbf{Rand.}
& \textbf{On.} & \textbf{Off.} & \textbf{Rand.}
& \textbf{On.} & \textbf{Off.} & \textbf{Rand.} \\
\midrule

LR
& .579 & .542 & .655
& .568 & .579 & .661
& .716 & .706 & .754 \\

HGB
& .636 & .583 & .720
& .655 & .623 & .707
& .768 & .747 & .810 \\

LApredict
& .637 & .596 & .667
& .641 & .643 & .679
& .750 & .775 & .769 \\

\bottomrule
\end{tabular*}
\end{table}

\subsection{Per-Commit Cost Relative to Commit Arrival Time}
\label{sec:commit-spacing}
The practical suitability of KG-Commit depends on whether its per-commit processing cost is compatible with the rate at which commits arrive in practice. KG-Commit requires approximately $1.33$~s to process an incoming commit (\autoref{sec:rq2}), with most of this cost arising from the incremental graph update. \autoref{tab:commit_spacing} reports the observed intervals between consecutive commits across the projects. Even among closely spaced commits, the shortest lower-tail interval is $8.0$~s on Cassandra, while the corresponding intervals for the remaining projects range from $25$~s to more than $27$~min. Median commit spacing is considerably larger, ranging from $1.20$~h on Camel to $45.55$~h on Zookeeper. These observations indicate that KG-Commit's per-commit processing cost is compatible with the commit arrival rates observed in the evaluated projects.

\begin{table}[!htbp]
\centering
\caption{Commit arrival time across all 11 projects. Mean, standard deviation (SD), median, and P95 are reported in hours, while P05 is reported in seconds to expose short intervals between consecutive commits.}
\label{tab:commit_spacing}
\scriptsize
\setlength{\tabcolsep}{2.6pt}
\renewcommand{\arraystretch}{0.95}

\begin{tabular*}{0.9\columnwidth}{
    @{}l @{\extracolsep{\fill}}
    r r r r@{}
}
\toprule
\textbf{Project} &
\makecell{\textbf{Mean $\pm$ SD}\\\textbf{(h)}} &
\makecell{\textbf{Median}\\\textbf{(h)}} &
\makecell{\textbf{P05}\\\textbf{(s)}} &
\makecell{\textbf{P95}\\\textbf{(h)}} \\
\midrule
ActiveMQ  & $20.03 \pm 44.95$   & 4.23  & 117.2  & 90.15  \\
Camel     & $4.93 \pm 9.95$     & 1.20  & 114.0  & 20.79  \\
Cassandra & $11.57 \pm 24.93$   & 2.88  & 8.0    & 53.38  \\
Flink     & $6.77 \pm 17.86$    & 1.32  & 25.0   & 26.45  \\
Groovy    & $17.72 \pm 37.57$   & 4.59  & 105.0  & 77.56  \\
HBase     & $12.77 \pm 20.93$   & 5.34  & 301.2  & 50.60  \\
Hive      & $14.45 \pm 30.73$   & 5.21  & 306.0  & 61.66  \\
Kafka     & $30.18 \pm 189.05$  & 8.67  & 574.7  & 88.45  \\
Spark     & $54.22 \pm 377.47$  & 19.19 & 1664.6 & 162.21 \\
Zeppelin  & $38.37 \pm 77.34$   & 16.08 & 360.6  & 147.91 \\
Zookeeper & $126.30 \pm 218.03$ & 45.55 & 429.7  & 501.36 \\
\bottomrule
\end{tabular*}
\end{table}

\subsection{S@200 Switching and Cold-Start Improvement}
\label{sec:switch-s200}
\begin{figure}[!htbp]
\centering
    
    \begin{subfigure}[t]{0.9\columnwidth}
        \centering
        \includegraphics[width=\linewidth]
            {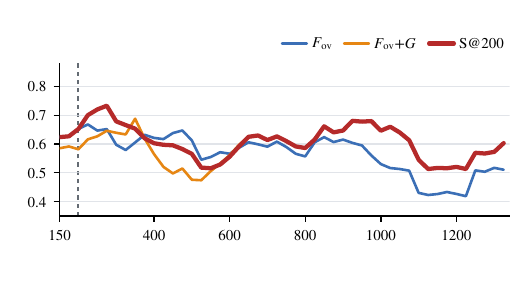}
        \caption{Spark}
        \label{fig:switch_spark}
    \end{subfigure}

    \vspace{-2pt}

    \begin{subfigure}[t]{0.9\columnwidth}
        \centering
        \includegraphics[width=\linewidth]
            {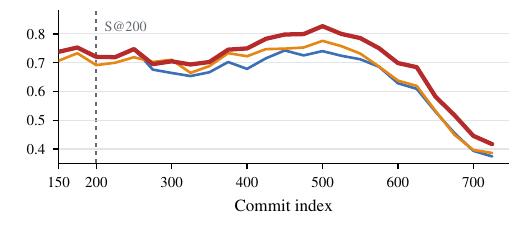}
        \caption{Zookeeper}
        \label{fig:switch_zookeeper}
\end{subfigure}

\caption{Effect of S@200 on Spark and Zookeeper. Curves report 150 rolling Macro-F1 with a stride of 25 commits; the displayed sequences therefore begin at the first complete window, at commit index 150.}

\label{fig:switch_examples}
\end{figure}

Unlike the inference channel $F$, the CSTG channel $G$ is not equally useful from the start. It learns from the vocabulary accumulated in commit messages and diffs (\autoref{sec:kg-cstg}), so after a short warm-up, its term and co-occurrence statistics may still be too sparse for stable predictions. On the other hand, $F$ operates on graph topology (\autoref{sec:kg-infer}) and does not depend on a fitted vocabulary. The two channels also capture fundamentally different patterns, which may explain why their combination performs better in~\autoref{fig:rq4_f1}. We therefore use $F_{\mathrm{ov}}$ for the first 200 commits (after warm-up) and add $G$ when more project-specific language has accumulated. On both Spark (\autoref{fig:switch_spark}) and Zookeeper (\autoref{fig:switch_zookeeper}), S@200 avoids the weaker early performance of $F{+}G$ and stays ahead through much of the later stream, even after $G$ is introduced, so the early choice of channel appears to shape later adaptation. Noisy $G$ channel scores from cold start can carry over into subsequent updates, whereas starting with $F_{\mathrm{ov}}$ gives later predictions a cleaner history.

\subsection{Sensitivity of the Online Protocol to Operational Constraints and Hyperparameters}
\label{sec:sensitivity}
We study KG-Commit's sensitivity to the operational constraints of the online protocol and to the choice of refit interval $M$ and switching point $S$ (\autoref{sec:setup-protocol-section}). The warm-up ratio $K$ and verification gap $g$ determine how much historical information is available and when labels become observable (\autoref{sec:operational-constraints}). These sensitivity experiments use $F_{\mathrm{ov}}{+}G$ under the S@200 policy. This isolates the effects of the online protocol and its hyperparameters from the additional traversal/global-context features in $T$. The selected values are subsequently used unchanged in the final $F_{\mathrm{ov}}{+}G{+}T$ configuration.

We test $K\in\{0.05,0.10,0.20,0.30,0.40,0.50\}$ and $g\in\{0,10,25,50,100,200\}$. Macro-F1 remains stable across $K$ for most projects, with the largest variation occurring in the smaller Spark, Zeppelin, and Zookeeper histories. It is even less sensitive to $g$, whose curves are nearly flat except for a modest change on Zookeeper. We also test $M\in\{25,50,100,200\}$ and $S\in\{0,100,200,250,300,400,500,600,750,1K,1.5K,2K,\allowbreak3K,\infty\}$, where project history permits. Most projects remain stable across $M$, while the larger changes at $M{=}200$ again occur in smaller projects. Performance also changes little for $S$ between 200 and 500. Moving from $S{=}0$ to $S{=}200$, however, improves Macro-F1 by $0.012$ on Spark and $0.044$ on Zookeeper, consistent with the effect discussed in~\autoref{sec:switch-s200}.

\begin{figure}[!htbp]
\centering
    \includegraphics[width=0.95\columnwidth]
    {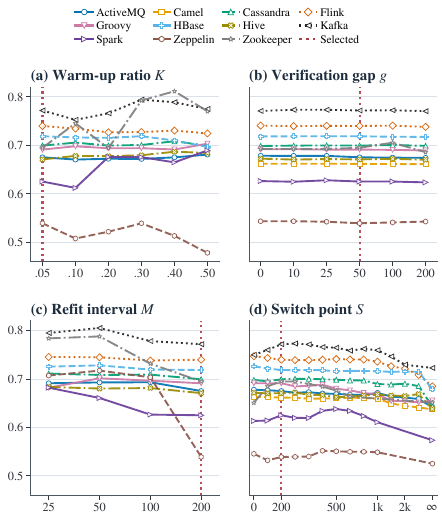}
\caption{Macro-F1 sensitivity of the $F_{\mathrm{ov}}{+}G$ configuration to the warm-up ratio $K$, verification gap $g$, refit interval $M$, and switch point $S$. Dotted lines mark the selected settings $K{=}0.05$, $g{=}50$, $M{=}200$, and $S{=}200$.}
\label{fig:sensitivity}
\end{figure}

The values used in our main experiments (\autoref{sec:results}) were fixed before obtaining the sensitivity results in~\autoref{fig:sensitivity}. They were chosen for practical deployment instead of aiming to maximize performance on each project. Using $K{=}0.05$ allows KG-Commit to begin predicting after only 5\% of the project history, although it trails the best tested $K$ by $0.064$ on Spark and $0.117$ on Zookeeper. On most other projects, the choice of $K$ has little effect. We similarly choose the largest refit interval, $M{=}200$, to reduce how often refitting occurs. This setting lowers Macro-F1 on some smaller projects, most clearly on Zeppelin, Zookeeper, and Spark, where performance decreases by $0.168$, $0.089$, and $0.057$ relative to $M{=}25$. Lowering $M$ would trigger refitting more frequently and increase its amortized $C_{\mathrm{refit}}/M$ cost. The selected setting therefore trades some performance on smaller projects for less frequent refitting, making our reported results for S@200 conservative.

\subsection{Online Threshold Adaptation}
\label{sec:online-threshold-adaptation}
\begin{figure}[!htbp]
\centering
\includegraphics[width=0.95\columnwidth]
    {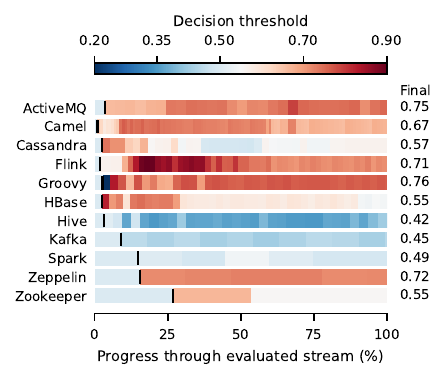}
\caption{Online decision thresholds across the evaluated commit streams. Color indicates the threshold in force, black marks the first retuning after 300 evaluated commits, and the final threshold is reported on the right.}
\label{fig:online-threshold}
\end{figure}

KG-Commit uses the default threshold of $0.5$ for the first 300 evaluated commits and then re-estimates it every 150 commits by maximizing Macro-F1 over the labels available at that point (\autoref{sec:setup-protocol-section}). The updated value is used from the next commit onward. As shown in~\autoref{fig:online-threshold}, the resulting operating point varies across projects. Cassandra and HBase remain close to $0.5$, whereas Camel, Flink, and Groovy move toward higher thresholds. Hive, Kafka, Spark, Zeppelin, and Zookeeper instead settle below $0.5$. Final thresholds range from $0.251$ on Hive to $0.748$ on Flink. A fixed cutoff would therefore impose different precision-recall trade-offs across projects. Periodic threshold tuning allows KG-Commit to adapt its decisions as the project’s score and class distributions evolve, without refitting the underlying scoring model.

\subsection{Limitations and Future Work}
\label{sec:limits}
KG-Commit currently runs entirely on CPU. This keeps deployment inexpensive but restricts the complexity of graph inference. The evaluated inference channels use mostly generic graph projections and are not customized to consider specific node types, relations, or more informative regions for the purpose of JIT-SDP. The $T$ channel exposes selected dependency, package, and graph-global context, but other aspects of the graph may remain unexploited, particularly fine-grained AST structure and richer typed or higher-order relational patterns. Therefore, although KG-Commit outperforms all evaluated baselines in predictive performance, the reported results should be viewed as a conservative estimate of what its representation can support under the current inference design, rather than as an upper bound on KG-Commit's inherent capability.

Furthermore, the current implementation is limited to Java because AST construction and delta extraction rely on \texttt{javalang}. Extending KG-Commit to other programming languages would require corresponding parsers and language-specific handling of source-code structure. In addition, each evaluated project is represented as a single repository. The current graph construction therefore does not model cross-repository dependencies or projects whose development history is distributed across multiple repositories.

Future work can examine other inference methods such as R-GCN, CompGCN, and HGT, together with projections designed around the entities and relations most relevant to a target commit for defect prediction. GPU-based inference can also make broader graph context practical, provided that the resulting end-to-end per-commit processing cost remains compatible with the commit rates expected in online JIT-SDP. Taken together, these directions suggest that the current results capture only an initial use of KG-Commit's representational capacity. Considerable room remains to exploit KG-Commit through richer inference and better graph projections. Beyond this, KG-Commit itself can be extended with additional information (e.g., caller/callee relations) to capture forms of project context not represented in the current design.

\section{Threats to Validity}
\label{sec:validity}
We discuss potential threats that may affect the credibility and generalizability of our results, grouped into internal, external, and construct validity.

\subsection{Internal Validity}
Temporal leakage is the main internal concern. The online protocol admits the current commit's unlabeled information but restricts labels and learned components to information available at prediction time (\autoref{sec:setup-protocol-section}). Baselines were adapted to the same protocol which may introduce implementation differences. We mitigate these risks through a common stream, fixed settings, and the reported sensitivity analyses.

\subsection{External Validity}
The evaluation uses projects from the Apache open-source ecosystem, so the observed performance may not transfer to projects with different development processes or defect distributions. Evaluating KG-Commit on industrial and other open-source repositories is needed to establish broader generalizability.

\subsection{Construct Validity}
ApacheJIT uses an SZZ-based procedure to identify bug-inducing commits, and these labels may contain incorrect or missing links (\autoref{sec:dataset}). KG-Commit also relies on heuristic choices in both the AST (\autoref{sec:kg-ast}) and CSTG layers (\autoref{sec:kg-cstg}). The AST layer uses heuristic matching to align nodes across edits, while CSTG uses rule-based term extraction, intent assignment, and term grounding. Moreover, efficiency and scalability are measured using execution time under the reported hardware and software environment, so absolute timings may differ on other systems.

\section{Conclusion}
\label{sec:conclusion}
This paper addressed a practical limitation of online JIT-SDP by making broader project context available at a computational cost compatible with real commit streams. KG-Commit maintains this context incrementally in an evolving knowledge graph. Its three layers capture repository history, within-file code structure, and commit semantics, while AST deltas maintain fine-grained changes through file evolution. The final $F_{\mathrm{ov}}{+}G{+}T$ predictor additionally incorporates traversal/global context from the full graph. Across 11 Apache projects, this configuration achieves the highest aggregate Macro-F1 ($0.704$), G-Mean ($0.706$), and AUC ($0.809$). Under a realistic online protocol, KG-Commit outperforms LR, HGB, RF, and DeepJIT on all 11 projects, LApredict on 10, and JITLine-online on 9 projects in Macro-F1, with the aggregate paired difference statistically significant in every case.

KG-Commit processes each commit in approximately $1.33$~s on CPU, with graph ingestion accounting for most of this cost. Lifecycle analysis shows that this per-commit processing remains stable as project history and graph size grow. Moreover, KG-Commit's processing time remains well below the observed time between consecutive commits. These results indicate that the computational cost of maintaining rich project context can be practical under the commit rates observed in real-world software projects. The current results are obtained using lightweight, CPU-only inference channels. Hence, they should be viewed as a conservative estimate of KG-Commit's potential. This leaves considerable room for richer inference methods to better exploit KG-Commit's representational capacity and make fuller use of the global structural and relational information maintained throughout project evolution.

\section{Declarations}

\subsection{Funding}
This research received no external funding.

\subsection{Ethical approval}
Ethical approval: Not applicable.

\subsection{Informed consent}
Informed consent: Not applicable.

\subsection{Data Availability Statement}
The datasets generated and/or analyzed during the current study are available at:
\begin{itemize}
    \item \textbf{Data repository:}
    \begin{itemize}
        \item Graph Dumps: 
        \url{https://zenodo.org/records/22348686}
        \item ApacheJIT Dataset: \url{https://zenodo.org/records/5907002}
    \end{itemize}
    \item \textbf{Code repository:}\\ \url{https://github.com/Knowledge4Software/kg-commit}
\end{itemize}

\subsection{Conflict of Interest}
The authors declare that they have no conflict of interest.

\subsection{Clinical trial number}
Clinical trial number: Not applicable.


\section{Declaration of generative AI and AI-assisted technologies in the manuscript preparation process}
During the preparation of this work, the authors used ChatGPT (OpenAI), using GPT-5.6 Sol, to improve the language, clarity, organization, and visual presentation of the manuscript, including the presentation of figures and other manuscript elements. After using this tool, the authors reviewed and edited the content as needed and take full responsibility for the content of the published article.

\bibliographystyle{elsarticle-harv}

\bibliography{cas-refs}

\appendix

\FloatBarrier

\section{Alternative Within-File Representations} \label{app:alternative-subgraphs} CFG, DFG, and PDG are constructed independently for each Java method using \code{CFGBuilder}, \code{DefUseBuilder}, and \code{CPGBuilder}, respectively. The CFG represents execution order and branching; the DFG represents definition-use value flow, and the PDG combines control and data dependencies.~\autoref{tab:alternative-subgraphs} summarizes their construction. Construct coverage therefore follows the nodes produced by the respective builder.

\begin{table*}[!htbp]
\centering
\caption{Construction and delta extraction for the alternative within-file representations.}
\label{tab:alternative-subgraphs}
\small
\setlength{\tabcolsep}{4pt}
\renewcommand{\arraystretch}{1.15}
\begin{tabularx}{\textwidth}{
    l
    l
    >{\raggedright\arraybackslash}X
    >{\raggedright\arraybackslash}X
    >{\raggedright\arraybackslash}X}
\toprule
\textbf{Rep.} &
\textbf{Builder} &
\textbf{Nodes / edges} &
\textbf{Matching and cycles} &
\textbf{Validation / cost} \\
\midrule

CFG & \code{CFGBuilder} & Nodes are typed by control-flow \code{kind}; directed edges encode execution successors and branches. & Two-pass ordinal matching within each method: exact $(\text{method},\text{type},\text{ordinal})$, followed by deterministic greedy same-type matching. Cycles require no special handling because matching does not recursively traverse graph edges. & Built and matched only for affected methods using deterministic construction and matching. \\

DFG & \code{DefUseBuilder} & Definition nodes use \code{def_type}; use nodes are typed \code{USE}; edges encode definition--use value flow. & Same ordinal matcher as CFG. Multiple predecessors do not require a unique tree parent; move detection uses the primary predecessor. & Same deterministic procedure and scope as CFG. \\

PDG & \code{CPGBuilder} & Nodes are typed by \code{kind}; edges combine control and data dependencies. & Same ordinal matcher as CFG and DFG; no canonical child ordering is assumed. & Same deterministic procedure and scope as CFG/DFG. \\

\bottomrule
\end{tabularx}
\end{table*}

\begin{figure*}[!htbp]
    \centering

    \tikzset{
        graphnode/.style={
            draw,
            rounded corners=1.5pt,
            align=center,
            minimum height=6.5mm,
            minimum width=15mm,
            inner sep=2pt,
            font=\scriptsize
        },
        valnode/.style={
            draw,
            circle,
            minimum size=6mm,
            inner sep=1pt,
            font=\scriptsize
        },
        flow/.style={
            ->,
            thick
        },
        data/.style={
            ->,
            thick,
            dashed
        },
        edgelabel/.style={
            font=\scriptsize,
            fill=white,
            inner sep=1pt
        }
    }

    \begin{subfigure}[t]{0.23\textwidth}
        \vspace{0pt}
        \centering

        \fbox{
            \begin{minipage}[t][5.05cm][t]{0.88\linewidth}
                \vspace{2mm}
                \normalsize
                \ttfamily

                if (x > 0) \{\\[1mm]
                \hspace*{1em}y = x + 1;\\[1mm]
                \} else \{\\[1mm]
                \hspace*{1em}y = 0;\\[1mm]
                \}\\[1mm]
                print(y);
            \end{minipage}%
        }

        \caption{Source fragment.}
        \label{fig:code-example}
    \end{subfigure}
    \hfill
    \begin{subfigure}[t]{0.23\textwidth}
        \vspace{0pt}
        \centering

        \begin{tikzpicture}[x=0.9cm,y=0.9cm]

            \node[graphnode] (entry) at (0,4.4)
                {Entry};

            \node[graphnode] (condition) at (0,3.25)
                {$x>0$};

            \node[graphnode] (then) at (-1.15,1.85)
                {$y\gets x+1$};

            \node[graphnode] (else) at (1.15,1.85)
                {$y\gets 0$};

            \node[graphnode] (print) at (0,0.50)
                {\texttt{print(y)}};

            \node[graphnode] (exit) at (0,-0.60)
                {Exit};

            \draw[flow]
                (entry) -- (condition);

            \draw[flow]
                (condition)
                -- node[edgelabel,above left] {T}
                (then);

            \draw[flow]
                (condition)
                -- node[edgelabel,above right] {F}
                (else);

            \draw[flow]
                (then) -- (print);

            \draw[flow]
                (else) -- (print);

            \draw[flow]
                (print) -- (exit);

        \end{tikzpicture}

        \caption{Control-flow graph.}
        \label{fig:cfg-example}
    \end{subfigure}
    \hfill
    \begin{subfigure}[t]{0.23\textwidth}
        \vspace{0pt}
        \centering

        \begin{tikzpicture}[x=0.9cm,y=0.9cm]

            \node[valnode] (x) at (-1.25,4.10)
                {$x$};

            \node[valnode] (one) at (0,4.10)
                {$1$};

            \node[valnode] (zero) at (1.25,4.10)
                {$0$};

            \node[graphnode] (addition) at (-0.55,2.85)
                {$x+1$};

            \node[graphnode] (ytrue) at (-0.95,1.60)
                {$y_{\mathrm{T}}$};

            \node[graphnode] (yfalse) at (0.95,1.60)
                {$y_{\mathrm{F}}$};

            \node[valnode] (merge) at (0,0.45)
                {$\phi_y$};

            \node[graphnode] (print) at (0,-0.65)
                {\texttt{print(y)}};

            \draw[data]
                (x) -- (addition);

            \draw[data]
                (one) -- (addition);

            \draw[data]
                (addition) -- (ytrue);

            \draw[data]
                (zero) -- (yfalse);

            \draw[data]
                (ytrue) -- (merge);

            \draw[data]
                (yfalse) -- (merge);

            \draw[data]
                (merge) -- (print);

        \end{tikzpicture}

        \caption{Data-flow graph.}
        \label{fig:dfg-example}
    \end{subfigure}
    \hfill
    \begin{subfigure}[t]{0.23\textwidth}
        \vspace{0pt}
        \centering

        \begin{tikzpicture}[x=0.9cm,y=0.9cm]

            \node[valnode] (x) at (-1.35,4.00)
                {$x$};

            \node[graphnode] (condition) at (0,3.15)
                {$x>0$};

            \node[graphnode] (then) at (-1.10,1.75)
                {$y\gets x+1$};

            \node[graphnode] (else) at (1.10,1.75)
                {$y\gets 0$};

            \node[graphnode] (print) at (0,0.25)
                {\texttt{print(y)}};

            \draw[flow]
                (condition)
                -- node[edgelabel,above left] {T}
                (then);

            \draw[flow]
                (condition)
                -- node[edgelabel,above right] {F}
                (else);

            \draw[data]
                (x) -- (condition);

            \draw[data]
                (x) -- (then);

            \draw[data]
                (then) -- (print);

            \draw[data]
                (else) -- (print);

            \draw[flow]
                (-1.45,-0.65) -- (-0.75,-0.65);

            \node[
                font=\tiny,
                anchor=west
            ] at (-0.70,-0.65)
                {control};

            \draw[data]
                (0.20,-0.65) -- (0.90,-0.65);

            \node[
                font=\tiny,
                anchor=west
            ] at (0.95,-0.65)
                {data};

        \end{tikzpicture}

        \caption{Program dependence graph.}
        \label{fig:pdg-example}
    \end{subfigure}

    \caption{The same conditional code fragment represented using alternative structural representations: (a) the source fragment, (b) the CFG, (c) the DFG, and (d) the PDG. In the PDG, solid edges denote control dependencies, whereas dashed edges denote data dependencies.}
\label{fig:alternative-subgraphs}
\end{figure*}
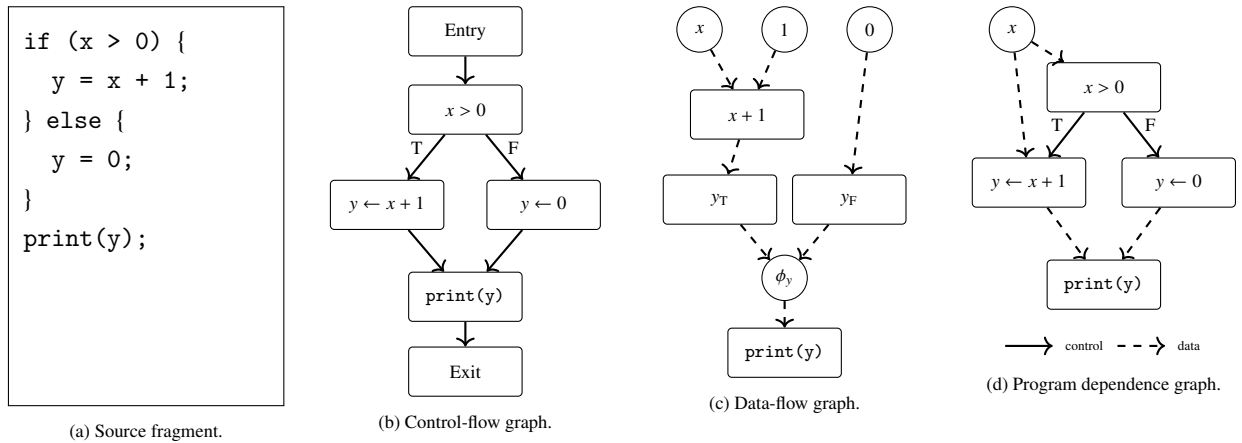

Unlike the AST, these graphs do not have canonically ordered children or a unique parent, so subtree hashing cannot be applied directly. Their shared matcher instead ranks same-type nodes within a method by $(\code{line},\code{id})$ and matches them by ordinal, with a deterministic same-type greedy pass for remaining nodes. A matched node is marked \edge{MOVES} when its primary predecessor changes to a different predecessor type; unmatched nodes yield \edge{ADDS} and \edge{REMOVES}, and changed matched nodes yield \edge{UPDATES}. If identity matching is unstable, the implementation provides a conservative typed-multiset fallback containing only additions and removals. Thus, CFG, DFG, and PDG differ in the program structure they encode, while their matching and downstream delta interface are held fixed.~\autoref{fig:alternative-subgraphs} illustrates the three representations.

\section{Layer-Wise Results for Individual Inference Channels}
\label{app:rq3_single_inference}
To complement the fixed-$F_{\mathrm{ov}}$ analysis in \autoref{sec:rq3}, we report the layer-wise evaluation for the five individual graph-inference channels in \autoref{tab:rq3_single_inference_matrix}. These results are kept separate from the main analysis because the contribution of a representation is not invariant to the inference mechanism. The magnitude, and occasionally the ordering, of layer-level gains can change across inference channels and projects. The main text therefore holds inference fixed at $F_{\mathrm{ov}}$ when attributing performance changes to the representation.

\begin{table*}[!htbp]
\centering
\caption{Layer-wise online performance of the five individual graph-inference channels across all 11 projects. Underlining marks the best representation within each inference row.}
\label{tab:rq3_single_inference_matrix}
\scriptsize
\setlength{\tabcolsep}{2.0pt}
\renewcommand{\arraystretch}{0.88}
\adjustbox{max width=\textwidth,max totalheight=0.90\textheight}{
\begin{tabular}{@{}ll ccc ccc ccc ccc ccc ccc@{}}
\toprule
Project & Inf. & \multicolumn{3}{c}{\textsc{Layer 1}} & \multicolumn{12}{c}{\textsc{Layer 2: within-file subgraph}} & \multicolumn{3}{c}{\textsc{Layer 3}} \\
\cmidrule(lr){3-5} \cmidrule(lr){6-17} \cmidrule(lr){18-20}
 & & \multicolumn{3}{c}{Core} & \multicolumn{3}{c}{+AST} & \multicolumn{3}{c}{+CFG} & \multicolumn{3}{c}{+DFG} & \multicolumn{3}{c}{+PDG} & \multicolumn{3}{c}{CSTG} \\
\cmidrule(lr){3-5} \cmidrule(lr){6-8} \cmidrule(lr){9-11} \cmidrule(lr){12-14} \cmidrule(lr){15-17} \cmidrule(lr){18-20}
 & & $F_1$ & $G$ & AUC & $F_1$ & $G$ & AUC & $F_1$ & $G$ & AUC & $F_1$ & $G$ & AUC & $F_1$ & $G$ & AUC & $F_1$ & $G$ & AUC \\
\midrule
\multirow{5}{*}{\texttt{ActiveMQ}} & RN & \underline{.483} & \underline{.563} & \underline{.675} & .340 & .389 & .555 & .433 & .509 & .593 & .432 & .508 & .614 & .435 & .511 & .592 & .348 & .401 & .561 \\
 & PPR & .532 & .604 & .671 & .589 & .663 & .704 & .569 & .646 & .699 & .571 & .647 & .696 & .573 & .648 & .700 & \underline{.660} & \underline{.702} & \underline{.778} \\
 & LP & \underline{.442} & \underline{.519} & .650 & .294 & .318 & .575 & .349 & .400 & .606 & .350 & .402 & .614 & .341 & .389 & .606 & .437 & .514 & \underline{.670} \\
 & DW & .505 & .573 & .602 & .579 & .629 & .682 & .547 & .612 & .648 & .537 & .606 & .636 & .544 & .607 & .646 & \underline{.609} & \underline{.665} & \underline{.711} \\
 & KGE & .530 & .593 & .624 & .583 & .651 & .701 & .527 & .603 & .639 & .516 & .592 & .644 & .528 & .607 & .653 & \underline{.616} & \underline{.670} & \underline{.719} \\
\midrule
\multirow{5}{*}{\texttt{Camel}} & RN & \underline{.588} & \underline{.648} & \underline{.706} & .371 & .524 & .626 & .435 & .595 & .663 & .473 & .632 & .683 & .433 & .593 & .662 & .379 & .533 & .629 \\
 & PPR & .535 & .654 & .696 & .508 & .648 & .687 & .522 & .664 & .690 & .528 & .665 & .693 & .519 & .660 & .689 & \underline{.635} & \underline{.715} & \underline{.772} \\
 & LP & .458 & .607 & \underline{.664} & .181 & .245 & .553 & .360 & .512 & .613 & .395 & .553 & .625 & .360 & .511 & .613 & \underline{.471} & \underline{.634} & .654 \\
 & DW & .478 & .570 & .608 & .554 & .635 & .697 & .514 & .621 & .662 & .510 & .612 & .646 & .530 & .626 & .667 & \underline{.569} & \underline{.669} & \underline{.718} \\
 & KGE & .464 & .596 & .624 & .558 & .661 & .705 & .485 & .620 & .645 & .516 & .642 & .673 & .506 & .631 & .658 & \underline{.590} & \underline{.689} & \underline{.748} \\
\midrule
\multirow{5}{*}{\texttt{Cassandra}} & RN & .389 & .330 & \underline{.621} & .396 & .338 & .570 & \underline{.459} & \underline{.434} & .609 & .429 & .392 & .609 & .455 & .428 & .606 & .405 & .352 & .577 \\
 & PPR & .489 & .481 & .635 & .563 & .571 & .668 & .525 & .525 & .657 & .532 & .534 & .658 & .526 & .526 & .657 & \underline{.610} & \underline{.621} & \underline{.737} \\
 & LP & \underline{.331} & \underline{.221} & .560 & .327 & .212 & .505 & .328 & .214 & .523 & .328 & .214 & .515 & .328 & .214 & .523 & .327 & .212 & \underline{.599} \\
 & DW & .589 & .602 & .662 & .611 & .625 & .671 & .613 & .626 & .667 & .572 & .584 & .641 & .612 & .626 & .670 & \underline{.629} & \underline{.643} & \underline{.704} \\
 & KGE & .479 & .471 & .596 & .597 & .610 & .672 & .555 & .562 & .621 & .530 & .534 & .592 & .527 & .530 & .598 & \underline{.643} & \underline{.657} & \underline{.697} \\
\midrule
\multirow{5}{*}{\texttt{Flink}} & RN & \underline{.581} & \underline{.649} & \underline{.731} & .389 & .431 & .546 & .406 & .453 & .570 & .406 & .454 & .562 & .406 & .454 & .570 & .391 & .434 & .554 \\
 & PPR & .646 & .660 & .755 & .622 & .679 & .722 & .626 & .682 & .715 & .620 & .679 & .714 & .620 & .680 & .715 & \underline{.715} & \underline{.740} & \underline{.800} \\
 & LP & \underline{.459} & \underline{.521} & .622 & .349 & .373 & .550 & .379 & .417 & .577 & .367 & .400 & .573 & .379 & .417 & .577 & .363 & .394 & \underline{.625} \\
 & DW & .555 & .610 & .658 & .590 & .639 & .696 & .602 & \underline{.645} & .688 & .581 & .629 & .673 & \underline{.603} & \underline{.645} & .689 & .578 & .642 & \underline{.731} \\
 & KGE & .506 & .568 & .602 & .629 & .666 & .721 & .562 & .605 & .639 & .506 & .569 & .603 & .534 & .593 & .626 & \underline{.645} & \underline{.685} & \underline{.759} \\
\midrule
\multirow{5}{*}{\texttt{Groovy}} & RN & \underline{.519} & \underline{.614} & \underline{.709} & .365 & .445 & .650 & .426 & .520 & .651 & .446 & .542 & .666 & .420 & .512 & .648 & .376 & .458 & .658 \\
 & PPR & .593 & .668 & .726 & .595 & .676 & .717 & .596 & .681 & .730 & .622 & .686 & .729 & .597 & .683 & .729 & \underline{.674} & \underline{.707} & \underline{.781} \\
 & LP & .469 & .566 & .698 & .342 & .414 & .654 & .382 & .466 & .673 & .390 & .476 & .672 & .388 & .474 & .672 & \underline{.486} & \underline{.589} & \underline{.714} \\
 & DW & .548 & .600 & .650 & .571 & .641 & .688 & .558 & .630 & .672 & .551 & .618 & .655 & .554 & .629 & .673 & \underline{.619} & \underline{.673} & \underline{.731} \\
 & KGE & .487 & .577 & .634 & .610 & \underline{.678} & .718 & .570 & .641 & .679 & .530 & .608 & .665 & .543 & .618 & .647 & \underline{.626} & .672 & \underline{.725} \\
\midrule
\multirow{5}{*}{\texttt{HBase}} & RN & .381 & .271 & \underline{.619} & .364 & .230 & .559 & .452 & .387 & .592 & \underline{.467} & \underline{.409} & .586 & .449 & .383 & .590 & .377 & .257 & .565 \\
 & PPR & .488 & .445 & .658 & .547 & .522 & .676 & .574 & .560 & .680 & .583 & .572 & .675 & .579 & .566 & .679 & \underline{.682} & \underline{.684} & \underline{.764} \\
 & LP & \underline{.377} & \underline{.268} & .554 & .360 & .225 & .493 & .363 & .232 & .518 & .363 & .232 & .515 & .363 & .232 & .517 & .368 & .243 & \underline{.587} \\
 & DW & .560 & .544 & .632 & .648 & .650 & .685 & .623 & .623 & .655 & .592 & .587 & .620 & .619 & .618 & .651 & \underline{.651} & \underline{.654} & \underline{.694} \\
 & KGE & .475 & .435 & .604 & .634 & .633 & .702 & .578 & .568 & .652 & .578 & .567 & .639 & .554 & .537 & .629 & \underline{.637} & \underline{.637} & \underline{.722} \\
\midrule
\multirow{5}{*}{\texttt{Hive}} & RN & .475 & .315 & .655 & .445 & .258 & .652 & .487 & .333 & .676 & \underline{.514} & \underline{.378} & \underline{.678} & .508 & .370 & .674 & .443 & .252 & .657 \\
 & PPR & .467 & .314 & .639 & .454 & .295 & .637 & .444 & .274 & .642 & .456 & .298 & .644 & .446 & .278 & .642 & \underline{.599} & \underline{.520} & \underline{.754} \\
 & LP & .465 & .291 & .638 & .383 & .049 & .520 & .398 & .127 & .593 & .401 & .138 & .572 & .399 & .132 & .593 & \underline{.494} & \underline{.342} & \underline{.679} \\
 & DW & .447 & .267 & .663 & .606 & .551 & .671 & .565 & .477 & .653 & .499 & .365 & .656 & .545 & .445 & .650 & \underline{.624} & \underline{.567} & \underline{.677} \\
 & KGE & .403 & .155 & .585 & .536 & .426 & .640 & .494 & .353 & .606 & .493 & .356 & .607 & .486 & .341 & .576 & \underline{.576} & \underline{.490} & \underline{.694} \\
\midrule
\multirow{5}{*}{\texttt{Kafka}} & RN & .533 & .494 & .689 & .464 & .395 & .677 & .539 & .500 & \underline{.712} & \underline{.570} & \underline{.540} & .711 & .548 & .513 & .709 & .465 & .396 & .689 \\
 & PPR & .428 & .373 & .644 & .535 & .509 & .669 & .530 & .505 & .671 & .528 & .502 & .669 & .511 & .478 & .673 & \underline{.681} & \underline{.679} & \underline{.803} \\
 & LP & .569 & .546 & .665 & .438 & .354 & .595 & .512 & .463 & .639 & .522 & .476 & .640 & .516 & .469 & .642 & \underline{.652} & \underline{.639} & \underline{.753} \\
 & DW & .621 & .616 & .719 & .682 & .684 & .726 & \underline{.706} & \underline{.709} & .729 & .687 & .689 & .726 & .705 & .708 & .730 & .698 & .700 & \underline{.748} \\
 & KGE & .487 & .446 & .594 & .591 & .573 & .602 & .532 & .503 & .564 & .547 & .519 & .604 & .576 & .556 & .601 & \underline{.664} & \underline{.662} & \underline{.675} \\
\midrule
\multirow{5}{*}{\texttt{Spark}} & RN & \underline{.415} & \underline{.379} & .545 & .338 & .214 & .581 & .364 & .273 & .567 & .365 & .279 & .534 & .370 & .285 & .565 & .337 & .214 & \underline{.589} \\
 & PPR & .402 & .324 & .612 & .361 & .246 & .566 & .382 & .283 & .573 & .374 & .276 & .558 & .398 & .312 & .571 & \underline{.540} & \underline{.520} & \underline{.699} \\
 & LP & \underline{.375} & \underline{.330} & .503 & .356 & .269 & .509 & .372 & .319 & .514 & .371 & .326 & .499 & .369 & .314 & .510 & .374 & .304 & \underline{.577} \\
 & DW & .432 & .370 & .564 & .488 & \underline{.464} & .602 & .372 & .285 & .539 & .372 & .290 & .546 & .376 & .283 & .546 & \underline{.490} & \underline{.464} & \underline{.606} \\
 & KGE & .348 & .254 & .486 & .462 & .425 & \underline{.599} & \underline{.508} & \underline{.491} & .583 & .429 & .382 & .531 & .432 & .387 & .549 & .496 & .481 & .593 \\
\midrule
\multirow{5}{*}{\texttt{Zeppelin}} & RN & .535 & .514 & \underline{.569} & .542 & .519 & .543 & .542 & .519 & .541 & \underline{.544} & \underline{.522} & .518 & .542 & .519 & .540 & .543 & .520 & .548 \\
 & PPR & .499 & .471 & .588 & .582 & .572 & .648 & .565 & .547 & .660 & .570 & .553 & .640 & .566 & .548 & .652 & \underline{.625} & \underline{.619} & \underline{.688} \\
 & LP & .542 & .519 & .525 & \underline{.543} & \underline{.520} & .517 & .542 & .519 & .519 & .541 & .518 & .510 & .541 & .518 & .517 & .541 & .519 & \underline{.573} \\
 & DW & .480 & .472 & .568 & .514 & .509 & .610 & .578 & .573 & \underline{.631} & .547 & .540 & .606 & \underline{.581} & \underline{.577} & .621 & .527 & .514 & .627 \\
 & KGE & .531 & .504 & .597 & \underline{.609} & \underline{.603} & \underline{.660} & .549 & .525 & .615 & .532 & .500 & .615 & .566 & .539 & .658 & .589 & .586 & .608 \\
\midrule
\multirow{5}{*}{\texttt{Zookeeper}} & RN & .424 & \underline{.420} & \underline{.567} & .429 & .373 & .475 & .456 & .418 & .524 & \underline{.457} & \underline{.420} & .558 & .455 & .415 & .520 & .445 & .399 & .497 \\
 & PPR & .443 & .448 & .588 & .491 & .492 & .560 & .430 & .417 & .532 & .404 & .381 & .511 & .416 & .395 & .530 & \underline{.655} & \underline{.665} & \underline{.756} \\
 & LP & \underline{.397} & \underline{.382} & \underline{.561} & .390 & .367 & .401 & .393 & .371 & .429 & .393 & .371 & .440 & .393 & .371 & .429 & .395 & .375 & .518 \\
 & DW & .338 & .254 & .592 & .487 & .475 & .640 & .377 & .329 & .543 & .363 & .314 & .523 & .375 & .327 & .545 & \underline{.605} & \underline{.616} & \underline{.653} \\
 & KGE & .385 & .332 & .512 & .552 & .544 & .545 & .503 & .489 & .537 & .462 & .436 & .521 & .518 & .507 & .572 & \underline{.616} & \underline{.626} & \underline{.625} \\
\bottomrule
\end{tabular}}
\end{table*}

\section{Commit--Hub Projection and Graph-Based Inference}
\label{app:commit-hub-projection}
The five graph-based inference channels operate on the commit--hub projection introduced in~\autoref{sec:kg-projection}. At commit $j$, this projection is a weighted bipartite graph
\begin{equation}
    \mathcal{P}^{(j)}
    =
    \left(
        \mathcal{C}^{(j)} \cup \mathcal{H}^{(j)},
        \mathcal{E}^{(j)}
    \right),
    \qquad
    \mathcal{E}^{(j)}
    \subseteq
    \mathcal{C}^{(j)} \times \mathcal{H}^{(j)},
\end{equation}
where $\mathcal{C}^{(j)}$ contains the commit nodes available at that point and $\mathcal{H}^{(j)}$ contains inference-level context hubs. Four hub types are used: \node{File}s, \node{Developer}s, change tokens, and CSTG \node{Term}s. A hub is therefore an entity that can be shared by multiple commits and provides a path between commits with related context.

A change token is a pair $(r,\tau)$, where $r\in\{\edge{ADDS},\edge{REMOVES},\edge{UPDATES},\edge{MOVES}\}$ is the structural edit operation and $\tau$ is the type of the affected \node{ASTNode} defined in~\autoref{sec:kg-ast}. The AST delta of a commit is represented as a multiset of these tokens. For example, if a commit adds three \texttt{MethodInvocation} nodes, the corresponding $(\edge{ADDS},\texttt{MethodInvocation})$ token occurs three times. File and developer hubs are obtained from the Core layer, while CSTG term hubs are obtained from the semantic layer.

Two commits that share a hub are connected through the path $c_i\rightarrow h\rightarrow c_k$ in $\mathcal{P}^{(j)}$. We call this one commit-to-commit hop, which corresponds to two edges in the bipartite projection. Two commit-to-commit hops therefore correspond to four bipartite edges. Relations that are present in the full knowledge graph but do not form commit--hub links, such as \edge{AST_CHILD}, commit ancestry, \edge{COOCCURS}, and file-to-file \edge{IMPORTS}, are not edges of this projection.

Only labels that have become available under the gap constraint are used during inference. As defined in~\autoref{sec:operational-constraints}, let $\mathcal{C}_{\mathrm{past}}$ denote these gap-resolved commits. For a commit $c$ and hub $h$, let $\operatorname{tf}(c,h)$ denote the number of observed occurrences of $h$ in $c$. The commit--hub incidence weight is
\begin{equation}
W_{ch}
=
\operatorname{tf}(c,h)
\log\left(
    1+
    \frac{
        |\mathcal{C}_{\mathrm{past}}|
    }{
        \operatorname{df}_{\mathrm{past}}(h)
    }
\right),
\end{equation}
where $\operatorname{df}_{\mathrm{past}}(h)$ is the number of gap-resolved past commits connected to $h$. The first term reflects how strongly a hub occurs in the current commit, while the logarithmic term reduces the influence of hubs that occur in many previous commits.

Let $W$ denote the resulting weighted commit--hub incidence matrix. The adjacency matrix of the bipartite projection is
\begin{equation}
A_{\mathcal{P}}
=
\begin{bmatrix}
    0 & W\\
    W^{\top} & 0
\end{bmatrix},
\end{equation}
and its column-normalized transition matrix is
\begin{equation}
P
=
A_{\mathcal{P}}
\operatorname{diag}
\left(
    A_{\mathcal{P}}^{\top}\mathbf{1}
\right)^{\dagger},
\end{equation}
where ${}^{\dagger}$ denotes the element-wise reciprocal with zero entries kept at zero.

Although RN, LP, PPR, DW, and KGE all use this same projection, they use it in different ways. RN performs a weighted vote from commits connected through shared hubs~\citep{macskassy2003probabilistic}. LP performs three commit--hub--commit propagation rounds and re-clamps the labels in $\mathcal{C}_{\mathrm{past}}$ after every round~\citep{zhu2003semi}. PPR performs class-seeded propagation separately for buggy and clean commits over 40 bipartite edges~\citep{haveliwala2002topic}.

DW and KGE use the same projection but learn a representation from it rather than repeatedly following explicit paths at prediction time. DW applies truncated SVD to a positive-PMI transformation of the past commit--hub incidence matrix~\citep{perozzi2014deepwalk}. KGE represents projected edges as typed triples $(c,r,h)$ and learns DistMult embeddings~\citep{yang2015embedding}. The relation $r$ distinguishes structural-token, file, developer, and CSTG-term hubs. These two methods can therefore capture patterns beyond a single shared hub through their learned representations.

\begin{table}[!htbp]
\centering
\caption{Inference scope of the five graph-based channels on the commit--hub projection.}
\label{tab:inference-reach}
\small
\setlength{\tabcolsep}{4pt}
\renewcommand{\arraystretch}{1.12}

\begin{tabularx}{\columnwidth}{l X l}
\toprule
\textbf{Method} &
\textbf{Mechanism} &
\textbf{Explicit reach} \\
\midrule

RN &
Weighted shared-hub vote &
2 edges (1 hop) \\

LP &
Clamped label diffusion &
6 edges (3 hops) \\

PPR &
Class-seeded personalized PageRank &
40 edges ($\leq20$ hops) \\

DW &
PPMI matrix factorization &
2 edges + global SVD \\

KGE &
DistMult over commit--hub triples &
1 edge + global embedding \\

\bottomrule
\end{tabularx}
\end{table}

\section{CSTG Semantic Channel Features}
\label{app:g-channel-features}
The feature-based $G$ channel uses the semantic information constructed by the CSTG layer in~\autoref{sec:kg-cstg}. Unlike the five graph-based channels, it does not produce its score from the commit--hub projection. Instead, it combines the textual-risk, term-type, intent, structural consistency, and hashed-text information of the current commit into a single feature vector and learns a commit-level defect probability.

For a commit $c$, let $\rho(c)$ denote the propagated textual-risk prior defined in~\autoref{sec:kg-cstg}. The five semantic-mass features are
\begin{equation}
\begin{aligned}
\mathbf{M}(c)
=
\big[
    &M_{\mathrm{code}}(c),
    M_{\mathrm{bug}}(c),
    M_{\mathrm{action}}(c),\\
    &M_{\mathrm{error}}(c),
    M_{\mathrm{natural-language}}(c)
\big].
\end{aligned}
\end{equation}
where each component gives the CSTG mass assigned to the corresponding term type. The intent assigned to the commit is represented by an indicator vector
\begin{equation}
    \mathbf{i}(c)\in\{0,1\}^{7},
\end{equation}
whose dimensions correspond to $\{\textsc{fix},\textsc{feat},\textsc{refactor}, \textsc{test},\textsc{docs},\textsc{perf},\textsc{revert}\}$. A commit assigned the \textsc{other} intent receives the all-zero vector.

In addition to the intent label itself, we use seven features that compare the stated intent of the commit with the size and structure of its actual change. Let
\begin{equation}
    \mathbf{q}(c)
    =
    [q_1(c),\ldots,q_7(c)]
\end{equation}
denote this intent-realization block. The seven features are defined in \autoref{tab:cstg-consistency-features}.

\begin{table}[!htbp]
\centering
\caption{Intent realization and cross-modal consistency features used by the CSTG semantic channel. $\mathbb{I}[\cdot]$ denotes the indicator function, $|\delta_c|$ is the size of the structural delta, and $a_c$ and $r_c$ denote the numbers of additions and removals, respectively.}
\label{tab:cstg-consistency-features}
\scriptsize
\setlength{\tabcolsep}{3pt}
\renewcommand{\arraystretch}{1.10}

\begin{tabularx}{\columnwidth}{@{}lX@{}}
\toprule
\textbf{Feature} & \textbf{Definition} \\
\midrule

Small-change mismatch
&
$q_1=\mathbb{I}[\text{minor/trivial/typo}]\log(1+|\delta_c|)$
\\

Refactor add/remove
&
$q_2=\mathbb{I}[\operatorname{intent}(c)=\textsc{refactor}]
\dfrac{a_c+r_c}{|\delta_c|+1}$
\\

Fix size
&
$q_3=\mathbb{I}[\operatorname{intent}(c)=\textsc{fix}]
\log(1+|\delta_c|)$
\\

Feature without additions
&
$q_4=\mathbb{I}[\operatorname{intent}(c)=\textsc{feat}]
\left(1-\dfrac{a_c}{|\delta_c|+1}\right)$
\\

Terse large change
&
$q_5=\mathbb{I}[|\operatorname{words}(c)|\leq4]
\log(1+\mathit{la}_c+\mathit{ld}_c)$
\\

Grounding gap
&
$q_6=\log(1+|\operatorname{codeTerms}(c)|)-\log(1+|\delta_c|)$
\\

Revert
&
$q_7=\mathbb{I}[\operatorname{intent}(c)=\textsc{revert}]$
\\

\bottomrule
\end{tabularx}
\end{table}
Finally, let
\begin{equation}
    \mathbf{h}(c)\in\mathbb{R}^{2^{18}}
\end{equation}
denote the sparse hashed-text representation of the commit. Each typed CSTG term is converted to a tagged token and mapped to one of $2^{18}$ feature positions using feature hashing~\citep{weinberger2009feature}. The term weights computed by CSTG are carried into this representation, so terms with greater semantic weight contribute more strongly.

The complete input to the semantic channel is
\begin{equation}
    \mathbf{z}_c
    =
    \left[
        \rho(c),
        \mathbf{M}(c),
        \mathbf{i}(c),
        \mathbf{q}(c),
        \mathbf{h}(c)
    \right].
\end{equation}
A class-balanced logistic-regression model is fitted using the currently available commits in $\mathcal{C}_{\mathrm{past}}$. The semantic-channel score is
\begin{equation}
    G(c)
    =
    \sigma\left(
        \beta_0+
        \boldsymbol{\beta}^{\top}\mathbf{z}_c
    \right).
\end{equation}
Thus, $G(c)$ is a single commit-level probability summarizing the semantic evidence extracted by CSTG and is passed to the common score-fusion model.

\section{Traversal and Global-Context Features}
\label{app:tgc-features}
The commit--hub projection in~\autoref{app:commit-hub-projection} gives the five graph-based inference methods a compact representation of each commit. However, not every relation stored in KG-Commit appears in this projection. In particular, the file-to-file \edge{IMPORTS} relations and package relationships remain in the full knowledge graph. We use a traversal/global-context (TGC) feature block to make a small part of this information available to the predictor without running the graph-based inference methods directly over the full graph.

For a target commit $c=c^{(j)}$, let $\mathcal{F}_0(c)$ denote the set of files touched by the commit. These are the files connected to $c$ through the file-change relations in the Core layer, including \edge{ADDED}, \edge{MODIFIED}, \edge{DELETED}, \edge{RENAMED_FROM}, and \edge{RENAMED_TO}. We use this set as the starting point for TGC. The corresponding feature block is denoted by $\mathbf{T}_0(c)$ and provides the reference information for the files directly involved in the change.

To obtain cross-file context, consider the file-level import graph available at commit $j$,
\begin{equation}
    \mathcal{D}^{(j)}
    =
    \left(
        \mathcal{F}^{(j)},
        \mathcal{I}^{(j)}
    \right),
\end{equation}
where $\mathcal{F}^{(j)}$ is the set of files and
\begin{equation}
    (f_a,f_b)\in\mathcal{I}^{(j)}
    \quad\Longleftrightarrow\quad
    f_a \xrightarrow{\edge{IMPORTS}} f_b
    \text{ in }\mathcal{K}^{(j)}.
\end{equation}
The direction of this relation is important. If a touched file imports another file, the second file is an \emph{importee}; if another file imports a touched file, that file is an \emph{importer}. We therefore define the two one-hop sets
\begin{align}
    \mathcal{F}^{\rightarrow}_1(c)
    &=
    \left\{
        f'\notin\mathcal{F}_0(c)
        \;\middle|\;
        \exists f\in\mathcal{F}_0(c):
        (f,f')\in\mathcal{I}^{(j)}
    \right\},\\
    \mathcal{F}^{\leftarrow}_1(c)
    &=
    \left\{
        f'\notin\mathcal{F}_0(c)
        \;\middle|\;
        \exists f\in\mathcal{F}_0(c):
        (f',f)\in\mathcal{I}^{(j)}
    \right\}.
\end{align}
Thus, $\mathcal{F}^{\rightarrow}_1(c)$ contains untouched files imported by the changed files, whereas $\mathcal{F}^{\leftarrow}_1(c)$ contains untouched files that import the changed files. TGC keeps these two directions separate rather than merging them into a single neighborhood. Their features form the two direction-specific blocks $\mathbf{T}_{1,\mathrm{dep}}(c)$ and $\mathbf{T}_{1,\mathrm{dry}}(c)$.

This distinction allows TGC to use information that the commit--hub projection does not directly contain. Suppose, for example, that commit $c$ changes file $f$, while an untouched file $f'$ still imports $f$. The projection describes $c$ through its file, developer, structural-token, and CSTG-term hubs, but it does not contain the edge $f'\xrightarrow{\edge{IMPORTS}}f$. The incoming dependency neighborhood therefore brings $f'$ into the context of $c$ even though $f'$ itself was not changed. The same applies in the other direction when a changed file depends on an untouched file.

TGC also includes package-level context. Let $\operatorname{pkg}(f)$ denote the package containing file $f$. We define
\begin{equation}
    \mathcal{F}_{P}(c)
    =
    \left\{
        f'\notin\mathcal{F}_0(c)
        \;\middle|\;
        \exists f\in\mathcal{F}_0(c):
        \operatorname{pkg}(f')
        =
        \operatorname{pkg}(f)
    \right\}.
\end{equation}
The corresponding block $\mathbf{T}_{P}(c)$ therefore describes untouched files that belong to the same package as at least one file changed by the commit. This supplies package-level context even when no direct \edge{IMPORTS} relation exists between the two files.

Finally, TGC records where the relevant files lie in the project dependency graph. We compute PageRank and $k$-core values on the file-level dependency graph $\mathcal{D}^{(j)}$. PageRank measures the relative centrality of a file in the import structure, while its $k$-core value indicates whether it belongs to a more densely connected part of that graph. These graph-position features form $\mathbf{M}_1(c)$.

Combining the five feature groups gives
\begin{equation}
    \mathbf{T}(c)
    =
    \left[
        \mathbf{T}_{0}(c),
        \mathbf{T}_{1,\mathrm{dep}}(c),
        \mathbf{T}_{1,\mathrm{dry}}(c),
        \mathbf{T}_{P}(c),
        \mathbf{M}_{1}(c)
    \right].
\end{equation}
Their roles are summarized in~\autoref{tab:tgc-feature-groups}.

\begin{table}[!htbp]
\centering
\caption{Feature groups used by the TGC representation.}
\label{tab:tgc-feature-groups}
\small
\setlength{\tabcolsep}{4pt}
\renewcommand{\arraystretch}{1.12}

\begin{tabularx}{\columnwidth}{l X}
\toprule
\textbf{Group} & \textbf{Graph information used} \\
\midrule

$\mathbf{T}_0(c)$ &
Files directly changed by the target commit. \\

$\mathbf{T}_{1,\mathrm{dep}}(c)$,
$\mathbf{T}_{1,\mathrm{dry}}(c)$ &
Untouched one-hop import neighbors of the changed files, with importers and importees kept as separate directions. \\

$\mathbf{T}_{P}(c)$ &
Untouched files in the same package as the changed files. \\

$\mathbf{M}_1(c)$ &
Position of the file context in the project dependency graph, represented through PageRank and $k$-core. \\

\bottomrule
\end{tabularx}
\end{table}

Because projects differ in size and dependency structure, the same raw feature value can have a different scale from one project to another. Each scalar TGC feature is therefore supplied in both raw and standardized forms. For a feature $x(c)$,
\begin{equation}
    z_x(c)
    =
    \frac{
        x(c)-\mu_x^{(j)}
    }{
        \sigma_x^{(j)}
    },
\end{equation}
where $\mu_x^{(j)}$ and $\sigma_x^{(j)}$ are the running mean and standard deviation of that feature in the current project. The model therefore sees both the original value and how large or small it is relative to the project observed so far.

TGC augments the graph and semantic inference outputs rather than replacing them. Let $s_m(c)$ be the score of graph channel $m$ and let $G(c)$ be the semantic-channel score defined in~\autoref{app:g-channel-features}. The complete $F+G+T$ model is
\begin{equation}
\widehat{y}(c)
=
\sigma\left(
    \theta_0
    +
    \sum_{m\in F}\theta_m s_m(c)
    +
    \theta_G G(c)
    +
    \boldsymbol{\theta}_{T}^{\top}\mathbf{T}(c)
\right).
\end{equation}
The $F$ configuration uses the selected graph-channel scores, $G$ evaluates the semantic channel alone, $F+G$ combines the graph and semantic scores, and $F+G+T$ additionally includes the TGC feature block. As with the other learned components, fitting uses only the gap-resolved commits in $\mathcal{C}_{\mathrm{past}}$. The TGC-augmented stacking head uses standardized, class-balanced logistic regression with an $L_1$ penalty, $C=0.2$, and the \texttt{liblinear} solver.

The additional context does not require a full KG traversal every time a commit is scored. The TGC information is extracted for each project and cached, and per-commit scoring reads the corresponding feature values from this cache. TGC can therefore add dependency, package, and architectural context without placing a Neo4j traversal on the online prediction path.

\end{document}